\documentclass[aps,prd,10pt,twocolumn,superscriptaddress,nofootinbib,showkeys,showpacs,altaffilletter]{revtex4-1}

\usepackage{graphicx}
\usepackage{dcolumn}
\usepackage{amssymb}
\usepackage{amsmath}
\usepackage{amsfonts}
\usepackage{amsbsy}
\usepackage{color}
\usepackage{rotating}
\usepackage[english]{babel}
\usepackage{multirow}

\usepackage{caption}

\begin{document}

\title{Breaking the Vainshtein screening in clusters of galaxies}

\date{\today}

\author{Vincenzo Salzano}
\email{enzo.salzano@wmf.univ.szczecin.pl}
\affiliation{Institute of Physics, University of Szczecin, Wielkopolska 15, 70-451 Szczecin, Poland}
\author{David F. Mota}
\email{D.F.Mota@astro.uio.no}
\affiliation{Institute of Theoretical Astrophysics, University of Oslo, 0315 Oslo, Norway}
\author{Salvatore Capozziello}
\email{capozzie@na.infn.it}
\affiliation{Dipartimento di Fisica ``E. Pancini'' , Universita' degli Studi di Napoli ``Federico II'' and  INFN, Sezione di Napoli, Complesso
Universitario di Monte S. Angelo, Via Cinthia, Edificio N, 80126 Napoli, Italy}
\author{Megan Donahue}
\email{donahue@pa.msu.edu}
\affiliation{Physics and Astronomy Dept., Michigan State University, East Lansing, MI, 48824 USA}

\begin{abstract}
In this work we will test an alternative model of gravity belonging to the large family of galileon models. It is characterized by an intrinsic breaking of the Vainshtein mechanism inside large astrophysical objects, thus having possibly detectable observational signatures. We will compare theoretical predictions from this model with the observed total mass profile for a sample of clusters of galaxies. The profiles are derived using two complementary tools: X-ray hot intra-cluster gas dynamics, and strong and weak gravitational lensing. We find that a dependence with the dynamical internal status of each cluster is possible; for those clusters which are very close to be relaxed, and thus less perturbed by possible astrophysical local processes, the galileon model gives a quite good fit to both X-ray and lensing observations. Both masses and concentrations for the dark matter halos are consistent with earlier results found in numerical simulations and in the literature, and no compelling statistical evidence for a deviation from general relativity is detectable from the present observational state. Actually, the characteristic galileon parameter $\Upsilon$ is always consistent with zero, and only an upper limit ($\lesssim0.086$ at $1\sigma$, $\lesssim0.16$ at $2\sigma$, and $\lesssim0.23$ at $3\sigma$) can be established. Some interesting distinctive deviations might be operative, but the statistical validity of the results is far from strong, and better data would be needed in order to either confirm or reject a potential tension with general relativity.
\end{abstract}



\maketitle

\section{Introduction}
\label{sec:introduction}

The latest, most precise sets of measurements concerning the dynamics of our Universe, the second release of the \textit{Planck} satellite \citep{PlanckCosmo,PlanckMod}\footnote{For a fully-comprehensive look into all \textit{Planck} results, see \url{http://www.cosmos.esa.int/web/planck/pla}.}, have confirmed that the $\Lambda$CDM model has to be considered as the best model to explain most of the phenomena occurring in it. Nevertheless, it is undisputable that it also has many problems and caveats; a non exhaustive list can be found in \citep{BeyondLCDM} and references therein.

For what we are interested in, the $\Lambda$CDM paradigm is based on: the cosmological constant (CC), introduced to explain the accelerated expansion of our Universe detected for the first by means of Type Ia Supernovae in \citep{Riess98,Perlmutter99}; dark matter (DM), as the main ingredient of large scale structure formation and evolution; and on the acceptance of General Relativity (GR) as the theory of gravity. The intrinsic simplicity of the CC makes it difficult to confirm or refute on a statistical base; and we still lack a direct laboratory detection of one of the many suitable candidates for DM.

GR endures any challenge and passes any test it is undergone \citep{GW}. But both the DM and the CC problem might be closely connected due to the adoption of GR; thus, overcoming GR might help to solve them. Unfortunately, extension or modification of GR can be performed in too many ways; an interesting summary of most of the approaches still on study, is in Fig.~3 of \citep{BeyondLCDM} and in \citep{Clifton12}. The main problem is that GR is a very well-tested theory at Solar System scales \citep{Will14}, and this poses very strong and limiting bounds on any possible extension. Among the plethora of models that have been proposed so far, theories which exhibit a \textit{screening mechanism} are gaining much interest. Basically, most of such scenarios require a scalar field coupled to matter, and mediating a so-called ``fifth-force'' which, in principle, might span the entire range from Solar System up to cosmological scales. In regions of high density, this force has to be self-suppressed, so that no deviation from GR should be operative or, at least, if there was any, it should be hardly detectable \citep{Hamilton15}. Where the density is lower, the modification to GR should start to be effective and possibly some observational signature arises (what kind of and what order it might be, depends on the model).

The screening can be accomplished in many ways \citep{Joyce15,Berti15}: with a weak coupling between the field and matter in regions of high density, thus inducing a weak fifth-force as in symmetron theories \citep{Hint10,Hint11}; the field can acquire a large mass in high density environments, being short-ranged and undetectable, and be light and long-ranged in lower density regions, as for chameleon fields \citep{Khoury04A,Khoury04B,Brax04,Mota06,mota1} and $f(R)$ models \citep{DeFelice10,Sotiriou10,Capozziello11,Capozziello12}; or one may change the kinetic contribution of the field to the Lagrangian, with first or second order derivatives becoming important in a certain range, as it happens with the Vainshtein screening mechanism \citep{Vainshtein72}.

Among all them, we have decided to focus on the Vainshtein screening partially-driven and/or partially-broken (first discovered by \citep{Kobayashi15} and further discussed in \citep{Saito15}) by the so-called \textit{galileon} fields which, as pointed out in \citep{KoyamaSakstein2015}, are only the most common approach for it, but not the only one. \textit{Galileon} fields are so defined because, by construction, they are invariant under the galilean shift symmetry
\begin{equation}
\phi(x) \mapsto \phi(x) + c + b_{\mu}x^{\mu} \; .
\end{equation}
Given this property, the peculiarity of galileon models is that, although being higher-derivative field theories, they still have second order equations of motion. Since their first introduction in \cite{Nicolis09}, galileons have been studied in many works \citep{Deffayet09,Deffayet11,Deffayet13}, so that we have now a fully-comprehensive analysis which has helped to give them the right place in the very extended branch of theoretical alternative to GR \citep{Horndeski1974,Gleyzes15A,Gleyzes15B}. In this work we will deal with a relatively new formulation of galileon fields, as presented in \citep{KoyamaSakstein2015}.

Galileon models (and Vainshtein screening) have also been tested against observational data \citep{Nesseris10,Barreira12,Barreira13A,Barreira13B,Barreira14B,Barreira14C,Barreira15A,Barreira15B,Brax15A}, and using numerical simulations \citep{Barreira13C,Li13A,Li13B,Barreira14A,Falck14,Barreira15C,Falck15}. This point is fundamental because when dealing with alternative models of gravity, the main problem is to find some clear signature which robustly discriminates between GR and a competing alternative candidate. It would be helpless to have a theory which includes GR, solves some of its problems, and fits data with the same statistical accuracy, but we had no smoking-gun to clearly differentiate it from GR. Examples of such possible probes are in \citep{Brax11,Davis12,Hui12,Hellwing14,mota2,Brax15B}. In this respect, clusters of galaxies are one of the most interesting tools to be used for such analysis: being at the border of the astrophysical and cosmological regimes \citep{Voit05,Clowe06,Brownstein07,Merten11,Jee12,Clowe12,Jee14}.

Through this work, when necessary, we will assume a fiducial cosmological background described by a $\Lambda$CDM model with $\Omega_{m} = 0.27$, $\Omega_{DE} = 0.73$, and $H_{0} = 70$ km s$^{-1}$ Mpc$^{-1}$ (as we will explain in next sections, this choice has very negligible impact on our study, and it is not in contrast with our adoption of an alternative theory of gravity).

\section{Theory}
\label{sec:theory}

In this section we will introduce the theoretical model we have chosen for our analysis. We will describe only the properties which are important for our goals, and the differences with previous similar approaches; the interested reader can find more details about it in \cite{KoyamaSakstein2015} and references therein.

Following \citep{Deffayet13}, the most general galileon Lagrangian in four dimensions and flat space-time can have only up to five different terms; using the notation of \citep{KoyamaSakstein2015} it can be written as:
\begin{equation}
\mathcal{L} = M^{2}_{Pl} \sum_{i} \frac{\mathcal{L}_{i}}{\Lambda^{2(i-2)}_{i}} + \alpha \phi T + \frac{T^{\mu\nu} \partial_{\mu}\phi\partial^{\nu}\phi}{\mathcal{M}^{4}} \; ,
\end{equation}
where $T^{\mu\nu}$ is the energy-moment tensor in the matter sector; $T$ its trace; $\phi$ is the galileon field; $\alpha$ is a possible coupling between the galileon field and matter; $\mathcal{M}$ is a possible coupling to the kinetic part of the field; $\Lambda$ is a mass dimension scale/constant \citep{KoyamaNiz}, which might be associated, for example, to the current accelerated expansion of the Universe (in which case $\Lambda \sim (M_{Pl} H^{2}_{0})^{1/3}$); and $\mathcal{L}_{i}$ are the five Lagrangian functions depending on the galileon field and its kinetic contribution:
\begin{eqnarray}\label{eq:galileon_lagrangian}
\mathcal{L}_{1} &\equiv& \phi \\
\mathcal{L}_{2} &\equiv& X \nonumber \\
\mathcal{L}_{3} &\equiv& X \square \phi - \phi_{\mu}\phi^{\mu\nu}\phi_{\nu} \nonumber \\
\mathcal{L}_{4} &\equiv& -X \left[ \left(\square \phi \right)^{2} - \phi_{\mu\nu} \phi^{\mu\nu} \right] \nonumber \\
&-& \left( \phi^{\mu} \phi^{\nu} \phi_{\mu\nu} \square \phi - \phi^{\mu} \phi_{\mu\nu} \phi_{\rho} \phi^{\rho\nu}\right) \nonumber \\
\mathcal{L}_{5} &\equiv& -2X \left[ \left(\square \phi \right)^{3} - 3\phi_{\mu\nu} \phi^{\mu\nu} \square \phi + 2\phi_{\mu\nu}\phi^{\nu\rho}\phi{\mu}_{\rho}\right] \nonumber \\
&& - \frac{3}{2} \left( \left(\square \phi \right)^{2} \phi^{\mu}\phi^{\nu}\phi_{\mu\nu} -2\phi_{\mu}\phi^{\mu\nu}\phi_{\nu\rho}\phi^{\rho} \right. \nonumber \\
&& - \left. \phi_{\mu\nu}\phi^{\mu\nu}\phi_{\rho}\phi^{\rho\sigma}\phi_{\sigma} + 2\phi_{\mu}\phi^{\mu\nu}\phi_{\nu\rho}\phi^{\rho\sigma}\phi_{\sigma}\right) \nonumber \;
\end{eqnarray}
where $\phi_{\mu_{1}\ldots\mu_{n}} \equiv \nabla_{\mu_{1}}\ldots\nabla_{\mu_{n}}\phi$ and $X \equiv -1/2 \partial_{\mu}\phi \partial^{\mu}\phi$ is the standard kinetic term. In \citep{KoyamaSakstein2015}, they do not assume any coupling $\alpha$, but note instead that, after the quartic term is covariantized, the galileon appears to be coupled to the curvature tensor; thus the total Lagrangian looks equivalent to the one analyzed in \citep{Barreira12} and companion papers.

In particular, in \citep{KoyamaSakstein2015} the authors focus on a relatively new sub-class of this family, defined by the Lagrangian
\begin{equation}
\frac{\mathcal{L}}{\sqrt{-g}} = M^2_{Pl} \left[ \frac{R}{2} - \frac{1}{2} \partial_{\mu}\phi \partial^{\mu}\phi + \frac{\mathcal{L}_{4}}{\Lambda^4} \right]\; ,
\end{equation}
where $g$ is the determinant of the metric and $R$ is the Ricci scalar. Following the notation introduced by \citep{Gleyzes15B,KoyamaSakstein2015}, we will call such a model $G^3$-galileon. It is important to point out that the reduced Planck mass appearing in the Lagrangian is defined as $M_{Pl} = (8\pi G)^{-1}$, where $G$ is the bare gravitational constant and differs from the usually measured one, $G_{N}$.

Assuming a metric signature $(-, +, +, +)$, and the Newtonian Gauge, the perturbed Friedmann-Lema\^{i}tre-Robertson-Walker metric can be written as
\begin{eqnarray}\label{eq:metric_perturb}
\mathrm{d} s^2 &=& -\left[1+2\frac{\Phi(r,t)}{c^2}\right] c^2 dt^2 \\
&+& a^2(t)\left[1-2\frac{\Psi(r,t)}{c^2}\right]\delta_{ij} \mathrm{d}x^i \mathrm{d}x^j\; ,
\end{eqnarray}
where $c$ is the speed of light; $a$ is the cosmological scale factor; and $\Phi$ and $\Psi$ are the gravitational and the metric potentials. After having defined the parameter
\begin{equation}
\Upsilon \equiv \left(\frac{\dot{\phi}_{0}}{\Lambda}\right)^{4} \; ,
\end{equation}
the model can be fully characterized by the following equations:
\begin{equation}\label{eq:potential_phi}
\frac{\mathrm{d} \Phi(r)}{\mathrm{d} r} = \frac{G_{N} M(r)}{r^2} + \frac{\Upsilon}{4} G_{N} M''(r) \; ,
\end{equation}
\begin{equation}\label{eq:potential_psi}
\frac{\mathrm{d} \Psi(r)}{\mathrm{d} r} = \frac{G_{N} M(r)}{r^2} - \frac{5\Upsilon}{4} \frac{G_{N} M'(r)}{r} \, ,
\end{equation}
where the measured gravitational constant $G_{N}$ is defined as
\begin{equation}
G_{N} = \frac{G}{1 + 5 \Upsilon} \; ,
\end{equation}
and the mass enclosed in a radius $r$, $M(r)$, and its derivatives with respect to the distance, are:
\begin{eqnarray}\label{eq:mass_derivatives}
M(r) &=& \int^{r}_{0} 4\pi r'^{2} \rho(r') dr' \nonumber \\
M'(r) &=& 4\pi r^2 \rho(r) \\
M''(r) &=& 8\pi r \rho(r) + 4\pi r^2 \frac{d \rho}{d r} \nonumber \,.
\end{eqnarray}

As it can be easily seen from Eqs.~(\ref{eq:potential_phi}),~(\ref{eq:potential_psi}) and (\ref{eq:mass_derivatives}), we have two main unknown quantities: the theoretical parameter $\Upsilon$, and the matter density $\rho$, which we need to define.

The parameter $\Upsilon$ can be recognized as quantifying the deviation of our $G^{3}$-galileon model from GR, which is recovered for $\Upsilon = 0$. It is worth to point out that the model under consideration has one intrinsic theoretical parameter $\Upsilon$, but this is only the simplest version: it can be generalized to the case of two constants $\Upsilon_1$ and $\Upsilon_2$, for $\Phi$ and $\Psi$ respectively (while writing this work, a paper on this topic has appeared \citep{Sakstein16}). Current bounds on such a parameter are given in: \citep{SaksteinPRL15,SaksteinPRD15}, where the specific model we are considering here is tested against red dwarf stars and their minimum mass for hydrogen burning; \citep{Jain16}, where white dwarf stars are used; and \citep{Babichev16} where the generalized model with two constants is considered.

For what concerns the matter density $\rho$, as we are going to study clusters of galaxies, we have to specify their three components: DM, gas, and galaxies. The DM distribution can only be inferred indirectly from observations, and is generally described by phenomenological profiles mainly derived from numerical simulations. The gas mass can contribute up to $10-15\%$ of the total mass budget; due to internal dynamics, gas is mainly detectable at X-ray wavelength, and can be described quite well, when in hydrostatic equilibrium, by the so-called $\beta$-model \citep{Cavaliere78}, which is often modified in order to account for some peculiar dynamical behaviours which take place in the inner regions of the clusters. Finally, galaxies cannot be properly described by a continuous density function \citep{Bahcall95}, at least, on the entire spatial scale of the cluster; in \cite{Newman13} the central profile of the brightest cluster galaxy is inferred mostly by stellar kinematics using long-slit spectroscopy.

We will follow the usual custom to model the total matter density of the cluster by with a Navarro-Frenk-White (NFW) profile \citep{NFW}, given by
\begin{equation}\label{eq:NFW}
\rho_{NFW} = \frac{\rho_s}{(r/r_s)(1+r/r_s)^2}\; ,
\end{equation}
where the only free parameters are a density $(\rho_{s})$ and a scale $(r_{s})$.

As an illustrative case, also discussed in \citep{KoyamaSakstein2015}, we calculate the gravitational and metric potentials, assuming a NFW profile, from Eqs.~(\ref{eq:potential_phi})~-~(\ref{eq:potential_psi}):
\begin{equation}\label{eq:grav_pot}
\Phi(r) = -\frac{4 G_{N} \pi  r_{s}^3 \rho_{s}}{r} \left[ \log \left(1+\frac{r}{r_{s}}\right) - \frac{\Upsilon}{4}\frac{r^2}{(r+r_{s})^2} \right]
\end{equation}
\begin{equation}\label{metric_pot}
\Psi(r) = -\frac{4 G_{N} \pi  r_{s}^3 \rho_{s}}{r} \left[ \log \left(1+\frac{r}{r_{s}}\right) - \frac{5\Upsilon}{4}\frac{r}{r+r_{s}} \right] .
\end{equation}
It is clear that the $G^{3}$-galileon model predicts ``less gravity'' than GR (the same conclusion is obtained for a - up to some extent - similar \textit{quartic} galileon model in \cite{Barreira14A}). The reason for such a behaviour is easily explained; as pointed out also in \citep{KoyamaSakstein2015}, the final corrective term in $\Phi$ is typically negative, provided that the density decreases outwards \textit{as} a NFW profile, and the parameter $\Upsilon$ is positive. We can check this even closer, if we calculate the first and second order derivative of the mass for a NFW profile, given by
\begin{eqnarray}
M'(r) &=& \frac{4 \pi r r_{s}^3 \rho_{s}}{(r + r_{s})^2} \\
M''(r) &=& \frac{4 \pi r_{s}^3 (r_{s}-r) \rho_{s}}{(r + r_s)^3} \, . \nonumber
\end{eqnarray}
It can be easily checked that $M''(r) >0$ for $r<r_{s}$: at least in principle, in this range we may need more dark matter in order to fit observations. The same is true for $\Psi$: $M'(r)$ is always greater than zero, leading to the same ``less gravity'' conclusion. In Sec.~(IV.A) of \citep{KoyamaSakstein2015}, the authors qualitatively discuss the possible implications of this property on rotation curves of spiral galaxies, which depend on $\mathrm{d}\Phi/\mathrm{d}r$. We add here that the same conclusions would apply to the case of mass reconstruction of clusters of galaxies using X-ray hot intra-cluster gas observations, because, as we will discuss in next section, this latter mass estimation is related to the derivative of the potential $\Phi$. And it also applies to lensing mass reconstruction, because it depends on the derivative of both $\Phi$ and $\Psi$. Then, we can finally conclude that the $G^{3}$-galileon model will require a larger amount of dark matter, with respect to GR, in order to match both X-ray mass and strong/weak lensing observations in clusters of galaxies. We can also check that our model has $\Phi \neq \Psi$, as long as $\Upsilon \neq 0$, differently from GR, for which the condition $\Phi = \Psi$ holds.

This is an interesting difference with respect to, for example, \citep{Barreira15B} where a galileon model is studied using the same lensing data we will consider in this work. From their Eq.~(20), from the bottom panel of their Fig.~(1), and from the right panel of their Fig.~(9), where the same quantity of our Eq.~(\ref{eq:potential_phi}) is shown, it can be easily checked that the extra terms in the gravitational potential give a negative contribution all over the testable distance range; thus, the model predicts a slightly stronger gravity, i.e., can possibly mimic a small amount of DM. Actually, this can be also verified by inspecting the mass values estimates given in their Table II, where the galileon-derived masses are slightly lower than the GR-derived ones; even if, after taking into account statistical errors, the galileon model finally does not predict a significant departure from GR, at least in the range of scales covered by data. Moreover, such model implies that $\Phi = \Psi$, exactly like in GR.

On the other hand, in \citep{TerukinaLombriser2015}, we have a galileon model which, in principle, may allow both positive and negative extra terms to the gravitational potential, and which has $\Phi \neq \Psi$, like in our case. At the end, after using a set of combined gas and lensing data, the model seems to favour positive values, which our model is forced to have by default. It is also true that the constraints are statistically very weak, maybe because the considered cluster, the Coma cluster, is not the best target for this kind of analysis, being not a relaxed system, with observations suggesting substructures and orientation dependence \citep{TerukinaLombriser2014}.

To summarize, in this work we will try to make some small further steps toward a more complete view:
\begin{itemize}
  \item we will consider a model with $\Phi \neq \Psi$; thus, a departure from GR, if any, might be found from the combined use of both X-ray observations and gravitational lensing;
  \item we will apply this model on an extended set of clusters of galaxies, in order to enforce the statistical validity of our results, and have a broader and more general picture;
\end{itemize}

\section{Data}
\label{sec:data}

We will approach the problem of reconstructing the mass distribution in clusters of galaxies using two different astrophysical probes:
\begin{itemize}
  \item X-ray observations of the hot intra-cluster gas;
  \item detection and analysis of strong and weak gravitational lensing.
\end{itemize}

In \citep{Donahue14} and \citep{Merten15} two samples of clusters, with a large overlap, have been analyzed, respectively, using properties of the X-ray emitting hot intra-cluster gas, and through the analysis of gravitational lensing events produced by each cluster. The sample has been observed within the survey program Cluster Lensing and Supernova survey with Hubble, CLASH \citep{Postman12}, a multi-cycle treasury program which, using $524$ Hubble Space Telescope (HST) orbits, has targeted $25$ galaxy clusters. Among them, $20$ clusters were selected mainly following the criterium of an approximately unperturbed and relatively symmetric X-ray morphology. Possibly, they constitute a reference sample of clusters with regular density profiles that might be proven to be an optimal ruler to compare models of cosmological structure formation, to test $\Lambda$CDM predictions and, eventually, to test possible departures from GR. These same X-ray selected clusters have been extensively studied in the context of weak-lensing and magnification analysis \citep{Umetsu14,Umetsu15}, and compared with identically-selected clusters derived from a $\Lambda$CDM simulation \citep{Meneghetti14}.

Thus, we have the possibility to analyze the same clusters with both, an X-ray and lensing approach, in a consistent way, and to compare the results. This is very important because these two observational tools are sensitive in different ways to the gravitational potential of the cluster: lensing delivers a nearly unbiased estimate of the total mass, although with significant scatter from line-of-sight matter distribution out of the cluster; X-ray gas dynamics has less statistical scatter, but can be disturbed by local dynamics, with gas not in hydrostatic equilibrium, mainly in the inner regions ($<100$ kpc), thus leading to wrong mass estimations.

Moreover: X-ray gas, as non-relativistic matter, is sensitive to the gravitational potential $\Phi$ only; while lensing (photons) is sensitive to the  combination of both the potentials, $\Phi+\Psi$. This is very important in our case, because we want to test a model for which $\Phi \neq \Psi$. In GR the two potentials are equal. This implies that in a combined analysis of X-ray and lensing observations, possible divergences between the two mass estimations %(lead, for example, by the shortcomings induced by perturbations to the gas equilibrium)
are automatically shared between the two probes. And, eventually, inconsistencies can become dominant. Instead, in models like the $G^{3}$-galileon, these problems might be alleviated: the information from X-ray gas only involves $\Phi$, with $\Psi$ still having some freedom to adjust lensing expectations to observations.

\subsection{X-ray hot gas}

All the CLASH clusters have been selected using \textit{Chandra} telescope observations; in \citep{Donahue14}, archival data from both \textit{Chandra} and XMM are used to estimate the total mass of the clusters from X-ray observations, and to compare them with gravitational-lensing estimations. These archival data (we will focus only on \textit{Chandra}-derived data because they are available for all the clusters in the sample) are reprocessed, re-calibrated and analyzed using the procedure outlined in \citep{Donahue14}.

When working with X-ray observations of clusters of galaxies, some preliminary hypothesis are generally needed, the main ones being the assumption of spherical symmetry and that the system is in hydrostatic equilibrium (HSE); such assumptions are made also in \citep{Donahue14}. Then, starting from these hypothesis, we use the collisionless Boltzmann equation
\begin{equation}\label{eq:boltzmann_equation}
-\frac{{\mathrm{d}}\Phi(r)}{{\mathrm{d}}r} = \frac{k T_{gas}(r)}{\mu m_{p} r}\left[\frac{{\mathrm{d}}\ln\rho_{gas}(r)}{{\mathrm{d}}\ln r} +\frac{{\mathrm{d}}\ln T_{gas}(r)}{{\mathrm{d}}\ln r}\right] \; ,
\end{equation}
from which, in GR, we can simply obtain:
\begin{eqnarray}\label{eq:mass_boltzmann_equation}
M_{tot}(r) &=& M_{gas}(r) + M_{gal}(r) + M_{DM}(r) = \\
&-& \frac{k T_{gas}(r)}{\mu m_{p} G_{N}} r \left[\frac{{\mathrm{d}}\ln\rho_{gas}(r)}{{\mathrm{d}}\ln r}+\frac{{\mathrm{d}}\ln T_{gas}(r)}{{\mathrm{d}}\ln r}\right] \nonumber \; .
\end{eqnarray}
From the right-hand-side, making direct use of observations (gas density and temperature profiles), one can obtain the total mass in the cluster, $M_{tot}$; of course, from the observed density $\rho_{gas}$, one can also derive the mass of the hot gas, $M_{gas}$. Thus, Eq.~(\ref{eq:boltzmann_equation}) is used to indirectly infer properties of the dark matter halo embedding the cluster, $M_{DM}$. It is better to highlight here that the mass estimated through Eq.~(\ref{eq:boltzmann_equation}) should be more properly named as \textit{thermal} pressure-supported HSE mass; but \textit{non-thermal} contributions might arise as, for example, among other, bulk motions, turbulences, cosmic rays, and magnetic fields. In GR, the \textit{non-thermal} contribution are derived and parameterized from numerical simulations; thus, in order to be taken in consideration when an alternative gravity scenario is studied, one should, in principle, run the same simulations and find for a new parametrization \citep{TerukinaLombriser2014,TerukinaLombriser2015,Wilcox2015}. %Moreover, in this work, in the original data which have been used, no \textit{non-thermal} contribution has been considered in the early stages; thus, we will focus only on the thermal mass, using Eq.~(\ref{eq:boltzmann_equation}) for our purposes.

The approach followed by \citep{Donahue14} is slightly different: they use the Joint Analysis of Cluster Observations (JACO) code from \citep{Mahdavi07}, which may provide a simultaneous fit of many kinds of observations related to clusters of galaxies, like X-ray, Sunyaev-Zeldovich and weak-lensing data, once parametric models for matter components are considered. In \citep{Donahue14}, only the X-ray ones are used. JACO starts from assuming separate matter components (in our case, DM and gas; stellar contribution might also be considered, but this is not done in our case), and from them it directly calculates synthetic multi-wavelength spectra which are then compared with the observed ones. Thus, it performs a fit of the spectra directly in terms of the interested theoretical parameters (NFW parameters, for example), and quantities like the temperature are not directly needed from measurements, but, once matter components are given, and the HSE is assumed, they can be calculated as pure theoretical quantities. Then, the constraints from JACO might be stronger than the approach based on the use of Eq.~(\ref{eq:boltzmann_equation}), but at the expense of a partial loss of model-independency, because some parametric model has to be assumed for the mass distribution.

In \citep{Donahue14}, as input for JACO, a NFW DM profile is assumed, as in Eq.~(\ref{eq:NFW}); while for the gas they use a triple $\beta$-model \citep{Cavaliere78}, with one component being truncated at low-radius by a power-law,
\begin{eqnarray}
\rho_{gas} &=& \rho_{0} \left( \frac{r}{r_{0}} \right)^{-\alpha}  \left[1 + \left(\frac{r}{r_{e,0}}\right)^{2} \right]^{-\frac{3\beta_{0}}{2}} \\
&+& \sum^{2}_{i=1}  \rho_{i} \left[1 + \left(\frac{r}{r_{e,i}}\right)^{2} \right]^{-\frac{3\beta_{i}}{2}}\; , \nonumber
\end{eqnarray}
where $\rho_{0},\rho_{i},r_{e,0},r_{e,i},\beta_{0},\beta_{i}$ are determined by fitting the spectra as described above.
The final JACO constraints on the total mass are obtained from a Monte Carlo Markov Chain (MCMC) procedure. The final output, the total cluster mass of the cluster, $M_{tot}$, can than be used to constrain our $G^{3}$-galileon model through the relation
\begin{equation}\label{eq:total_mass_mod}
M_{tot} = \frac{r^2}{G_{N}} \frac{\mathrm{d}\Phi}{\mathrm{d}r}\; ,
\end{equation}
which is valid regardless the theory of gravity. What is going to change is, of course, the content of the right hand side which, in our alternative scenario, is given by Eq.~(\ref{eq:potential_phi}), and the possible interpretation of some ``modified gravity aspects'' as a sort of \textit{effective} mass (if it is the case).

\subsection{Gravitational lensing}

In \citep{Merten15} many clusters in common with the sample of \citep{Donahue14} are used for a lensing-based study; in particular, the lensing analysis is extended by combining weak-lensing constraints from the HST and from ground-based wide-field data with strong lensing constraints from HST.

In\citep{Schneider92,Bartelmann01}, the typical configuration of a gravitational lensing system comprises a source, positioned at a (angular diameter) distance from the observer, $D_{s}$, and a lens (in this case a cluster), situated at a distance, $D_{l}$, with the distance between the lens and the source generally indicated as $D_{ls}$. In the cluster regime, the non-relativistic gravitational potential $\Phi$ and the peculiar velocity of the lens are small, and one can presume that a locally flat space-time is being disturbed by the potential $\Phi$, with a metric given by Eq.~(\ref{eq:metric_perturb}). Moreover, given that the distances observer-lens and lens-source are much larger than the physical extension of the lens, the latter can be thus approximated as a two-dimensional system (``thin-lens'' approximation). The main effect of the lens, in such regime, is to deflect light rays from the source by a certain angle $\hat{\vec{\alpha}}$, which, in GR, can be defined as
\begin{equation}
\hat{\vec{\alpha}} = \frac{2}{c^2} \int^{+\infty}_{-\infty} \vec{\nabla}_{\perp} \Phi \, \mathrm{d} z \;
\end{equation}
with $\vec{\nabla}_{\perp}$ being the two-dimensional gradient operator perpendicular to light propagation. The relation between the real (unknown) position of the source and the apparent one, then, can be obtained by simple geometrical considerations to be
\begin{equation}
\vec{\beta} = \vec{\theta} - \frac{D_{ls}}{D_{s}} \hat{\vec{\alpha}}(\vec{\theta}) = \vec{\theta} - \vec{\alpha}(\vec{\theta}) \; ,
\end{equation}
where $\vec{\beta}$ is the original (two-dimensional) vector position of the source, $\vec{\theta}$ is the new apparent position, and $\hat{\vec{\alpha}}$ ($\vec{\alpha}$) is the deflection (scaled) angle. In a more general (relativistic) context, this equation can be generalized to large deflection angles too, as pointed out in \citep{Dabrowski00}.

Finally, the angle deflection can be expressed in terms of the \textit{effective lensing potential}, $\Phi_{lens}$, i.e. the line-of-sight projection of the full three-dimensional gravitational potential of the cluster on the lens plane, properly rescaled:
\begin{equation}
\Phi_{lens}(R) = \frac{2}{c^2} \frac{D_{ls}}{D_{l}D_{s}}\int^{+\infty}_{-\infty} \Phi(R=D_{l}\theta,z) \, \mathrm{d} z \; ,
\end{equation}
where $R$ is the two-dimensional projected radius and $z$ is the line of sight direction.

It is easy to verify that the gradient of $\Phi_{lens}$ gives the scaled deflection angle $\vec{\alpha}$, i.e. $\vec{\nabla} \Phi_{lens} = \vec{\alpha}$. Another important relation, is given by the Laplacian of the same potential, which results to be equal to the \textit{convergence} $\kappa$,
\begin{equation}\label{eq:kappa_general}
\kappa(R) =  \frac{1}{c^{2}} \frac{D_{l}D_{ls}}{D_{s}} \int^{+\infty}_{-\infty} \nabla_{r} \Phi(R,z)\, dz \, ,
\end{equation}
where, again, $R$ is the two-dimensional projected radius; $z$ is the line of sight direction; $r = \sqrt{R^2+z^2}$ is the three-dimensional radius; $\nabla_{r}$ is the Laplacian operator in spherical coordinates; and $c$ is the speed of light. The potential $\Phi$ satisfies the Poisson equation,
\begin{equation}
\nabla^{2}_{r} \Phi(r) = 4 \pi G_{N} \rho(r) \; ,
\end{equation}
and, using it in Eq.~(\ref{eq:kappa_general}), we obtain the final expression for the convergence in GR:
\begin{equation}\label{eq:kappa_GR}
\kappa(R) = \int^{+\infty}_{-\infty} \frac{4\pi G_{N}}{c^2}\frac{D_{l}D_{ls}}{D_{s}} \rho(R,z) dz \quad \equiv \quad \frac{\Sigma}{\Sigma_{crit}}\; ,
\end{equation}
with the surface density of the lens defined by
\begin{equation}
\Sigma = \int^{+\infty}_{-\infty}  \rho(R,z) dz\;
\end{equation}
and the \textit{critical} surface mass density for lensing defined as
\begin{equation}
\Sigma_{crit} = \frac{c^2}{4\pi G_{N}}\frac{D_{s}}{D_{l}D_{ls}}\; .
\end{equation}
Finally, it results that the convergence is nothing more than the two-dimensional projected total matter density of the lens. Actually, all the previous relations have been obtained assuming GR; but the most general expression for the convergence, irrespective of the gravity theory, is
\begin{equation}\label{eq:kappa_general_mod}
\kappa(R) =  \frac{1}{c^{2}} \frac{D_{l}D_{ls}}{D_{s}} \int^{+\infty}_{-\infty} \nabla_{r} \left( \frac{\Phi(R,z) + \Psi(R,z)}{2}\right)\, dz \, ,
\end{equation}
with $\Phi$ and $\Psi$ the total gravitational and metric potentials. In GR, as known, $\Phi = \Psi$, and we obtain Eq.~(\ref{eq:kappa_GR}); but in general, they can be different, as it is the case of our $G^3$-galileon model. Thus, for our analysis, we will use directly the more general definition, Eq.~(\ref{eq:kappa_general_mod}).

The importance of gravitational lensing, in the context of searching for confirmation or rebuttal of a model alternative to GR, is thus strikingly plain: it is sensitive to both the potentials and, in principle, could help to detect if they are equal or not. Future planned surveys like ESA satellite mission \textit{Euclid}\footnote{\url{http://sci.esa.int/euclid}} \citep{Laureijs09,Laureijs11,Refregier10,Amendola13}, will take advantage of it. In our case, such use would be even stronger and more decisive if combined to other complementary \textit{independent} probes (like dynamics of hot gas) which are, instead, sensitive to only the gravitational potential $\Phi$. The combined use of both might help to disentangle the contributions from both potentials and state, with more or less statistical evidence, if GR or an alternative theory is feasible or not.

In order to use Eq.~(\ref{eq:kappa_general_mod}), we need the convergence from the data and the two potentials from theory. For what concerns the latter point we can rely on the definition of the Laplacian operator in spherical coordinates, $\nabla_{r} = \frac{\partial^{2}}{\partial r^{2}} + \frac{2}{r} \frac{\partial}{\partial r}$, and use directly Eqs.~(\ref{eq:potential_phi})-(\ref{eq:potential_psi}); than, after providing a functional form for the matter density (in our case we will use a NFW DM profile), the integral in Eq.~(\ref{eq:kappa_general_mod}) can be calculated numerically.

The data are provided by \citep{Merten15}, where the cluster selected by CLASH are analyzed. Lensing events are retrieved from the HST field of the CLASH program,, which provides lensing constraints both for strong and weak lensing; these weak lensing maps are also combined with ground-based catalogs, mostly from Suprime-Cam on the Subaru Telescope. Given the properties of the CLASH survey, new lensed galaxies can be identified and their redshifts measured with greater accuracy, with a net improvement of the signal with respect to previously available literature. In order to infer matter distribution from lensing events, the Strong-and Weak Lensing (SaWLens) algorithm is used \citep{Merten09}. The main property of this approach is that no \textit{a priori} assumption is made about the mass distribution (by contrast, in JACO, you need to input a functional form). More details about the application of this method to the CLASH data are in \citep{Merten15}.

On last point should be addressed here: in principle, in order to calculate the convergence, we need some information from the \textit{cosmological background}, as it involves the calculation of the angular diameter distances observer-lens, observer-source, and lens-source. Angular diameter distance are generally defined as:
\begin{equation}
D_{A}(z) = \frac{1}{1+z}\int^{z}_{0} \frac{c \; \mathrm{d}z'}{H(z',\boldsymbol{\theta})} \; ,
\end{equation}
where the Hubble function $H(z,\boldsymbol{\theta})$ depends on the particular model one considers through the set of vectors $\boldsymbol{\theta}$. As for galileon models, in \citep{Barreira15B} there is a comparison between the $\Lambda$CDM and the galileon version. Then, one should first consider how the $G^{3}$-galileon model behaves at cosmological scales, and use its $H(z)$ expression in the lensing analysis in order to calculate $D_{l},D_{s}$ and $D_{ls}$. We lack such cosmological scale analysis for our model; but one could also question if these distances might have some influence on putting bounds on the theoretical parameters $\boldsymbol{\theta}$. In principle it would be possible, but we are highly confident this is not the case.

First of all, \citep{Barreira14C} show that the final expression for $H(z,\boldsymbol{\theta})$ for a galileon scenario does not depend on galileon parameters, but only on the total matter content, $\Omega_{m}$ (the dimensionless matter density parameter today). Furthermore, \citep{Barreira15B} show that the value of $\Omega_{m}$, one of the parameters which enter in $H(z)$, is unchanged when moving from $\Lambda$CDM (GR) to galileon, and is basically non influential both in the determination of the mass profiles and in the constraints on the galileon parameters. Moreover, we do not believe that the distances might help to constrain such parameters: cosmological geometrical probes are generally very weak when used to compare $\Lambda$CDM (GR) with other alternative models; we have no \textit{observational errors} on these distances which, thus, in principle, could re-scale in a completely free and un-physical way. Finally, in \citep{Barreira14C} it is shown that the Hubble function $H(z)$ derived from their galileon model can differ from the expected $\Lambda$CDM behaviour for $\lesssim5\%$ in the redshift range covered by our data; a variation which is smaller than present observational errors and dispersion and, thus, still not detectable in a statistically valid way nowadays. It is worth to stress that results from \citep{Barreira14C} are obtained using only \textit{Planck} CMB and some BAO data, which very likely pinpoint high redshifts regime behaviour quite well, but not equally well the lower one. The fit would have surely benefit (and maybe reduced the deviation from the baseline $\Lambda$CDM background) from considering two further elements: SNeIa, which are well known to play a complementary role with respect to BAO in fixing many cosmological parameters; and by applying a prior on $H_{0}$ from independent observations, given that present errors on $H_{0}$ are $\sim 2\%$ \citep{Riess16}, which is less than half the deviation from $\Lambda$CDM depicted in the same \citep{Barreira14C}. For all these reasons, we will use the fiducial cosmological background we have defined in the Introduction to calculate the critical surface mass density.

\subsection{Methodology}

In order to perform the statistical analysis of our model, we have built the respective $\chi^2$ function for each set of observations and for each cluster. In the case of hot X-ray gas mass profiles, the $\chi^{2}_{gas}$ is defined as
\begin{equation}
\chi^2_{gas}= \sum_{i=1}^{\mathcal{N}} \frac{\left( M^{theo}_{tot}(r_{i},\boldsymbol{\theta})-M^{obs}_{tot}(r_{i}) \right)^{2}}{\sigma^2_{M^{obs}_{tot}}(r_{i})} \; ,
\end{equation}
where: $\mathcal{N}$ is the number of points, for each cluster, for which measurements of the total mass as a function of radius are provided; $r_{i}$ is the distance from the center of the cluster; $M^{obs}_{tot}$ is the total mass finally obtained from JACO (the left hand side of Eq.~(\ref{eq:total_mass_mod})); $\sigma_{M^{obs}_{tot}}$ are the observational errors on the total mass; $M^{theo}_{tot}$ is the total mass calculated from our model from the right hand side of Eq.~(\ref{eq:total_mass_mod}); and $\boldsymbol{\theta}$ is the vector of the model parameters. As we are going to consider that all the mass in the cluster is described in terms of a NFW DM profile, this vector will be, $\boldsymbol{\theta}=\{\rho_{s},r_{s}\}$, for GR, and $\boldsymbol{\theta}=\{\rho_{s},r_{s}, \Upsilon\}$ for the $G^3$-galileon. We could have used the gas mass estimated by JACO from \citep{Donahue14}, but found it problematic because the gas density is very likely to depend not only on the global gravitational potential $\Phi$, but also on local dynamics. We have checked that, even when taking into account gas density, the results do not change in a statistically significant way.

For the lensing, the $\chi^{2}_{lens}$ is defined as
\begin{equation}
\chi^2_{lens} = \boldsymbol{(\kappa^{theo}(\theta)-\kappa^{obs})}\cdot\mathbf{C}^{-1}\cdot\boldsymbol{(\kappa^{theo}(\theta)-\kappa^{obs})} \; ,
\end{equation}
where: $\boldsymbol{\kappa^{obs}}$ is the vector of the observationally measured convergence; $\boldsymbol{\kappa^{theo}(\theta)}$ is the theoretical convergence obtained from the right hand side of Eq.~(\ref{eq:kappa_general_mod}); and $\mathbf{C}$ is the related observational covariance matrix.

The total $\chi^2$, defined as the sum of the gas and lensing $\chi^2$, is minimized by a Markov Chain Monte Carlo (MCMC) method, and its convergence is checked using the method developed by \citep{Dunkley05}. The main outputs of the MCMC are used to recover the $68\%$ and $95\%$ marginalized constraints of the theoretical parameters, and can also help us to assess how much the $G^{3}$-galileon model compares to GR. This is done by calculating the Bayesian Evidence for both GR and $G^{3}$-galileon for each cluster, using the algorithm described in \citep{Mukherjee06}. As this algorithm is stochastic, in order to reduce the statistical noise we run it $\sim 100$ times obtaining a distribution of values from which we extract the best value of the evidence as the median of the distribution. The Evidence, $\mathcal{E}$, is defined as the probability of the data $D$ given the model $M$ with a set of parameters $\boldsymbol{\theta}$, $\mathcal{E}(M) = \int \mathrm{d}\boldsymbol{\theta} L(D|\boldsymbol{\theta},M)\pi(\boldsymbol{\theta}|M)$: $\pi(\boldsymbol{\theta}|M)$ is the prior on the set of parameters, normalized to unity, and $L(D|\boldsymbol{\theta},M)$ is the likelihood function. There are many other tools to compare models but the Bayesian Evidence is considered the most reliable, even if it is not completely immune to problems, like its dependence on the choice of priors \citep{Nesseris13}. In order to minimize such problems, we have always used the same uninformative flat priors on the parameters, and over sufficiently wide ranges (much wider than the physically acceptable ones), so that further increasing them has no impact on the results. In particular, we only assumed the obvious priors: $\rho_{s}>0$, $r_{s}>0$ and $\Upsilon>0$. Note that the parameter $\Upsilon$ can have any sign, in general. In the peculiar case we are considering, through its definition Eq.~(7), it is clear it has to be positive.

Once the Bayesian Evidence is calculated, one can obtain the Bayes Factor, defined as the ratio of evidences of two models, $M_{i}$ and $M_{j}$, $\mathcal{B}^{i}_{j} = \mathcal{E}_{i}/\mathcal{E}_{j}$. If $\mathcal{B}^{i}_{j} > 1$,  model $M_i$ is preferred over $M_j$, given the data. For each cluster we have used the NFW-GR case as reference model $M_j$. Even with the Bayes Factor $\mathcal{B}^{i}_{j} > 1$, one is still not able to say how much better is model $M_i$ with respect to model $M_j$. For this, we choose the widely-used Jeffreys' Scale \citep{Jeffreys98}. In general, Jeffreys' Scale states that: if $\ln \mathcal{B}^{i}_{j} < 1$, the evidence in favor of model $M_i$ is not significant; if $1 < \ln \mathcal{B}^{i}_{j} < 2.5$, the evidence is substantial; if $2.5 < \ln \mathcal{B}^{i}_{j} < 5$, is strong; if $\mathcal{B}^{i}_{j} > 5$, is decisive. Negative values of $\ln \mathcal{B}^{i}_{j}$ can be easily
interpreted as evidence against model $M_i$ (or in favor of model $M_j$).
In \citep{Nesseris13}, it is shown that the Jeffreys' scale is not a fully-reliable tool for model comparison, but at the same time the statistical validity of the Bayes factor as an efficient model-comparison tool is not questioned: a Bayes factor $\mathcal{B}^{i}_{j}>1$ unequivocally states that the model $i$ is more likely than model $j$. We present results in both contexts for reader's interpretation.

Finally, in order to quantify the relative contribution of X-ray gas and lensing to the total $\chi^2$, we have used the so-called $\sigma$-distance, $d_{\sigma}$, i.e. the distance in units of $\sigma$ between the best fit points of the total sample and the best fit points of X-ray and lensing separately. Following \citep{numerical}, the $\sigma$-distance is calculated by solving
\begin{equation}
1- \Gamma(1,\vert\Delta \chi_{\sigma,X}^2/2\vert)/\Gamma(1) = \mathrm{erf}(d^{X}_{\sigma}/\sqrt{2})\, ,
\end{equation}
where $X$ stands for X-ray or lensing, and $\Delta \chi_{\sigma,X}^2$ is defined as $\chi^2(\boldsymbol{\theta}_{\mathbf{tot}}) - \chi^2(\boldsymbol{\theta}_{\mathbf{X}})$, i.e. the difference between the total chi-square function evaluated at $\boldsymbol{\theta}_{\mathbf{tot}}$ and $\boldsymbol{\theta}_{\mathbf{X}}$ which are, respectively, the best fit parameters from the joint analysis and the X-ray or lensing one. Of course, the condition $d^{Gas}_{\sigma}<d^{Lens}_{\sigma}$ would imply that X-ray gas is the leading term in the $\chi^2$, and viceversa if $d^{Lens}_{\sigma}<d^{Gas}_{\sigma}$.

\section{Results}
\label{sec:results}

In this section we are going to discuss the main results we obtained from our analysis.

\subsection{NFW and GR analysis}

First of all, we have performed separate fits using both X-ray gas and lensing data in the classical GR scenario and present the results in Table~\ref{tab:separate_fits}. For now, we will focus on the primary fit parameters, the NFW parameters $\rho_{s}$ and $r_{s}$, and we compare our results with \citep{Donahue14,Merten15} as cross-check. They are in full agreement, with small differences only with respect to the lensing estimations from \citep{Merten15}, mainly due to the stochastic nature of the MCMC we have used to perform the fit, in contrast with the Levenberg-Marquardt code used in that work. Anyway, both estimations are statistically consistent at $1\sigma$ level; the same level of consistency is with the analysis from \citep{Donahue14}.

From the values given in Table~\ref{tab:separate_fits}, it is clear that for some clusters there is a tension between the mass estimated using X-ray data and that derived from gravitational lensing. Given our discussion in previous sections, this is quite expected but, as we are interested in joining the two data sets, we will pay some more attention to these results, and will discuss them in more detail.

In \citep{Donahue14}, their Fig.~(7) shows the mass bias between the total cluster \textit{HSE thermal} mass that can be derived from X-ray observations, and the total mass derived from gravitational lensing. In the right panel of the same figure are shown the results from using \textit{Chandra} telescope data for the X-ray (which we have used in this work) and the algorithm $SaWLens$ used in \citep{Merten15} to process lensing observations. Actually, it is natural and possible to expect an intrinsic deviation of $\approx 10\%$ in this mass bias (due to hydrostatic equilibrium hypothesis, or projection effects), as obtained from numerical simulations; in this range of uncertainty the two mass estimations would be statistically consistent with each other. Even if the average behaviour derived from the CLASH sample is within this range or, at least, not in strikingly contrast, it is also clear that many clusters, taken individually, can exhibit larger deviations, mainly in the inner regions. Even if non-thermal effects are not considered explicitly, in \citep{Donahue14} it is argued that they can account, maximum, for about $10-20\%$ of the total pressure (or, equivalently, mass). Thus, in general, astrophysical processes are excluded as main source of such large deviations, otherwise they would imply that the gravitational potential at small radii is actually unable to balance the astrophysically-generated pressure of the gas \citep{Donahue14}. As a possible solution, the authors suggest that one should take into account the central galaxies stellar contribution, which is not considered there, and which actually results to be dominant exactly in the $<100$ kpc region. It is worth to mention that in \cite{Donahue14} also analyze X-ray data from XMM, which finally result in mass profiles with a larger tension than \textit{Chandra}, in terms of shape and normalization, with respect to lensing profiles. Moreover, in \citep{Umetsu15} lensing-lensing cross-checks show a generally good agreement despite some exceptions.

The tension between the X-ray estimations and the lensing ones is made also visually more clear in Fig.~(\ref{fig:NFW_regions}), where we plot the likelihood contours for the primary NFW parameters derived using only lensing data (blue), only X-ray observations (red), and joining the two data sets (black). From this figure, we are led to divide the clusters from our sample in three different groups:
\begin{itemize}
  \item \textit{group 1}: clusters whose $1\sigma$ confidence regions from both the approaches, taken separately, overlap or coincide. The clusters A209, RXJ2129, A611, MACSJ1720, MACSJ0429, MACSJ0329, MACSJ1311, MACSJ1423 and MACSJ0744, show a full overlap of their likelihood contours; while MACSJ1206 a partial overlap;
  \item \textit{group 2}: clusters whose likelihood contours overlap only at $2\sigma$: A2261, RXJC2248, MACSJ1931, RXJ1532;
  \item \textit{group 3}: clusters whose X-ray and lensing likelihoods are in tension at more than $2\sigma$: MACSJ1115, RXJ1347; and A383 and MS2137, which play a special role, as we will explain later in this section.
\end{itemize}
When considering results from the joint use of both data sets, the final results for the NFW parameters are clearly dependent on what group the cluster belongs to. As it is possible to visually check from Fig.~(\ref{fig:NFW_regions}), and from values of $\sigma$-distances in Table~\ref{tab:evidence}, in general, the fits are in some way led by X-ray data, which give more stringent constraints on the mass estimations (without forgetting all caveats about possible astrophysical interferences in this case)%, and thus rule in the $\chi^2$ sum.
Anyway, clusters from the first group, generally exhibit joint estimations which are consistent with the separate-data approaches, respectively, and also show some improvement in the final errors on the parameters. Lesser improvements are from clusters of group two; while no improvement at all is for group three.

How the different degrees of tension between the separate approaches influence the final comparison with observations, can be visually inspected in Fig.~(\ref{fig:final_plot}); in dashed blue lines, we show the best fit from using gas-only (left panels) and lensing-only (right panels) data; in solid blue lines we represent the best fit from a joint use of both sets of data. While from gas (left) possible differences between the models are barely distinguishable, more information can be derived focussing on lensing data (right). If we consider the joint analysis (solid blue lines), we can easily verify how the clusters from the first group show a very good agreement with data. For clusters of groups two and three, we can note how there is a progressive degradation in the goodness of the fit. In particular, the joint fit translate in an overestimation of the mass (convergence) in the central regions. Clusters A383 and MS2137 are the really-problematic cases in this sample, because the NFW + GR fit from joint data sets is unable to provide a satisfactory match to data, producing a global and extended underestimation of the mass all over the range covered by the lensing analysis. Unfortunately, neither in \citep{Donahue14} nor in \citep{Umetsu15} one can find some peculiar hint for such misbehaviour.

Again, these different levels of tension and their relation with a specific cluster, may be not unexpected: if we compare our classification with Fig.~(7) of \citep{Donahue14} it clearly emerges that all the clusters belonging to the second and third group are those which exhibit the largest departures from the average mass bias in the central region. From now on we will discuss the results for the full cluster sample, but we want to stress that our main statistical conclusions will be centered only on clusters from group one, which constitute half of the sample.

\subsection{NFW and $G^{3}$-galileon analysis}

If we now move to our $G^{3}$-galileon model, the first thing to be said is that we have problems to constrain our model using only lensing data. For the sake of clarity, in Table~\ref{tab:joint_fits} we will only report results from the joint fit, and not those derived from the separate use of both the chosen sets of observations, but we will discuss them here.

Basically, the lensing-only analysis is unable to put any bounds on our alternative model. From a ``conservative'' perspective, one would expect small deviations from GR. Instead, the statistics derived from the $\chi^2$ minimization only marginally contains the GR limit, being the minimum very far (in a statistical sense) from what expected for a ``small deviation from GR''. Just as illustrative examples, we report here the best fits for three clusters, in the form $(\rho_{s},r_{s},\Upsilon)$, with the same units of Table~\ref{tab:joint_fits}:
\begin{eqnarray}
\mathrm{A209}&:& (26.5,141,0.97)\nonumber \\
\mathrm{MACSJ1311}&:& (99.8,\;\,72,1.09) \\
\mathrm{RXJ1347}&:& (21.1,181,0.97) \nonumber \; .
\end{eqnarray}
We generally obtain a characteristic radius $r_{s} \approx 100$ kpc, well constrained at the lower limit, but poorly at the higher end. For what concerns the central DM NFW density distribution it exhibits a minimum for very high values (much larger than what expected from GR simulations), but is basically unconstrained, being its likelihood function almost flat on a large range of values (from $\sim 10^{15}$ up to $\sim10^{17} \, M_{\odot}$ Mpc$^{-3}$). Finally the galileon parameter $\Upsilon$ is well constrained at $\approx 1$, and it is \textit{never} consistent with the GR limit. It is interesting to note that the $\chi^2$ minimum can be as lower as $35,\,50$ and $5\%$ (for A209, MACSJ1311 and RXJ1347 respectively) with respect to the NFW $+$ GR scenario. This net improvement is mainly due to a better match of the theoretical convergence with data at $\approx 100-200$ kpc, while the fit remains as poor as GR at larger distances from the center, which have much larger errors and thus much less weight from a statistical point of view. However, we lack a strong theoretical and physical motivation in support of such results and, eventually, their statistical validity seems to be low. %In general, the theoretical model can also have some influence: the potentials $\Phi$ and $\Psi$ depend in different ways on the density and the freedom given by lensing (through the observational errors) makes it possible for them to ``compensate each other'' effects in a way different from the $\Phi=\Psi$ condition which holds for GR. As we will discuss below, with joined data, the X-ray data anchors in some way $\Phi$ at least, reducing such freedom. We will come back to this point at the end of this section with more details. If we would like to compare our results with previous literature, unfortunately we have no quantitative information about the values of the theoretical parameters for the galileon model considered in \citep{Barreira15B}; while, although with the due differences, the same conclusion is obtained in \citep{TerukinaLombriser2015}, where the parameters which are mostly and more directly related to the galileon model are basically unconstrained and (statistically) misbehaving.

On the other hand, our constraints from X-ray observation are much more stringent and appear to be as small deviations from the GR case but still consistent with GR, i.e., with $\Upsilon =0$. And also the total $\chi^2$ obtained from the joint analysis is much more consistent. As discussed above, still the gas contribution is leading the $\chi^2$ sum; but the more interesting considerations can be derived from the values assumed by the parameter $\Upsilon$.  In Table~\ref{tab:joint_fits}, in the $\Upsilon$ columns, we indicate in parenthesis the level of consistency of the best fits with the zero value (i.e. with GR): $<1\sigma$ means that the zero value falls within the $1\sigma$ confidence level ($<2\sigma$ and $>3\sigma$ are self-explanatory); $\neq 0$ means that the probability distribution does not reach the zero value at all. It is straightforward to check that all the clusters which have only upper limit on $\Upsilon$, thus being consistent with GR, are from group one. The only exception is A2261 which, while being consistent with GR, is from group two. It is also interesting to highlight that some clusters from group one (RXJ2129, A611, MACSJ0429 and MACSJ1206), and thus, have no statistically relevant discrepancy between X-ray and lensing observations, %and which we are assuming to have no ``pathological'' problem and being quite reliable,
may exhibit a more clear departure from GR, at $2\sigma$ level maximum. All clusters with a departure from GR greater than $2\sigma$ fall in the groups two and three; in this case, such departures from GR may be due to inconsistencies between mass estimations obtained from X-ray observations and lensing events. The only exceptions are A383 and MS2137 which have a very poor fit to data, but have also been proved to be the most problematic clusters even in the classical GR scenario.

In order to test if such departures from GR are real or how much they might be due to the tension between X-ray and lensing observable, we have performed some checks. We initially focused on one cluster only, RXJ1347, which exhibits the largest deviation from zero for the $\Upsilon$ parameter and one of the largest tension between the observations. First, we have checked if the value of $\Upsilon$ might depend on the choice of the dark matter density model. Thus, we have compared three models for DM distribution \citep{Umetsu15}: the classical $NFW$, already described in previous sections; a generalized $NFW$ \citep{Zhao96}, $gNFW$, given by the density profile:
\begin{equation}
\rho_{gNFW} = \frac{\rho_s}{(r/r_s)^{\gamma_c}(1+r/r_s)^{3-\gamma_c}}\, ;
\end{equation}
and a $DARKexp$ model \citep{Hjorth15},
\begin{equation}
\rho_{DARK} = \frac{\rho_s}{(r/r_s)^{\gamma_c}(1+r/r_s)^{4-\gamma_c}}\;.
\end{equation}
The first two models are phenomenological, and the $gNFW$ reduces to $NFW$ if $\gamma_c = 1$; the third one has some theoretical basis, as it can describe the distribution of particle energies in finite, self-gravitating, collisionless, isotropic systems, and has the interesting property that it cannot be reduced to a $NFW$ profile. Finally we have:
\begin{eqnarray}
NFW &:& \Upsilon = 0.554^{+0.090}_{-0.093}\; , \nonumber \\
gNFW &:& \Upsilon = 0.550^{+0.094}_{-0.095}\; , \\
DARKexp &:& \Upsilon = 0.533^{+0.095}_{-0.096}\; . \nonumber
\end{eqnarray}
First, it comes out that, from a statistical point of view, the $NFW$ model is definitively the best model (that is consistent with results from \citep{Umetsu15}). Then, it is also plain that there is no dependence of $\Upsilon$ on the used dark matter model, at least for this cluster. We are aware that, in principle, one should check the same for all other clusters, but we think this kind of analysis is out of the main topic of this work, and we postpone it to a dedicated future work.

Following our previous discussion about Fig.~(7) in \citep{Donahue14}, we have performed a further check: first, we have added a constant $10\%$ mass budget to the total mass estimated from X-ray, as a crude way to take into account possible non-thermal pressure and other astrophysical effects; second, we have extracted the radial behaviour of the mass bias for RXJ1347 from the same figure and re-scaled the masses from X-ray observations (and the related errors) with it. Then, we have repeated the fit using these new \textit{normalized} mass profiles for X-ray observations, obtaining
\begin{eqnarray}
\mathrm{original\; X-ray} &:& \Upsilon = 0.554^{+0.090}_{-0.093}\; ; \\
\mathrm{constant\; bias} &:& \Upsilon = 0.497^{+0.098}_{-0.099}\; ; \nonumber \\
\mathrm{radial\; bias} &:& \Upsilon = 0.189^{+0.141}_{-0.113}\; . \nonumber
\end{eqnarray}
It is plain that the constant bias does not lower the value of $\Upsilon$ in a statistically significant way. Instead, the radial bias lowers the value of $\Upsilon$ by almost a $70\%$, and the final estimation is consistent with zero, and then with GR, at $2\sigma$, thus alleviating in a very considerable way the tension with GR we have detected in the previous cases of \textit{non-normalized} mass profiles from X-ray observations. Based on this preliminary result, we have extracted the radial mass biases from Fig.~(7) of \citep{Donahue14} for all the clusters in our sample, and used them for a new analysis; finally, all the results are shown in Table \ref{tab:joint_fits} as the \textit{normalized} case. It is clear that in this case we obtain a general decrease of $\Upsilon$ for all the clusters, and now they are mostly consistent with GR at a higher statistical level (less than $1\sigma$ for most of them).

It is good to stress that this check was performed only in order to better quantify the validity of our main results, but it is not realistic: it would resemble the case of having only lensing measurements (because the X-ray-derived profiles are scaled in order to match lensing estimations) but with an increased precision, as the error on the masses derived from X-ray observations are better than those from lensing analysis. In fact, this radial bias is an \textit{a posteriori} quantity, that can be quantified only if one has both X-ray and lensing mass estimates, and uses them separately to infer information about radial mass profiles.

Assuming that X-ray observations can suffer problems from some astrophysical effects of varied origin, while lensing not, one is basically stating that the \textit{real} and \textit{complete} mass estimation comes from lensing; and  one could think about this mass bias as an indirect tool to quantify how much such perturbing astrophysical effects can alter the X-ray-based mass estimations. Anyway, at least for clusters from group one, we find no reason for such particular treatment: when gas and lensing outputs coincide, in general, we find agreement with GR. Even if \textit{this is not the rule}: as pointed out above, some of these clusters, which have consistent mass estimation from both sides, still can have $\Upsilon \neq 0$ at $2\sigma$.

It is also interesting to stress another point: with the exception of A383 and MS2137, for all the clusters the $G^{3}$-galileon model works well even when the GR mass estimations seem to be in tension and show a bad fit for the joint analysis (mainly visible in the convergence profiles). In some way, the extra terms introduced by the $G^{3}$-galileon model, and related to a breaking of the Vainshtein mechanism, can \textit{mimic the physics} behind the discrepancy between gas and lensing (whatever it is the cause of such discrepancy). If such extra terms from the $G^{3}$-model had been dependent in an explicit way on the gas density distribution (as it happens, for example, in the case of non-thermal effects in GR), then the best performance of the $G^{3}$-galileon model with respect to GR would have been quite obvious. But that is not the case: we have extra-terms depending \textit{only} on the NFW profile which, in principle, carries no information at all about the internal gas dynamics and/or properties.

Maybe the reason for such behaviour is in the introduction of the parameter $\Upsilon$: for GR we have $\Upsilon=0$, thus resulting in the gravitational and metric potential being equal, $\Phi = \Psi$. In other words: in GR we have much less freedom to accomodate the tension between X-ray and lensing observables in the mass estimation than in the $G^{3}$-galileon model where, even if the potentials are still strictly correlated, they both depend on $\Upsilon$ but through $M''$ (through $\Phi$) and $M'$ (through $\Psi$), thus, with a different quantitative contribution. We might think that, with $\Phi$ playing some main role in the gas dynamics, while $\Psi$ working only for lensing, the $G^{3}$-galileon can, in some way, re-adjust the tension with gas observations. Of course, the difference between the classical and the alternative scenario is expected to be small; this is also clearly shown in Table \ref{tab:evidence}, where we report the minimum $\chi^2$, for both GR and $G^{3}$-galileon, and the Bayesian ratio, both in its pure form and in the logarithmic Jeffreys' scale units. It is clear that while, in general, the classical and alternative approach are basically equivalent (the $\log \mathcal{B}^{gal}_{GR}$ is always in the range $[-1,1]$ implying no evidence in favor of one model or another), all the clusters for which the tension is predominant (group two and three) show a clear statistical preference for the $G^{3}$-galileon, which provide a much better fit to data than GR (except for A383, MS2137).

%The final question to be answered could be: \textit{may the $G^{3}$-galileon potential (or any other equally valid alternative) be seen as just a ``transitory phase'' in a more complex evolutionary scenario of clusters of galaxies, where gravity is not led by GR for all time, but only tends to be GR-type in a possible ``close-to-virialized'' final state?} \vs{to check, not fully convinced}

\subsection{NFW Mass-concentration}

In Tables \ref{tab:separate_fits} and \ref{tab:joint_fits} we also show some parameters which are typically used in literature as comparison rulers: after fixing an over-density level, $\Delta$, relative to the critical density of the Universe at the cluster redshift, $\rho_{c}(z) = 3 H^{2}(z)/ 8\pi G$, we can calculate the DM mass in a sphere of radius $r_{\Delta}$, $M_{\Delta}$, as
\begin{equation}
M_{\Delta} = \frac{4}{3} \pi r^{3}_{\Delta} \Delta \rho_{c} \equiv 4 \pi \rho_{s} r^{3}_{s}\left[ \ln(1+c_{\Delta}) - \frac{c_{\Delta}}{1+c_{\Delta}}\right] \; ,
\end{equation}
where the second expression on the right hand side is explicitly derived when a NFW profile is used. By comparing the two expressions, one can numerically derive the concentration parameter $c_{\Delta}$, defined as,
\begin{equation}
c_{\Delta} = \frac{r_{\Delta}}{r_{s}} \; ,
\end{equation}
and from it to calculate the radius $r_{\Delta}$. The overdensity value which is typically used as a standard ruler is $\Delta = 200$; other possible choices are $\Delta = 500, 2500$ (corresponding to progressively inner regions within the cluster), and $\Delta_{vir}$ (virialized), whose value depends on the redshift of the cluster and, what is  more important, can be numerically derived from simulations based on GR. For our sample, it is in the range $[110;140]$ assuming $\Lambda$CDM cosmology; given that many alternative models of gravity can produce notable changes in the structure formation history, one should run simulations in a modified context and derive the corresponding new values for $\Delta_{vir}$ in order to check differences.

If we first consider the GR analysis, a quick comparison between the mass values $M_{200}$ given in Table \ref{tab:separate_fits} for the gas-only and lensing-only analysis, with those in Table \ref{tab:joint_fits} for the joint fits, clearly show how the mass estimation are statistically consistent (we strictly refer to clusters from group one). %As it might be expected, the joint analysis exhibit smaller errors than the separate cases, and, mainly led by X-ray observations, tends to select the higher-mass region inside the larger confidence levels from the lensing-only case.
We add that the results are very sensitive to the tension between the X-ray and the lensing observables; even a slight shift in the likelihood contours, toward a not complete overlap, as it is in the case of MACSJ1206, can suddenly lead to lower statistical consistency between the two approaches. This is made even more obvious when considering clusters from groups two and three. The same conclusion can be driven for the concentration parameter evaluated at the same over-density, $c_{200}$: we have statistically consistent results from both the X-ray, the lensing and the joint approach.

A visual summary is also given in Fig.~(\ref{fig:Mass_concentratio}), where observationally derived values for masses and concentrations are compared with theoretical expectations. Clusters with higher statistical significance from group one, are shown as black points; clusters from groups two and three are in grey. Colored lines are the expected relations for relaxed clusters (as most of our clusters are) derived from numerical simulations: dashed cyan is derived from \citep{Duffy08}; dashed yellow is from \citep{Bhattacharya13}; dashed green is from \citep{Meneghetti14}. Red-style lines are not obtained from simulations, but from fitting the $M_{200}-c_{200}$ relation when a NFW profile is used with lensing observations: the dot-dashed relation is derived in \citep{Merten15}, while the dashed one is from \citep{Umetsu15}. It is clear that even after joining lensing data with X-ray observations, the clusters are still consistent with expectations from simulations. It is interesting to note how in \citep{Merten15} and \citep{Umetsu15} the same sample of clusters is analyzed, but using different reconstruction methods; even if the final $M_{200}-c_{200}$ relation appears to be somewhat different, their are still statistical consistent with each other, with a little underestimation of the normalization factor, i.e., on average, slightly smaller values for both masses and concentrations. Anyway, in general, as shown in Fig.~(10) of \citep{Umetsu15}, the convergence profiles from both works are in good agreement, except only three cases (MACSJ1931, RXJ1347 and, even if at a lesser level, MACSJ0744) where the reconstruction is systematically lower. The two first clusters belong to our groups two and three; thus, in this case, the tension between X-ray and lensing measurements might be, possibly, an intrinsic hidden source for the discrepancy.

If we now move to the $G^{3}$-galileon model, we can appreciate what has been discussed in previous sections: the model needs more DM to match observations. This is clear in the (small) shift in Fig.~(\ref{fig:Mass_concentratio}) toward the right-end side of the plot which, however, does not alter in any consistent way the accordance of our results with numerical simulations relations. Moreover, in most of the cases, the change in the mass is within the statistical errors, thus having a general equivalence, in this sense, between GR and the $G^{3}$-galileon model. At this stage, once again, we have that GR and galileon are indistinguishable. At the same time, we may note a shift toward smaller values of concentrations in Fig.~(\ref{fig:Mass_concentratio}) when moving from GR to the alternative scenario.

About how much more DM is required from our chosen model, in Fig.~(\ref{fig:Mass_ratio}) we plot (only for clusters from group one) the relative difference between the mass enclosed in a sphere of radius $r$ derived from using a NFW mass profile in the context of our alternative scenario, and the same quantity but derived in GR. As pointed out in \citep{KoyamaSakstein2015}, the net effect of a Galileon inside a cluster or galaxy would be to reduce the gravity strength (see their Figs.~7-8). From our Fig.~(\ref{fig:Mass_ratio}) it is clear that, in order to work inside a cluster, the $G^{3}$-model requires from $2$ to $7.5\%$ more DM at $r_{200} \sim 2$ Mpc (maximum range covered by our data). But this trend is not uniform at all scales: it can be seen that below some scale ranging from $200$ to $500$ kpc (it depends on the cluster), the model can require up to $10\%$ less DM than GR. This means that, on that scales, Galileon induces an attractive force which can mimic dark matter effects. Thus, summarizing previous considerations, we might state more correctly that the $G^{3}$-galileon model predicts \textit{less} concentrated \textit{and} slightly \textit{more} massive haloes in clusters of galaxies than GR.

\section{Conclusions}

In this work we aimed to test an alternative model of gravity, proposed for the first time in \citep{KoyamaSakstein2015}. It is one of the many versions in which the galileon fields can be accomplished but, apart from being a quite new approach, its main attraction is the intrinsic breaking of the Vainshtein mechanism inside large astrophysical objects. The main consequence of such a breaking is that, while a fully operative screening would make this alternative model almost completely equivalent to GR in astrophysical objects with departures only at cosmological scales, in this case, instead, we have the interesting consequence of possible peculiar signatures imprinted in, for example, the internal dynamics and mass profiles of clusters of galaxies. Then, at least in principle, one should have a useful tool for differentiating this model from GR. The model we have considered, in particular, has been recently considered in \citep{SaksteinPRL15,SaksteinPRD15,Sakstein16}. In \citep{SaksteinPRL15,SaksteinPRD15}, in particular, it has been applied to stellar scales and found to be basically consistent with GR, with an upper limit of $\Upsilon < 0.027$. In \citep{Sakstein16} a particular version of the same model, with two constants $\Upsilon_1$ and $\Upsilon_2$ for $\Phi$ and $\Psi$ respectively is considered. In this case they might have both positive and negative sign, and the best fits are $\approx -0.1$ and $\approx -0.2$; but still there is full consistency with GR.

We have tried to address this point by focussing on the total mass profiles of clusters of galaxies. We have compared the theoretical predictions from this model with the observations. The profiles are derived using two complementary tools: X-ray hot intra-cluster gas dynamics, mainly derived from a re-analysis of archival data in \citep{Donahue14}; and strong and weak gravitational lensing, obtained updating archival data with novel observations run through the \textit{CLASH} survey program \citep{Merten15}. One of the main goals of this work is just to check in detail the compatibility of these a two approaches, and if those discrepancies which are generally ascribed to astrophysical process in the context of GR, might instead by due to an alternative gravity scenario.

As main result, we have to point out that, if we look at global mass estimations, or concentration parameters for dark matter haloes, they are consistent with GR results. Said in a different way: given present observational accuracies, it is impossible to state a clear difference between the two approaches; the two models are statistically equivalent.

What might sound more interesting, is that there might be a dependence of the outcomes of our analysis with the dynamical internal status of a cluster, which would make the galileon approach more viable than the classical GR to match some observations. It is well known that a tension between X-ray and lensing observations (in terms of mass estimations) is present \citep{Donahue14}. Due to this tension, we have classified all the clusters from our sample in three groups: clusters for which the separate analysis from X-ray and lensing are consistent at $1\sigma$ level, belong to group one; when the tension is at least at $2\sigma$ level, we define the group two; finally, in group three we consider clusters with a tension higher than $3\sigma$. It comes out that clusters which are more relaxed and, thus, whose X-ray profiles are less perturbed by possible astrophysical local processes, belong to group one. In this case, the galileon model gives a good fit to both X-ray and lensing observations. Generally, the parameters which quantify the deviation from GR, $\Upsilon$, is consistent with the GR value being $\lesssim0.086$ at $1\sigma$, $\lesssim0.16$ at $2\sigma$, and $\lesssim0.23$ at $3\sigma$; such values are also considered in \citep{SaksteinPRL15,SaksteinPRD15} as a safe limit under which deviation from GR are healthy. Anyway, statistically speaking, using tools like the Bayesian ratio, there is not a clear evidence in favor of this model, with respect to GR. At least, we can assert that the galileon model is as good as GR in order to explain observations.

Clusters from other groups exhibit a much more positive and striking evidence in favor of the galileon. In particular, it seems that the galileon is more able to reduce the tension between X-ray and lensing data than GR by mimicking in some way, through the terms that lead the Vainshtein breaking, the physics behind it. But even in this case, in order to be statistically confident about such results, and state in a more confident way that a real possible deviation from GR is operative, better data are needed (reducing some of the systematic uncertainty from calibration; exact choice of modeling methods; larger samples to limit scatter from relaxation state of the clusters or their asymmetry). It is also true that there could be astrophysical phenomena at nonlinear scales from baryonic physics that could be degenerate with Galileon effects. Such possible degeneracies with nonlinear baryonic effects should be studied and considered; nevertheless, we believe that at cluster scales astrophysical effects from baryons are most probably small and may not alter our effects. That would no be the case if we look at galaxy-scales for instance.

\section*{Acknowledgments}

V.S. was financed by the Polish National Science Center Grant DEC-2012/06/A/ST2/00395. V. S. warmly thanks Jeremy Sakstein for invaluable discussions and clarifications about his model; and Julian Merten, for having kindly provided lensing data and carefully read the draft giving interesting hints. D.F.M. thanks the Research Council of Norway and the NOTUR computer facilities. S.C. acknowledges the support of \textit{Istituto Nazionale di Fisica Nucleare} IS$-$QGSKY and IS$-$TEONGRAV. M.D. was partially supported by the NASA ADAP award NASA-NNX13AI41G and the STScI/NASA award HST-GO-12065.07-A. This article is also based upon work from COST action CA15117 (CANTATA), supported by COST (European Cooperation in Science and Technology).

%%%%%%%%%%%%%%%%%%%%%%%% Separate fits for X-ray and lensing  %%%%%%%%%%%%%%%%%%%%%%%%%%%%%%%%%%%%%%%%%%%%%%%%%%
%%%%%%%%%%%%%%%%%%%%%%%% Non-"normalized" X-ray mass profiles %%%%%%%%%%%%%%%%%%%%%%%%%%%%%%
%%%%%%%%%%%%%%%%%%%%%%%% NFW+GR - primary parameters %%%%%%%%%%%%%%%%%%%%%%%%%%%%%%%%%%%

{\renewcommand{\tabcolsep}{1.mm}
{\renewcommand{\arraystretch}{1.75}
\begin{table*}
\begin{minipage}{\textwidth}
\centering
\caption{Primary NFW parameters from separate fits for X-ray and lensing data. Units: densities are in $10^{15} \, M_{\odot}$ Mpc$^{-3}$; masses are in $10^{14}\, M_{\odot}$; radii are in Mpc.}\label{tab:separate_fits}
\resizebox*{0.925\textwidth}{!}{
\begin{tabular}{c|c|ccccc|ccccc}
  \hline
  \hline
  \multicolumn{12}{c}{\textbf{Separate fits for X-ray and lensing: non-``normalized'' X-ray mass profiles}} \\
  \hline
  name  & $z$  & \multicolumn{5}{c|}{Lensing} & \multicolumn{5}{c}{X-ray Gas} \\
        &      &$\rho_{s}$ & $r_{s}$ & $M_{200}$ & $c_{200}$ & $r_{200}$ & $\rho_{s}$ & $r_{s}$ & $M_{200}$ & $c_{200}$ & $r_{200}$ \\
  \hline
  \hline
  A209 & $0.206$ & $0.639^{+0.417}_{-0.257}$ & $0.601^{+0.212}_{-0.152}$ & $12.6^{+3.5}_{-2.8}$ & $3.49^{+0.87}_{-0.79}$ & $2.09^{+0.18}_{-0.17}$ & $0.415^{+0.220}_{-0.157}$ & $0.739^{+0.299}_{-0.183}$ & $12.8^{+5.2}_{-2.9}$ & $2.86^{+0.62}_{-0.60}$ & $2.10^{+0.25}_{-0.17}$ \\
%  \hline
  RXJ2129 & $0.234$ & $1.227^{+0.747}_{-0.463}$ & $0.386^{+0.125}_{-0.098}$ & $7.92^{+2.33}_{-2.07}$ & $4.66^{+1.07}_{-0.88}$ & $1.78^{+0.16}_{-0.17}$ & $1.703^{+0.119}_{-0.117}$ & $0.333^{+0.015}_{-0.013}$ & $7.98^{+0.39}_{-0.33}$ & $5.35^{+0.16}_{-0.15}$ & $1.78^{+0.03}_{-0.03}$ \\
%  \hline
  A611 & $0.288$ & $0.752^{+0.426}_{-0.269}$ & $0.541^{+0.180}_{-0.138}$ & $11.2^{+3.5}_{-2.9}$ & $3.60^{+0.81}_{-0.66}$ & $1.96^{+0.19}_{-0.18}$ & $0.786^{+0.107}_{-0.094}$ & $0.554^{+0.052}_{-0.046}$ & $12.7^{+1.4}_{-1.0}$ & $3.70^{+0.21}_{-0.23}$ & $2.04^{+0.07}_{-0.06}$ \\
%  \hline
  MACSJ1720 & $0.391$ & $1.480^{+1.219}_{-0.682}$ & $0.389^{+0.151}_{-0.112}$ & $9.80^{+2.70}_{-2.16}$ & $4.70^{+1.33}_{-1.05}$ & $1.81^{+0.15}_{-0.14}$  & $2.254^{+0.401}_{-0.370}$ & $0.295^{+0.036}_{-0.032}$ & $7.57^{+0.96}_{-0.95}$ & $5.64^{+0.39}_{-0.41}$ & $1.66^{+0.07}_{-0.07}$ \\
%  \hline
  MACSJ0429 & $0.399$ & $1.277^{+2.251}_{-0.859}$ & $0.424^{+0.307}_{-0.170}$ & $10.4^{+2.6}_{-2.8}$ & $4.35^{+2.31}_{-1.69}$ & $1.84^{+0.14}_{-0.18}$ & $1.975^{+0.727}_{-0.578}$ & $0.316^{+0.093}_{-0.065}$ & $7.91^{+2.92}_{-2.12}$ & $5.28^{+0.80}_{-0.75}$ & $1.68^{+0.19}_{-0.17}$ \\
%  \hline
  MACSJ0329 & $0.450$ & $1.246^{+0.919}_{-0.536}$ & $0.416^{+0.153}_{-0.116}$ & $9.33^{+2.95}_{-2.57}$ & $4.19^{+1.18}_{-0.92}$ & $1.74^{+0.17}_{-0.18}$  & $1.664^{+0.348}_{-0.291}$ & $0.325^{+0.048}_{-0.040}$ & $6.68^{+1.11}_{-0.94}$ & $4.77^{+0.41}_{-0.38}$ & $1.56^{+0.08}_{-0.08}$ \\
%  \hline
  MACSJ1311 & $0.494$ & $1.662^{+0.786}_{-0.511}$ & $0.315^{+0.081}_{-0.071}$ & $5.86^{+1.50}_{-1.49}$ & $4.77^{+0.85}_{-0.73}$ & $1.47^{+0.12}_{-0.14}$ & $1.494^{+0.514}_{-0.408}$ & $0.332^{+0.083}_{-0.061}$ & $6.14^{+1.60}_{-1.22}$ & $4.42^{+0.66}_{-0.55}$ & $1.49^{+0.12}_{-0.11}$ \\
%  \hline
  MACSJ1423 & $0.545$ & $2.472^{+2.363}_{-1.248}$ & $0.283^{+0.125}_{-0.091}$ & $7.06^{+2.76}_{-2.18}$ & $5.40^{+1.75}_{-1.46}$ & $1.53^{+0.18}_{-0.18}$ & $3.454^{+0.390}_{-0.399}$ & $0.234^{+0.022}_{-0.017}$ & $6.24^{+0.73}_{-0.56}$ & $6.28^{+0.29}_{-0.32}$ & $1.47^{+0.05}_{-0.05}$ \\
%  \hline
  MACSJ0744 & $0.686$ & $1.921^{+1.362}_{-0.731}$ & $0.346^{+0.111}_{-0.093}$ & $8.73^{+2.47}_{-2.11}$ & $4.52^{+1.20}_{-0.78}$ & $1.56^{+0.14}_{-0.14}$ & $1.715^{+0.744}_{-0.522}$ & $0.357^{+0.086}_{-0.067}$ & $8.49^{+1.41}_{-1.28}$ & $4.29^{+0.75}_{-0.61}$ & $1.54^{+0.08}_{-0.08}$ \\
%  \hline
  MACSJ1206 & $0.439$ & $1.561^{+1.301}_{-0.694}$ & $0.401^{+0.141}_{-0.113}$ & $11.5^{+2.5}_{-2.5}$ & $4.67^{+1.44}_{-1.02}$ & $1.87^{+0.13}_{-0.15}$  & $0.923^{+0.278}_{-0.225}$ & $0.602^{+0.118}_{-0.093}$ & $19.0^{+3.5}_{-2.7}$ & $3.70^{+0.47}_{-0.47}$  & $2.22^{+0.13}_{-0.11}$ \\
  \hline
  \hline
  A2261 & $0.225$ & $0.649^{+0.565}_{-0.312}$ & $0.634^{+0.292}_{-0.192}$ & $14.9^{+4.6}_{-4.5}$ & $3.48^{+1.18}_{-0.95}$ & $2.20^{+0.21}_{-0.25}$ & $2.764^{+0.338}_{-0.335}$ & $0.270^{+0.024}_{-0.020}$ & $8.00^{+0.71}_{-0.56}$ & $6.60^{+0.40}_{-0.36}$ & $1.79^{+0.05}_{-0.04}$ \\
%  \hline
  RXCJ2248 & $0.348$ & $0.923^{+1.029}_{-0.530}$ & $0.547^{+0.315}_{-0.186}$ & $14.5^{+4.6}_{-4.3}$ & $3.92^{+1.44}_{-1.33}$ & $2.09^{+0.20}_{-0.23}$  & $0.462^{+0.082}_{-0.074}$ & $1.054^{+0.143}_{-0.117}$ & $40.7^{+5.6}_{-4.6}$ & $2.78^{+0.23}_{-0.22}$ & $2.95^{+0.13}_{-0.12}$ \\
%  \hline
  MACSJ1931 & $0.352$ & $1.242^{+1.907}_{-0.743}$ & $0.359^{+0.260}_{-0.152}$ & $6.34^{+3.26}_{-2.44}$ & $4.40^{+2.13}_{-1.43}$ & $1.59^{+0.24}_{-0.24}$  & $2.663^{+0.128}_{-0.124}$ & $0.270^{+0.009}_{-0.008}$ & $7.31^{+0.30}_{-0.27}$ & $6.15^{+0.13}_{-0.12}$ & $1.66^{+0.02}_{-0.02}$ \\
%  \hline
  RXJ1532 & $0.363$ & $0.692^{+0.517}_{-0.315}$ & $0.485^{+0.198}_{-0.136}$ & $6.93^{+1.76}_{-1.84}$ & $3.38^{+0.95}_{-0.78}$ & $1.63^{+0.13}_{-0.16}$  & $1.135^{+0.042}_{-0.043}$ & $0.443^{+0.013}_{-0.012}$ & $10.4^{+0.3}_{-0.3}$ & $4.21^{+0.07}_{-0.07}$ & $1.86^{+0.02}_{-0.02}$ \\
  \hline
  \hline
  MACSJ1115 & $0.352$ & $0.341^{+0.156}_{-0.101}$ & $0.818^{+0.197}_{-0.174}$ & $12.0^{+2.7}_{-2.4}$ & $2.38^{+0.52}_{-0.38}$ & $1.96^{+0.14}_{-0.14}$  & $1.118^{+0.168}_{-0.149}$ & $0.468^{+0.054}_{-0.045}$ & $12.1^{+1.7}_{-1.4}$ & $4.20^{+0.27}_{-0.27}$ & $1.97^{+0.09}_{-0.08}$ \\
%  \hline
  RXJ1347 & $0.451$ & $0.867^{+0.621}_{-0.381}$ & $0.572^{+0.204}_{-0.148}$ & $15.1^{+3.3}_{-3.0}$ & $3.59^{+1.02}_{-0.83}$ & $2.04^{+0.14}_{-0.14}$ & $2.912^{+0.180}_{-0.164}$ & $0.394^{+0.017}_{-0.016}$ & $24.7^{+1.2}_{-1.3}$ & $6.10^{+0.16}_{-0.14}$ & $2.41^{+0.04}_{-0.04}$ \\
%  \hline
  A383 & $0.188$ & $1.405^{+0.986}_{-0.638}$ & $0.434^{+0.162}_{-0.108}$ & $14.2^{+3.7}_{-3.2}$ & $4.86^{+1.40}_{-1.06}$ & $2.19^{+0.17}_{-0.17}$ & $2.630^{+0.142}_{-0.135}$ & $0.211^{+0.007}_{-0.006}$ & $3.61^{+0.11}_{-0.10}$ & $6.57^{+0.14}_{-0.15}$ & $1.39^{+0.01}_{-0.01}$ \\
%  \hline
  MS2137 & $0.313$ & $0.564^{+0.349}_{-0.223}$ & $0.728^{+0.256}_{-0.189}$ & $18.1^{+4.8}_{-4.7}$ & $3.17^{+0.74}_{-0.66}$ & $2.28^{+0.18}_{-0.22}$ & $5.573^{+0.569}_{-0.598}$ & $0.158^{+0.013}_{-0.010}$ & $3.74^{+0.44}_{-0.29}$ & $8.55^{+0.35}_{-0.41}$ & $1.35^{+0.05}_{-0.04}$ \\
  \hline
  \hline
\end{tabular}}
\end{minipage}
\end{table*}}}

%%%%%%%%%%%%%%%%%%%%%%%% Joint fits for X-ray and lensing  %%%%%%%%%%%%%%%%%%%%%%%%%%%%%%%%%%%%%%%%%%%%%%%%%%
%%%%%%%%%%%%%%%%%%%%%%%% Non-"normalized" and "normalized" X-ray mass profiles %%%%%%%%%%%%%%%%%%%%%%%%%%%%%%
%%%%%%%%%%%%%%%%%%%%%%%% NFW+ GR and NFW + galileon - primary parameters %%%%%%%%%%%%%%%%%%%%%%%%%%%%%%%%%%%

{\renewcommand{\tabcolsep}{1.mm}
{\renewcommand{\arraystretch}{1.75}
\begin{table*}
\begin{minipage}{\textwidth}
\centering
\caption{Primary NFW and galileon parameters from joint fits for X-ray and lensing data. Units: densities are in $10^{15} \, M_{\odot}$ Mpc$^{-3}$; masses are in $10^{14}\, M_{\odot}$; radii are in Mpc.}\label{tab:joint_fits}
\resizebox*{0.875\textwidth}{!}{
\begin{tabular}{c|ccccc|cccccc}
  \hline
  \hline
  \multicolumn{12}{c}{\textbf{Joint fits for X-ray and lensing: non-``normalized'' X-ray mass profiles}} \\
  \hline
  name  & \multicolumn{5}{c|}{NFW (GR)} & \multicolumn{6}{c}{NFW (galileon)} \\
        & $\rho_{s}$ & $r_{s}$ & $M_{200}$ & $c_{200}$ & $r_{200}$ & $\rho_{s}$ & $r_{s}$ & $\Upsilon$ & $M_{200}$ & $c_{200}$ & $r_{200}$ \\
  \hline
  \hline
  A209 & $0.419^{+0.156}_{-0.107}$ & $0.743^{+0.165}_{-0.139}$ & $13.2^{+2.7}_{-2.3}$ & $2.86^{+0.47}_{-0.37}$ & $2.13^{+0.14}_{-0.13}$ & $0.373^{+0.149}_{-0.107}$ & $0.804^{+0.211}_{-0.163}$ & $<0.098\; (<1\sigma)$ & $14.2^{+3.4}_{-2.7}$ & $2.70^{+0.45}_{-0.42}$ & $2.18^{+0.16}_{-0.15}$ \\
%  \hline
  RXJ2129 & $1.692^{+0.117}_{-0.117}$ & $0.335^{+0.016}_{-0.014}$ & $7.99^{+0.40}_{-0.33}$ & $5.33^{+0.15}_{-0.16}$ & $1.78^{+0.03}_{-0.03}$ & $1.451^{+0.160}_{-0.156}$ & $0.368^{+0.027}_{-0.023}$ & $0.144^{+0.094}_{-0.079}\; (<2\sigma)$ & $8.69^{+0.58}_{-0.51}$ & $4.97^{+0.24}_{-0.26}$ & $1.83^{+0.04}_{-0.04}$ \\
%  \hline
  A611 & $0.785^{+0.101}_{-0.091}$ & $0.554^{+0.049}_{-0.044}$ & $12.7^{+1.3}_{-1.0}$ & $3.70^{+0.21}_{-0.21}$ & $2.05^{+0.07}_{-0.06}$ & $0.701^{+0.103}_{-0.092}$ & $0.595^{+0.059}_{-0.053}$ & $0.102^{+0.066}_{-0.057}\; (<2\sigma)$ & $13.5^{+1.3}_{-1.2}$ & $3.51^{+0.23}_{-0.22}$ & $2.08^{+0.07}_{-0.07}$ \\
%  \hline
  MACSJ1720 & $2.078^{+0.391}_{-0.307}$ & $0.312^{+0.035}_{-0.034}$ & $7.99^{+1.04}_{-0.92}$ & $5.43^{+0.42}_{-0.39}$ & $1.69^{+0.07}_{-0.07}$  & $1.758^{+0.353}_{-0.311}$ & $0.346^{+0.047}_{-0.038}$ & $<0.184\; (<1\sigma)$ & $8.73^{+1.33}_{-0.91}$ & $5.06^{+0.40}_{-0.42}$ & $1.74^{+0.08}_{-0.06}$ \\
%  \hline
  MACSJ0429 & $1.647^{+0.498}_{-0.350}$ & $0.362^{+0.070}_{-0.063}$ & $9.38^{+2.28}_{-2.06}$ & $4.86^{+0.63}_{-0.46}$ & $1.78^{+0.13}_{-0.14}$ & $1.156^{+0.436}_{-0.356}$ & $0.453^{+0.134}_{-0.091}$ & $0.271^{+0.199}_{-0.167}\; (<2\sigma)$ & $11.2^{+3.8}_{-2.6}$ & $4.19^{+0.66}_{-0.69}$ & $1.89^{+0.19}_{-0.16}$ \\
%  \hline
  MACSJ0329 & $1.509^{+0.284}_{-0.239}$ & $0.348^{+0.046}_{-0.039}$ & $7.26^{+1.20}_{-0.93}$ & $4.58^{+0.34}_{-0.34}$ & $1.60^{+0.08}_{-0.07}$  & $1.290^{+0.277}_{-0.244}$ & $0.385^{+0.058}_{-0.048}$ & $<0.179\; (<1\sigma)$ & $7.84^{+1.24}_{-0.96}$ & $4.27^{+0.36}_{-0.37}$ & $1.64^{+0.08}_{-0.07}$ \\
%  \hline
  MACSJ1311 & $1.484^{+0.392}_{-0.291}$ & $0.334^{+0.054}_{-0.049}$ & $6.13^{+1.14}_{-0.94}$ & $4.45^{+0.47}_{-0.40}$ & $1.49^{+0.09}_{-0.08}$ & $1.376^{+0.372}_{-0.273}$ & $0.352^{+0.058}_{-0.052}$ & $<0.086\; (<1\sigma)$ & $6.41^{+1.24}_{-0.96}$ & $4.33^{+0.46}_{-0.43}$ & $1.51^{+0.09}_{-0.08}$ \\
%  \hline
  MACSJ1423 & $3.384^{+0.381}_{-0.378}$ & $0.238^{+0.021}_{-0.017}$ & $6.35^{+0.70}_{-0.60}$ & $6.22^{+0.32}_{-0.28}$ & $1.48^{+0.05}_{-0.05}$ & $2.594^{+0.522}_{-0.467}$ & $0.280^{+0.037}_{-0.032}$ & $<0.285\; (<1\sigma)$ & $7.42^{+0.96}_{-0.97}$ & $5.56^{+0.47}_{-0.47}$ & $1.56^{+0.07}_{-0.07}$ \\
%  \hline
  MACSJ0744 & $1.642^{+0.555}_{-0.415}$ & $0.369^{+0.070}_{-0.057}$ & $8.81^{+1.31}_{-1.18}$ & $4.23^{+0.59}_{-0.51}$ & $1.56^{+0.07}_{-0.07}$ &   $1.599^{+0.539}_{-0.408}$ & $0.378^{+0.072}_{-0.059}$ & $<0.093\; (<1\sigma)$ & $9.14^{+1.40}_{-1.16}$ & $4.19^{+0.56}_{-0.55}$ & $1.58^{+0.08}_{-0.07}$ \\
%  \hline
  MACSJ1206 & $1.189^{+0.258}_{-0.221}$ & $0.507^{+0.067}_{-0.058}$ & $16.2^{+1.9}_{-1.6}$ & $4.12^{+0.42}_{-0.36}$ & $2.10^{+0.08}_{-0.07}$  & $0.930^{+0.259}_{-0.226}$ & $0.588^{+0.109}_{-0.082}$ & $0.225^{+0.14}_{-0.129}\; (<2\sigma)$ & $18.2^{+2.6}_{-2.1}$ & $3.69^{+0.43}_{-0.43}$  & $2.18^{+0.10}_{-0.09}$ \\
  \hline
  \hline
  A2261 & $2.633^{+0.358}_{-0.322}$ & $0.279^{+0.023}_{-0.021}$ & $8.24^{+0.68}_{-0.67}$ & $6.48^{+0.40}_{-0.35}$ & $1.81^{+0.05}_{-0.05}$ & $2.253^{+0.364}_{-0.331}$ & $0.307^{+0.029}_{-0.025}$ & $<0.191\; (<1\sigma)$ & $8.98^{+0.74}_{-0.64}$ & $6.06^{+0.40}_{-0.43}$ & $1.86^{+0.05}_{-0.05}$ \\
%  \hline
  RXCJ2248 & $0.610^{+0.084}_{-0.073}$ & $0.857^{+0.081}_{-0.075}$ & $32.2^{+3.3}_{-3.0}$ & $3.19^{+0.20}_{-0.18}$ & $2.73^{+0.09}_{-0.09}$  & $0.293^{+0.083}_{-0.064}$ & $1.356^{+0.248}_{-0.206}$ & $0.517^{+0.132}_{-0.145}\; (\neq0)$ & $44.8^{+6.3}_{-5.9}$ & $2.25^{+0.28}_{-0.26}$ & $3.05^{+0.14}_{-0.14}$ \\
%  \hline
  MACSJ1931 & $2.672^{+0.123}_{-0.125}$ & $0.270^{+0.009}_{-0.008}$ & $7.27^{+0.25}_{-0.25}$ & $6.16^{+0.12}_{-0.12}$ & $1.66^{+0.02}_{-0.02}$  & $1.945^{+0.231}_{-0.206}$ & $0.326^{+0.022}_{-0.020}$ & $0.295^{+0.112}_{-0.108}\; (>3\sigma)$ & $8.57^{+0.41}_{-0.44}$ & $5.36^{+0.28}_{-0.25}$ & $1.75^{+0.03}_{-0.03}$ \\
%  \hline
  RXJ1532 & $1.145^{+0.042}_{-0.043}$ & $0.439^{+0.012}_{-0.011}$ & $10.3^{+0.3}_{-0.3}$ & $4.22^{+0.07}_{-0.07}$ & $1.86^{+0.02}_{-0.02}$  & $0.784^{+0.102}_{-0.088}$ & $0.551^{+0.041}_{-0.038}$ & $0.342^{+0.112}_{-0.116}\; (>3\sigma)$ & $12.2^{+0.6}_{-0.6}$ & $3.56^{+0.22}_{-0.19}$ & $1.96^{+0.03}_{-0.04}$ \\
  \hline
  \hline
  MACSJ1115 & $1.108^{+0.142}_{-0.127}$ & $0.465^{+0.043}_{-0.039}$ & $11.8^{+1.4}_{-1.2}$ & $4.19^{+0.24}_{-0.22}$ & $1.95^{+0.07}_{-0.07}$  & $0.732^{+0.123}_{-0.107}$ & $0.608^{+0.073}_{-0.064}$ & $0.340^{+0.071}_{-0.067}\; (\neq0)$ & $14.9^{+2.0}_{-1.7}$ & $3.46^{+0.27}_{-0.23}$ & $2.11^{+0.09}_{-0.08}$ \\
%  \hline
  RXJ1347 & $3.029^{+0.196}_{-0.161}$ & $0.382^{+0.015}_{-0.016}$ & $23.5^{+1.1}_{-1.1}$ & $6.21^{+0.17}_{-0.14}$ & $2.37^{+0.04}_{-0.04}$ & $1.684^{+0.193}_{-0.158}$ & $0.540^{+0.033}_{-0.035}$ & $0.554^{+0.090}_{-0.093}\; (\neq0)$ & $31.0^{+1.8}_{-1.8}$ & $4.81^{+0.24}_{-0.21}$ & $2.60^{+0.05}_{-0.05}$ \\
%  \hline
  A383 & $2.582^{+0.137}_{-0.133}$ & $0.214^{+0.007}_{-0.006}$ & $3.65^{+0.12}_{-0.10}$ & $6.53^{+0.15}_{-0.15}$ & $1.39^{+0.02}_{-0.01}$ & $2.262^{+0.232}_{-0.260}$ & $0.233^{+0.017}_{-0.013}$ & $<0.158\; (<1\sigma)$ & $3.97^{+0.24}_{-0.21}$ & $6.15^{+0.27}_{-0.33}$ & $1.43^{+0.03}_{-0.03}$ \\
%  \hline
  MS2137 & $5.302^{+0.578}_{-0.541}$ & $0.163^{+0.014}_{-0.011}$ & $3.90^{+0.44}_{-0.32}$ & $8.37^{+0.39}_{-0.37}$ & $1.37^{+0.05}_{-0.04}$ & $4.008^{+0.591}_{-0.496}$ & $0.197^{+0.017}_{-0.016}$ & $0.186^{+0.108}_{-0.099}\; (<2\sigma)$  & $4.84^{+0.41}_{-0.41}$ & $7.44^{+0.45}_{-0.39}$ & $1.47^{+0.04}_{-0.04}$ \\
  \hline
  \hline
  \multicolumn{12}{c}{\textbf{Joint fits for X-ray and lensing: ``normalized'' X-ray mass profiles}} \\
  \hline
  name  & \multicolumn{5}{c|}{NFW (GR)} & \multicolumn{6}{c}{NFW (galileon)} \\
        & $\rho_{s}$ & $r_{s}$ & $M_{200}$ & $c_{200}$ & $r_{200}$ & $\rho_{s}$ & $r_{s}$ & $\Upsilon$ & $M_{200}$ & $c_{200}$ & $r_{200}$ \\
  \hline
  \hline
  A209 & $0.624^{+0.207}_{-0.163}$ & $0.615^{+0.134}_{-0.105}$ & $13.0^{+2.7}_{-2.1}$ & $3.45^{+0.46}_{-0.45}$ & $2.12^{+0.14}_{-0.12}$ & $0.548^{+0.206}_{-0.155}$ & $0.670^{+0.166}_{-0.125}$ & $<0.140\; (<1\sigma)$ & $14.2^{+2.9}_{-2.3}$ & $3.23^{+0.49}_{-0.46}$ & $2.18^{+0.14}_{-0.12}$ \\
%  \hline
  RXJ2129 & $1.062^{+0.083}_{-0.076}$ & $0.427^{+0.022}_{-0.021}$ & $8.95^{+0.52}_{-0.46}$ & $4.34^{+0.15}_{-0.15}$ & $1.85^{+0.04}_{-0.03}$ & $0.946^{+0.099}_{-0.101}$ & $0.458^{+0.034}_{-0.028}$ & $<0.141\; (<1\sigma)$ & $9.44^{+0.62}_{-0.56}$ & $4.12^{+0.20}_{-0.20}$ & $1.89^{+0.04}_{-0.04}$ \\
%  \hline
  A611 & $0.729^{+0.095}_{-0.083}$ & $0.558^{+0.050}_{-0.045}$ & $11.7^{+1.1}_{-1.0}$ & $3.58^{+0.21}_{-0.20}$ & $1.99^{+0.06}_{-0.06}$ & $0.671^{+0.097}_{-0.086}$ & $0.588^{+0.057}_{-0.051}$ & $<0.099\; (<1\sigma)$ & $12.2^{+1.2}_{-1.0}$ & $3.44^{+0.22}_{-0.21}$ & $2.02^{+0.07}_{-0.06}$ \\
%  \hline
  MACSJ1720 & $1.318^{+0.224}_{-0.192}$ & $0.416^{+0.048}_{-0.044}$ & $10.5^{+1.3}_{-1.2}$ & $4.43^{+0.32}_{-0.30}$ & $1.85^{+0.07}_{-0.08}$ & $1.097^{+0.227}_{-0.198}$ & $0.466^{+0.066}_{-0.056}$ & $<0.185\; (<1\sigma)$ & $11.6^{+1.5}_{-1.5}$ & $4.09^{+0.35}_{-0.35}$ & $1.91^{+0.08}_{-0.08}$ \\
%  \hline
  MACSJ0429 & $0.701^{+0.177}_{-0.136}$ & $0.574^{+0.105}_{-0.094}$ & $11.6^{+2.8}_{-2.4}$ & $3.32^{+0.36}_{-0.30}$ & $1.90^{+0.14}_{-0.14}$ & $0.475^{+0.183}_{-0.148}$ & $0.741^{+0.231}_{-0.158}$ & $<0.352\; (<1\sigma)$ & $14.4^{+4.6}_{-3.4}$ & $2.77^{+0.45}_{-0.45}$ & $2.05^{+0.20}_{-0.18}$ \\
%  \hline
  MACSJ0329 & $1.112^{+0.215}_{-0.174}$ & $0.442^{+0.060}_{-0.054}$ & $9.77^{+1.57}_{-1.48}$ & $4.00^{+0.35}_{-0.29}$ & $1.77^{+0.09}_{-0.09}$ & $0.946^{+0.205}_{-0.182}$ & $0.489^{+0.080}_{-0.063}$ & $<0.177\; (<1\sigma)$ & $10.6^{+2.0}_{-1.5}$ & $3.70^{+0.36}_{-0.35}$ & $1.82^{+0.11}_{-0.09}$ \\
%  \hline
  MACSJ1311 & $1.719^{+0.452}_{-0.361}$ & $0.308^{+0.052}_{-0.046}$ & $5.80^{+1.03}_{-0.94}$ & $4.76^{+0.51}_{-0.48}$ & $1.46^{+0.08}_{-0.08}$ & $1.578^{+0.403}_{-0.322}$ & $0.326^{+0.054}_{-0.046}$ & $<0.099\; (<1\sigma)$ & $6.19^{+1.11}_{-0.91}$ & $4.57^{+0.45}_{-0.43}$ & $1.49^{+0.08}_{-0.08}$ \\
%  \hline
  MACSJ1423 & $1.672^{+0.214}_{-0.188}$ & $0.363^{+0.035}_{-0.032}$ & $9.04^{+1.21}_{-1.01}$ & $4.57^{+0.26}_{-0.24}$ & $1.66^{+0.07}_{-0.06}$ & $1.337^{+0.265}_{-0.245}$ & $0.417^{+0.060}_{-0.049}$ & $<0.239\; (<1\sigma)$ & $10.1^{+1.6}_{-1.2}$ & $4.15^{+0.34}_{-0.39}$ & $1.73^{+0.09}_{-0.07}$ \\
%  \hline
  MACSJ0744 & $1.788^{+0.621}_{-0.443}$ & $0.369^{+0.068}_{-0.059}$ & $9.91^{+1.47}_{-1.37}$ & $4.35^{+0.65}_{-0.49}$ & $1.63^{+0.08}_{-0.08}$ & $1.740^{+0.568}_{-0.420}$ & $0.380^{+0.068}_{-0.057}$ & $<0.138\; (<1\sigma)$ & $10.4^{+1.5}_{-1.3}$ & $4.31^{+0.58}_{-0.48}$ & $1.65^{+0.08}_{-0.07}$ \\
%  \hline
  MACSJ1206 & $1.326^{+0.373}_{-0.280}$ & $0.434^{+0.065}_{-0.059}$ & $11.7^{+1.6}_{-1.3}$ & $4.35^{+0.51}_{-0.44}$ & $1.88^{+0.08}_{-0.07}$ & $1.149^{+0.336}_{-0.269}$ & $0.473^{+0.081}_{-0.066}$ & $<0.208\; (<1\sigma)$ & $12.7^{+1.5}_{-1.5}$ & $4.04^{+0.53}_{-0.45}$ & $1.94^{+0.07}_{-0.08}$ \\
  \hline
  \hline
  A2261 & $0.529^{+0.077}_{-0.066}$ & $0.725^{+0.073}_{-0.067}$ & $16.9^{+1.9}_{-1.8}$ & $3.17^{+0.21}_{-0.20}$ & $2.30^{+0.08}_{-0.08}$ & $0.441^{+0.085}_{-0.081}$ & $0.810^{+0.117}_{-0.089}$ & $<0.200\; (<1\sigma)$ & $18.3^{+2.5}_{-1.8}$ & $2.90^{+0.24}_{-0.27}$ & $2.36^{+0.10}_{-0.08}$ \\
%  \hline
  RXCJ2248 & $0.614^{+0.099}_{-0.090}$ & $0.686^{+0.078}_{-0.065}$ & $16.8^{+1.9}_{-1.6}$ & $3.20^{+0.23}_{-0.24}$ & $2.20^{+0.08}_{-0.07}$ & $0.491^{+0.113}_{-0.102}$ & $0.786^{+0.124}_{-0.098}$ & $<0.260\; (<1\sigma)$ & $18.4^{+2.3}_{-1.9}$ & $2.88^{+0.28}_{-0.30}$ & $2.26^{+0.09}_{-0.08}$ \\
%  \hline
  MACSJ1931 & $0.587^{+0.029}_{-0.027}$ & $0.594^{+0.022}_{-0.022}$ & $10.2^{+0.4}_{-0.5}$ & $3.12^{+0.07}_{-0.07}$ & $1.86^{+0.03}_{-0.03}$ & $0.490^{+0.057}_{-0.056}$ & $0.663^{+0.052}_{-0.044}$ & $0.152^{+0.109}_{-0.089}\; (<2\sigma)$ & $10.9^{+0.6}_{-0.6}$ & $2.88^{+0.15}_{-0.17}$ & $1.90^{+0.03}_{-0.04}$ \\
%  \hline
  RXJ1532 & $0.563^{+0.024}_{-0.023}$ & $0.563^{+0.018}_{-0.017}$ & $8.16^{+0.32}_{-0.28}$ & $3.05^{+0.06}_{-0.06}$ & $1.72^{+0.02}_{-0.02}$ & $0.481^{+0.053}_{-0.059}$ & $0.621^{+0.051}_{-0.040}$ & $<0.184\; (<1\sigma)$ & $8.70^{+0.50}_{-0.43}$ & $2.83^{+0.15}_{-0.16}$ & $1.75^{+0.03}_{-0.03}$ \\
  \hline
  \hline
  MACSJ1115 & $0.308^{+0.046}_{-0.038}$ & $0.871^{+0.101}_{-0.092}$ & $12.5^{+1.7}_{-1.5}$ & $2.31^{+0.16}_{-0.15}$ & $1.99^{+0.09}_{-0.08}$ & $0.267^{+0.049}_{-0.044}$ & $0.959^{+0.145}_{-0.120}$ & $<0.107\; (<1\sigma)$ & $13.6^{+2.5}_{-1.9}$ & $2.13^{+0.19}_{-0.19}$ & $2.05^{+0.12}_{-0.10}$ \\
%  \hline
  RXJ1347 & $1.058^{+0.066}_{-0.063}$ & $0.533^{+0.024}_{-0.023}$ & $16.0^{+0.9}_{-0.9}$ & $3.90^{+0.11}_{-0.11}$ & $2.08^{+0.04}_{-0.04}$ & $0.868^{+0.116}_{-0.117}$ & $0.606^{+0.056}_{-0.046}$ & $0.172^{+0.124}_{-0.101}\; (<2\sigma)$ & $17.5^{+1.4}_{-1.2}$ & $3.56^{+0.23}_{-0.24}$ & $2.15^{+0.06}_{-0.05}$ \\
%  \hline
  A383 & $1.192^{+0.084}_{-0.080}$ & $0.473^{+0.026}_{-0.023}$ & $14.5^{+1.0}_{-0.9}$ & $4.66^{+0.15}_{-0.14}$ & $2.21^{+0.05}_{-0.05}$ & $1.020^{+0.126}_{-0.140}$ & $0.522^{+0.050}_{-0.038}$ & $<0.176\; (<1\sigma)$ & $15.7^{+1.4}_{-1.2}$ & $4.36^{+0.23}_{-0.28}$ & $2.27^{+0.06}_{-0.06}$ \\
%  \hline
  MS2137 & $0.543^{+0.076}_{-0.071}$ & $0.703^{+0.091}_{-0.075}$ & $15.4^{+2.8}_{-2.1}$ & $3.08^{+0.18}_{-0.19}$ & $2.16^{+0.12}_{-0.10}$ & $0.477^{+0.082}_{-0.075}$ & $0.767^{+0.115}_{-0.092}$ & $<0.108\; (<1\sigma)$ & $16.6^{+3.2}_{-2.5}$ & $2.89^{+0.22}_{-0.22}$ & $2.22^{+0.13}_{-0.12}$ \\
  \hline
  \hline
\end{tabular}}
\end{minipage}
\end{table*}}}

%%%%%%%%%%%%%%%%%%%%%%%% Joint fits for X-ray and lensing  %%%%%%%%%%%%%%%%%%%%%%%%%%%%%%%%%%%%%%%%%%%%%%%%%%
%%%%%%%%%%%%%%%%%%%%%%%% Non-"normalized" and "normalized" X-ray mass profiles %%%%%%%%%%%%%%%%%%%%%%%%%%%%%%
%%%%%%%%%%%%%%%%%%%%%%%% Chi-squares and Bayesian Evidence %%%%%%%%%%%%%%%%%%%%%%%%%%%%%%%%%%%%%%%%%%%%%%%%%%

{\renewcommand{\tabcolsep}{1.25mm}
{\renewcommand{\arraystretch}{1.75}
\begin{table*}
\begin{minipage}{\textwidth}
\centering
\caption{$\chi^2$ and Bayesian evidence ratio comparison.}\label{tab:evidence}
\resizebox*{0.75\textwidth}{!}{
\begin{tabular}{c||ccc|ccc||ccc|ccc}
  \hline
  \hline
   & \multicolumn{6}{c||}{\textbf{Joint fits for X-ray and lensing:}} &
     \multicolumn{6}{c}{\textbf{Joint fits for X-ray and lensing:}} \\
   & \multicolumn{6}{c||}{\textbf{non-``normalized'' X-ray mass}} &
     \multicolumn{6}{c}{\textbf{``normalized'' X-ray mass}} \\
  \hline
  name  & $\chi^{2}_{GR}$ & $d^{Gas}_{\sigma}$ & $d^{Lens}_{\sigma}$ & $\chi^{2}_{Gal}$ & $\mathcal{B}^{Gal.}_{GR}$ & $\ln \mathcal{B}^{Gal.}_{GR}$ &
        $\chi^{2}_{GR}$ & $d^{Gas}_{\sigma}$ & $d^{Lens}_{\sigma}$  & $\chi^{2}_{Gal}$ & $\mathcal{B}^{Gal.}_{GR}$ & $\ln \mathcal{B}^{Gal.}_{GR}$ \\
  \hline
  \hline
  A209 & $3.66$ & $0.04$ & $2.17$ & $3.66$ & $0.53$ & $-0.63$ & $2.89$ & $2.49$ & $0.09$ & $2.87$ & $0.69$ & $-0.37$ \\
%  \hline
  RXJ2129 & $7.01$ & $0.03$ & $>3$ & $5.78$ & $1.40$ & $0.34$ & $5.70$ & $>3$ & $0.90$ & $5.68$ & $0.74$ & $-0.30$ \\
%  \hline
  A611 & $5.57$ & $0.05$ & $>3$ & $4.61$ & $1.24$ & $0.22$ & $4.62$ & $>3$ & $0.49$ & $4.59$ & $0.70$ & $-0.36$ \\
%  \hline
  MACSJ1720 & $5.81$ & $0.16$ & $>3$ & $5.87$ & $0.70$ & $-0.36$ & $4.47$ & $>3$ & $0.46$ & $4.47$ & $0.70$ & $-0.36$ \\
%  \hline
  MACSJ0429 & $3.29$ & $0.47$ & $>3$ & $3.09$ & $0.84$ & $-0.17$ & $1.99$ & $>3$ & $0.24$ & $1.98$ & $0.77$ & $-0.26$ \\
%  \hline
  MACSJ0329 & $8.10$ & $0.22$ & $1.89$ & $8.18$ & $0.62$ & $-0.48$ & $6.53$ & $0.42$ & $2.18$ & $6.52$ & $0.67$ & $-0.40$ \\
%  \hline
  MACSJ1311 & $4.70$ & $0.007$ & $0.39$ & $4.68$ & $0.61$ & $-0.49$ & $4.79$ & $0.58$ & $0.12$ & $4.79$ & $0.68$ & $-0.39$\\
%  \hline
  MACSJ1423 & $7.85$ & $0.007$ & $>3$ & $7.61$ & $0.88$ & $-0.13$ & $7.24$ & $>3$ & $0.64$ & $7.15$ & $0.67$ & $-0.40$ \\
%  \hline
  MACSJ0744 & $3.81$ & $0.09$ & $0.84$ & $3.83$ & $0.56$ & $-0.58$ & $3.51$ & $1.87$ & $0.36$ & $3.30$ & $0.85$ & $-0.16$ \\
%  \hline
  MACSJ1206 & $8.32$ & $>3$ & $1.02$ & $7.46$ & $1.19$ & $0.17$ & $4.88$ & $>3$ & $0.07$ & $4.89$ & $0.65$ & $-0.43$ \\
  \hline
  \hline
  A2261 & $9.61$ & $0.13$ & $>3$ & $8.36$ & $1.44$ & $0.36$ & $3.89$ & $>3$ & $0.31$ & $3.88$ & $0.69$ & $-0.37$ \\
%  \hline
  RXCJ2248 & $15.38$ & $2.13$ & $>3$ & $5.91$ & $80.36$ & $4.39$ & $2.03$ & $>3$ & $0.06$ & $2.02$ & $0.68$ & $-0.39$ \\
%  \hline
  MACSJ1931 & $12.15$ & $0.10$ & $>3$ & $5.42$ & $20.80$ & $3.03$ & $7.18$ & $>3$ & $>3$ & $6.52$ & $1.08$ & $0.08$\\
%  \hline
  RXJ1532 & $18.34$ & $0.07$ & $>3$ & $9.68$ & $54.80$ & $4.00$ & $7.87$ & $>3$ & $>3$ & $7.78$ & $0.74$ & $-0.30$ \\
  \hline
  \hline
  MACSJ1115 & $32.30$ & $0.67$ & $>3$ & $6.57$ & $\sim 10^{5}$ & $\sim 11.5$ & $5.63$ & $>3$ & $0.0002$ & $5.63$ & $0.63$ & $-0.46$\\
%  \hline
  RXJ1347 & $39.11$ & $0.51$ & $>3$ & $4.85$ & $\sim 10^{7}$ & $\sim 16$ & $4.54$ & $>3$ & $>3$ & $3.79$ & $1.13$ & $0.12$ \\
%  \hline
  A383 & $22.69$ & $0.25$ & $>3$ & $22.95$ & $0.47$ & $-0.76$ & $2.26$ & $>3$ & $1.98$ & $2.27$ & $0.62$ & $-0.48$ \\
%  \hline
  MS2137 & $19.74$ & $0.08$ & $>3$ & $18.99$ & $1.09$ & $0.09$ & $3.10$ & $>3$ & $2.39$ & $3.12$ & $0.58$ & $-0.54$ \\
  \hline
  \hline
\end{tabular}}
\end{minipage}
\end{table*}}}

%\clearpage
%\newpage

\begin{figure*}[htbp]
\centering
\includegraphics[width=5.5cm]{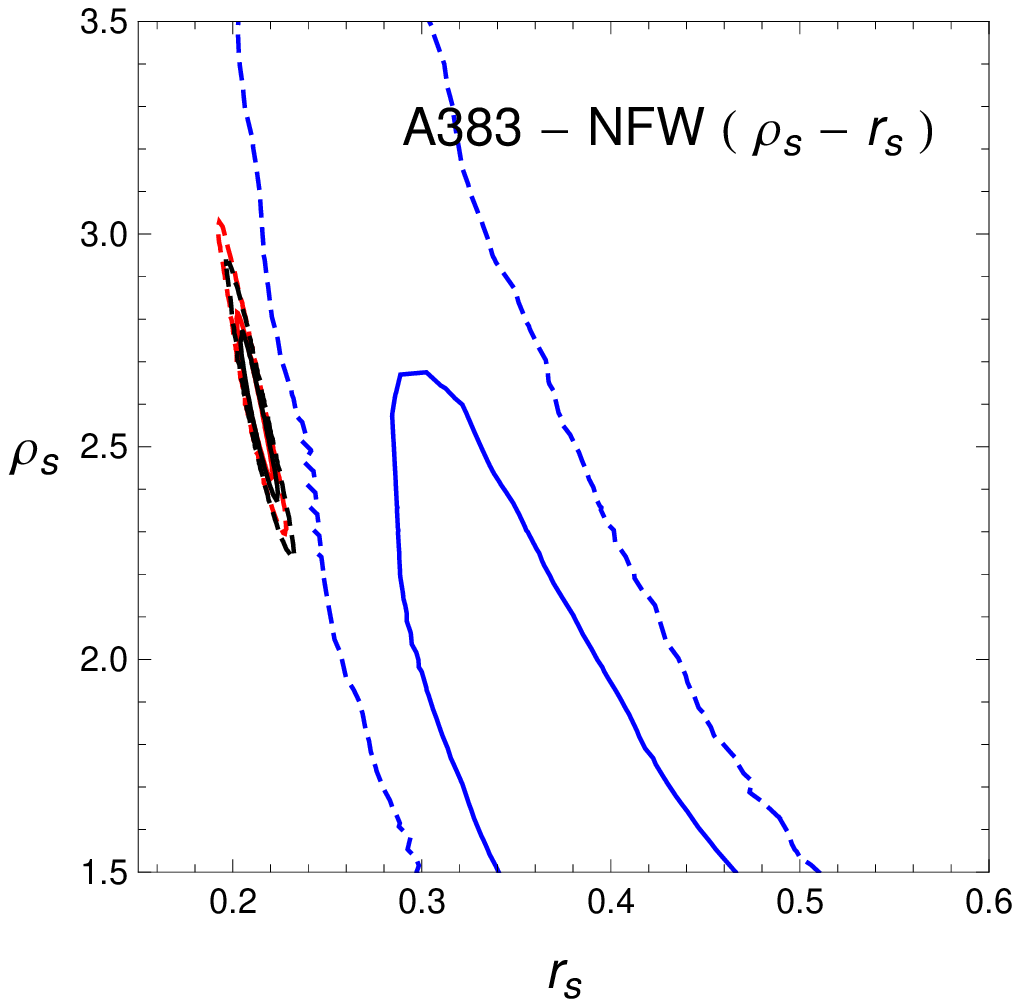}~~~
\includegraphics[width=5.5cm]{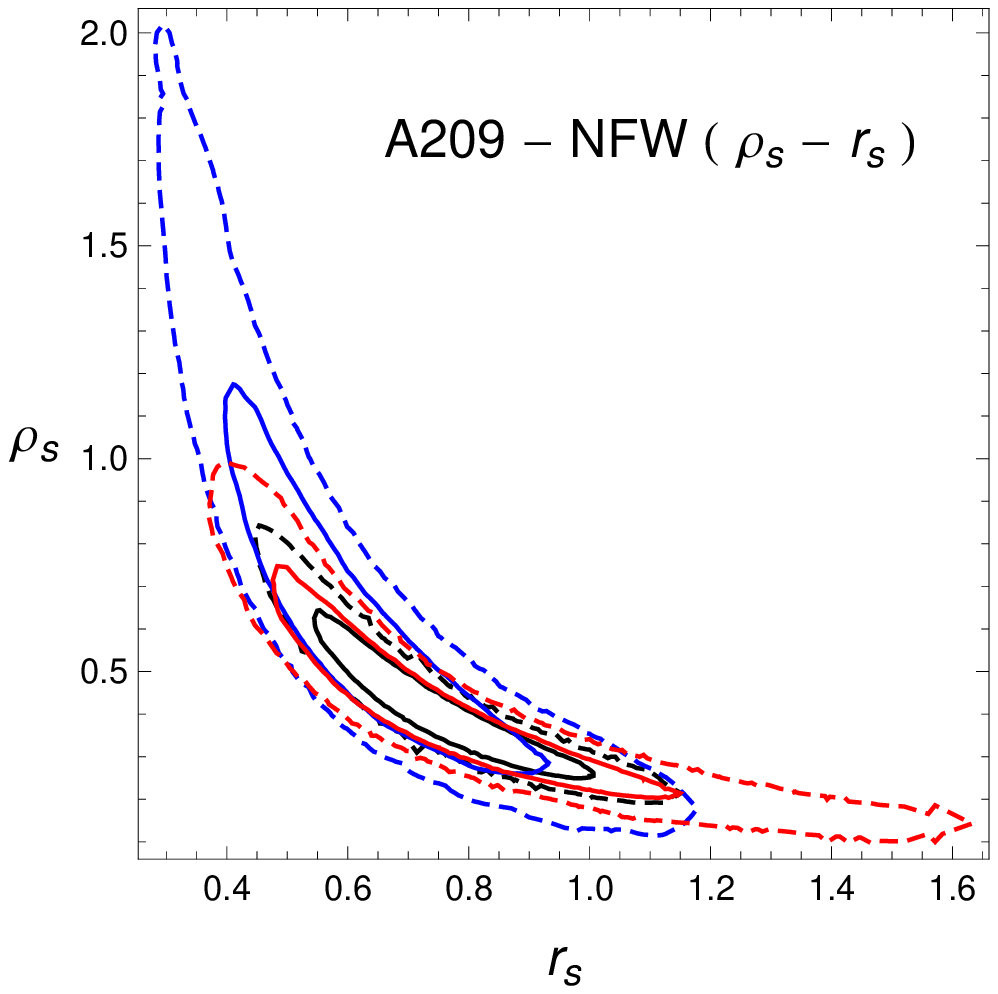}~~~
\includegraphics[width=5.5cm]{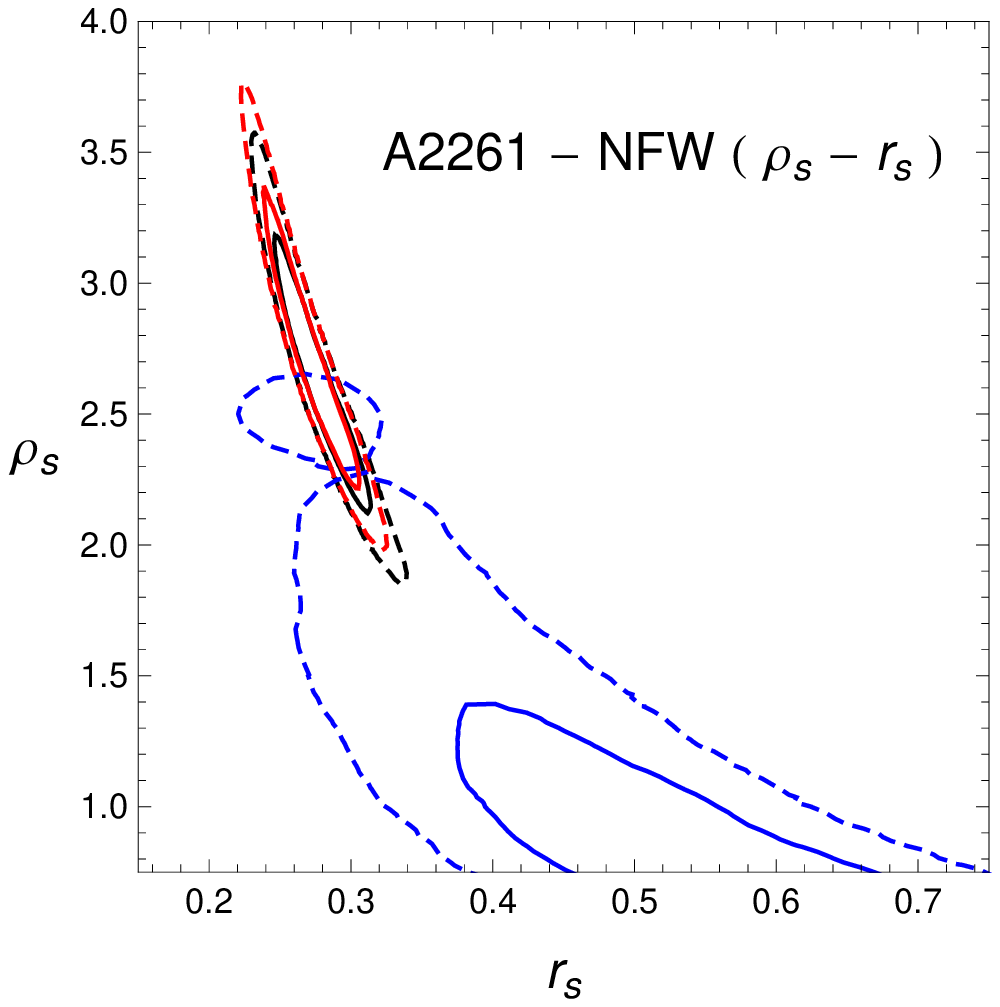}\\
~~~\\
\includegraphics[width=5.5cm]{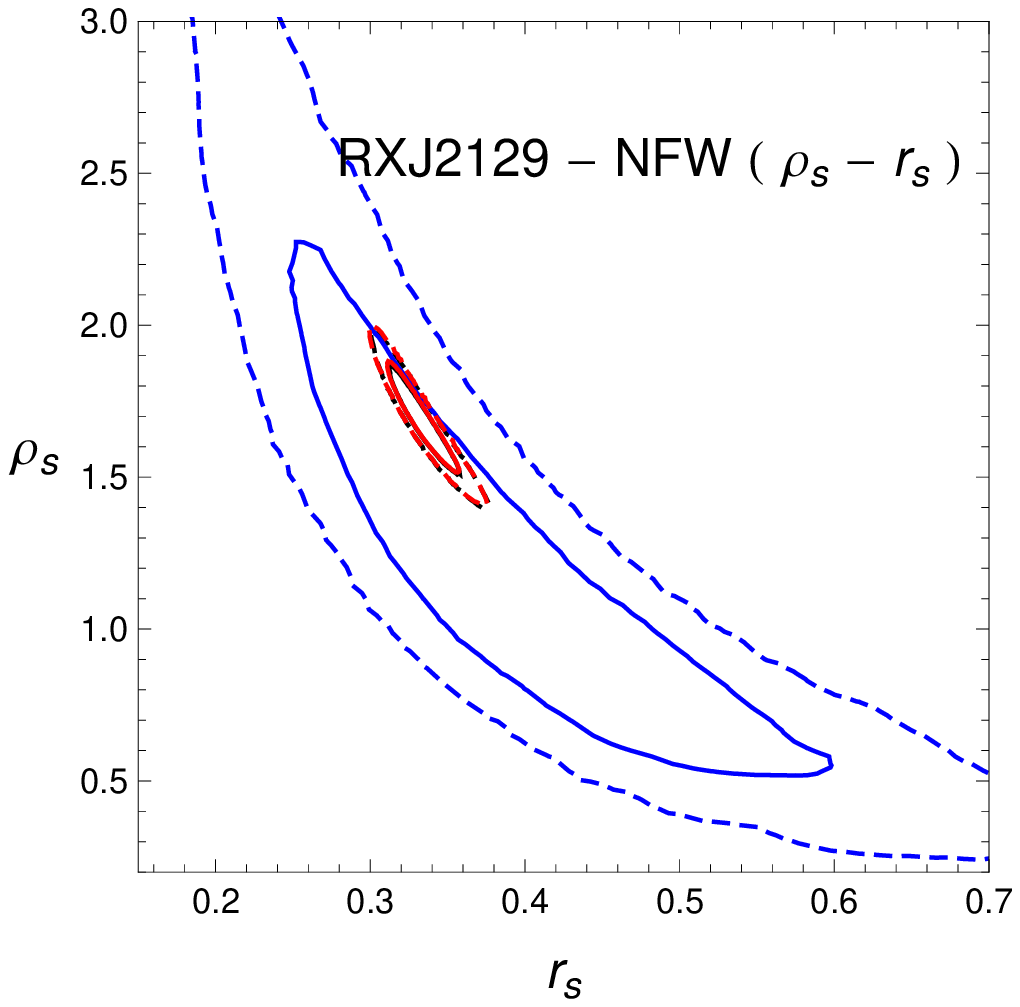}~~~
\includegraphics[width=5.5cm]{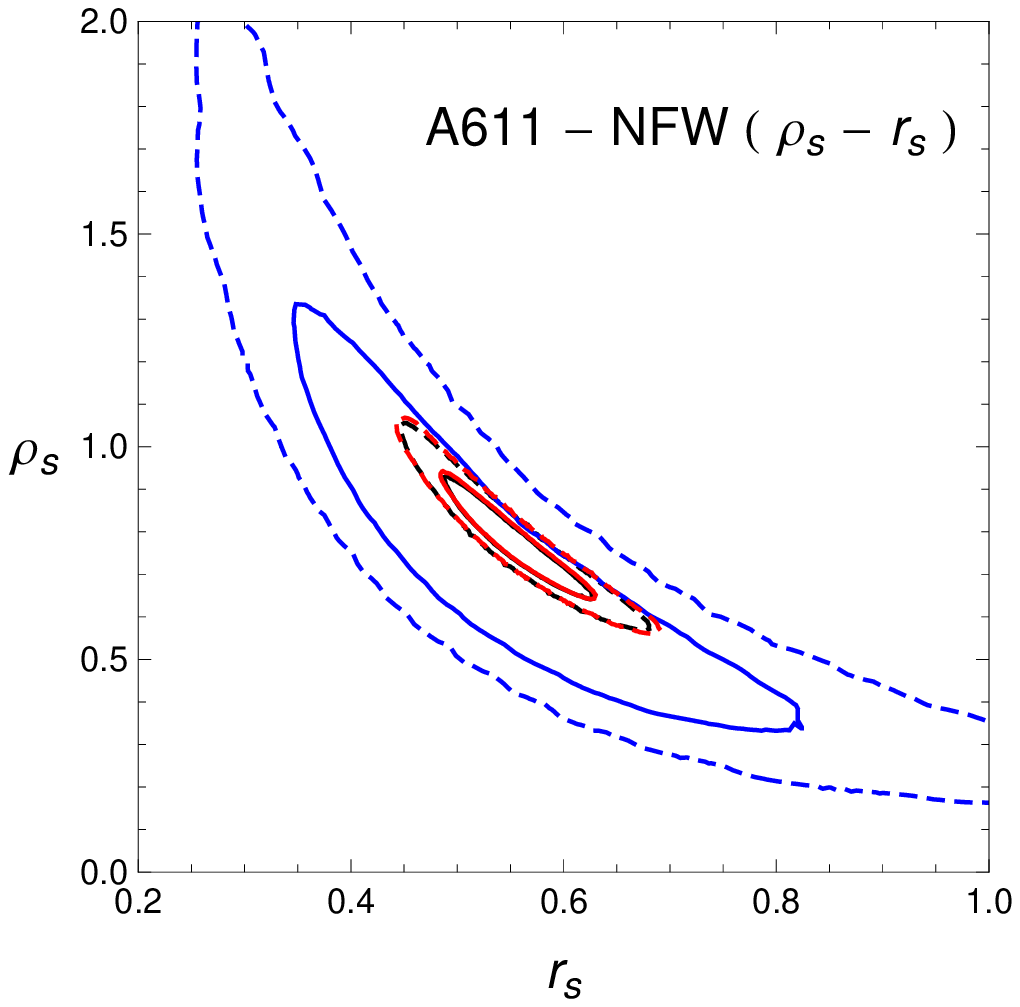}~~~
\includegraphics[width=5.5cm]{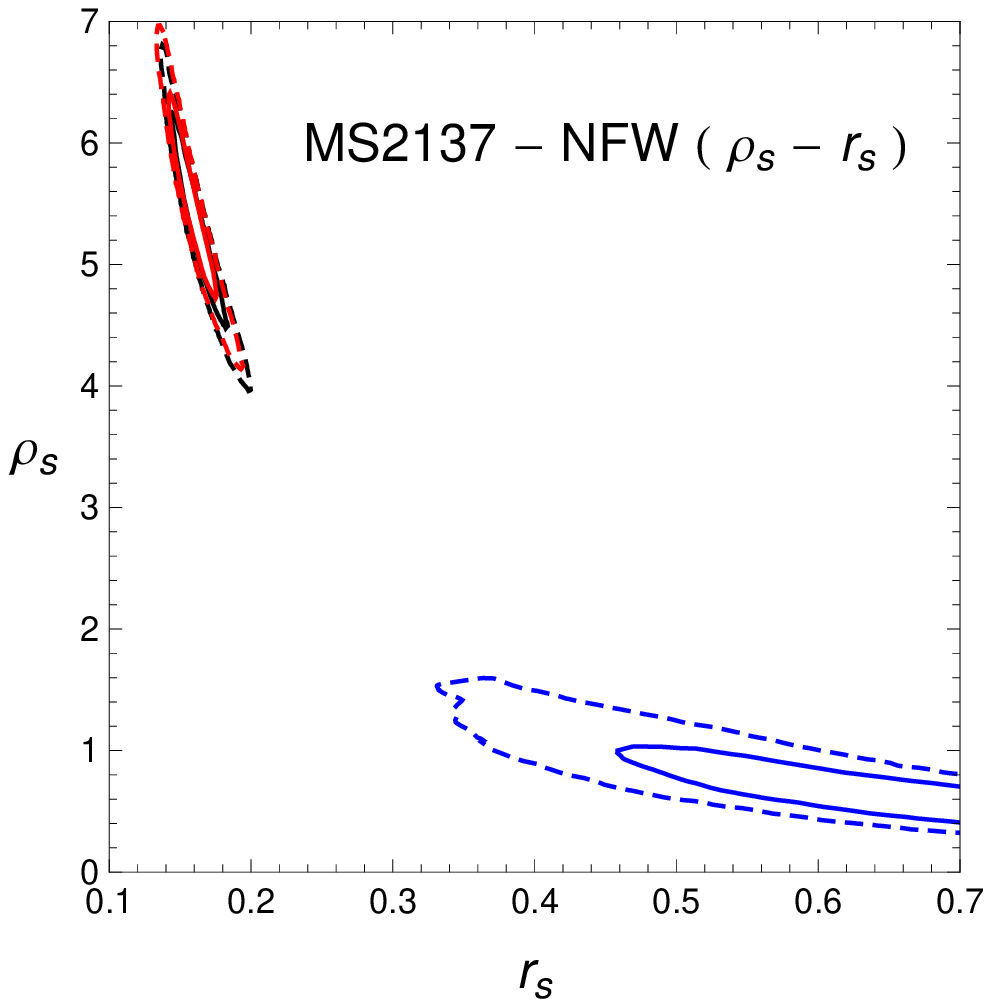} \\
~~~\\
\includegraphics[width=5.5cm]{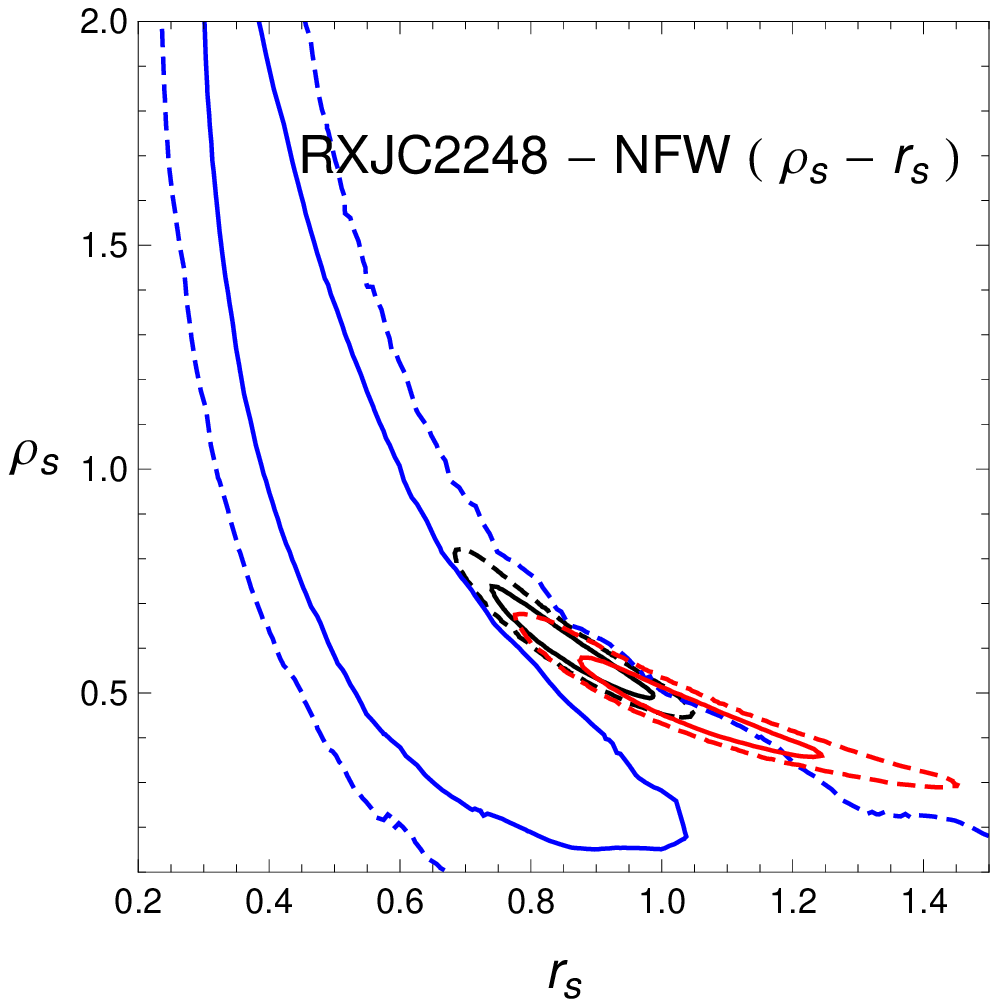}~~~
\includegraphics[width=5.5cm]{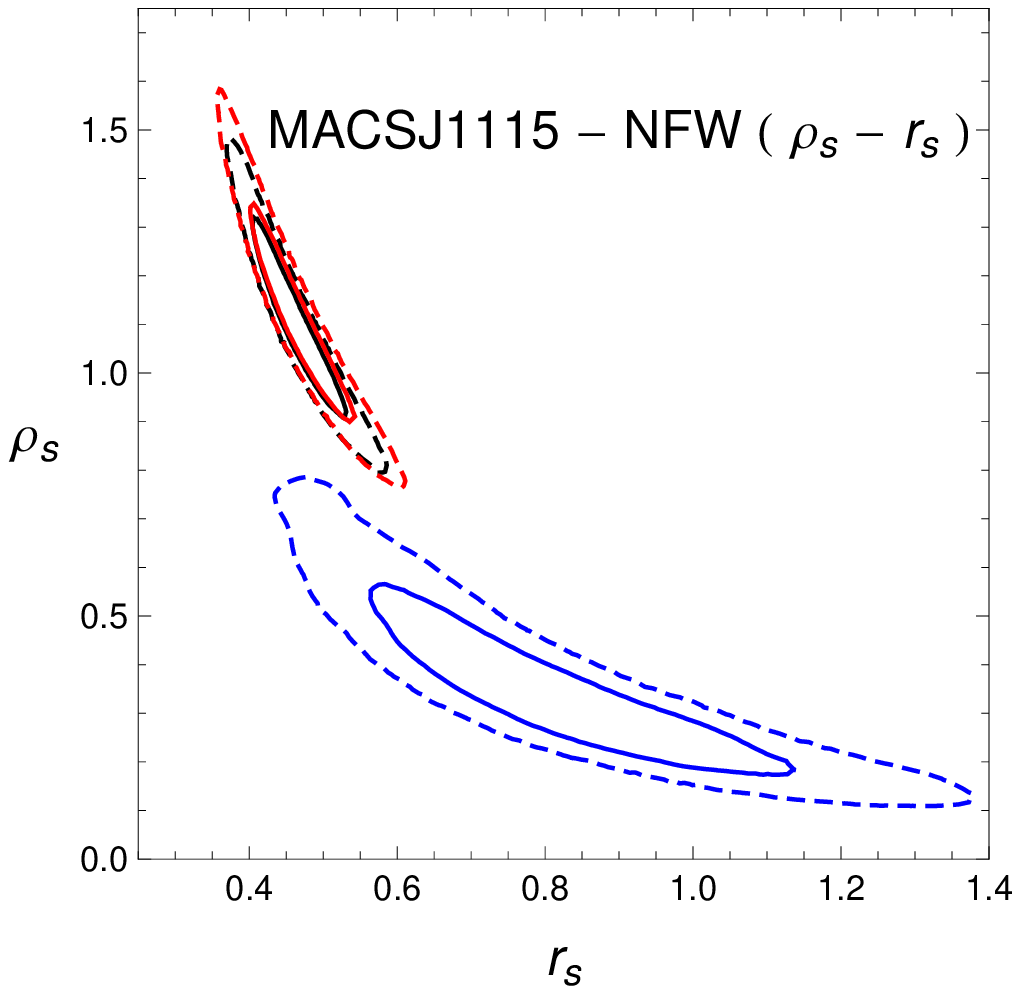}~~~
\includegraphics[width=5.5cm]{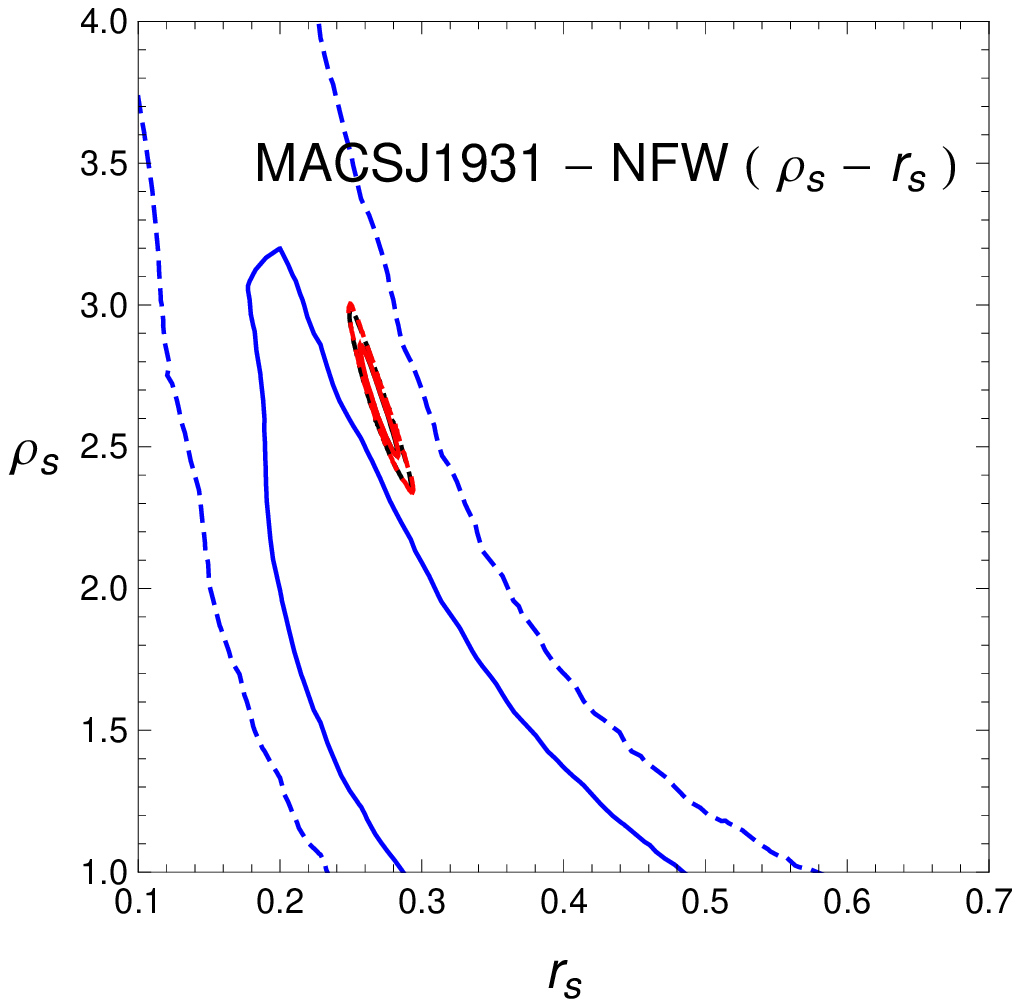}
\caption{NFW parameters likelihood, with scale $r_{s}$ in Mpc and density $\rho_{s}$ in $10^{15} \, M_{\odot}$ Mpc$^{-3}$. Blue: separate fit for lensing data; red: separate fit for X-ray gas data; black: joint analysis.}\label{fig:NFW_regions}
\end{figure*}

\begin{figure*}[htbp]
\ContinuedFloat
%\captionsetup{list=off,format=cont}
\centering
\includegraphics[width=5.5cm]{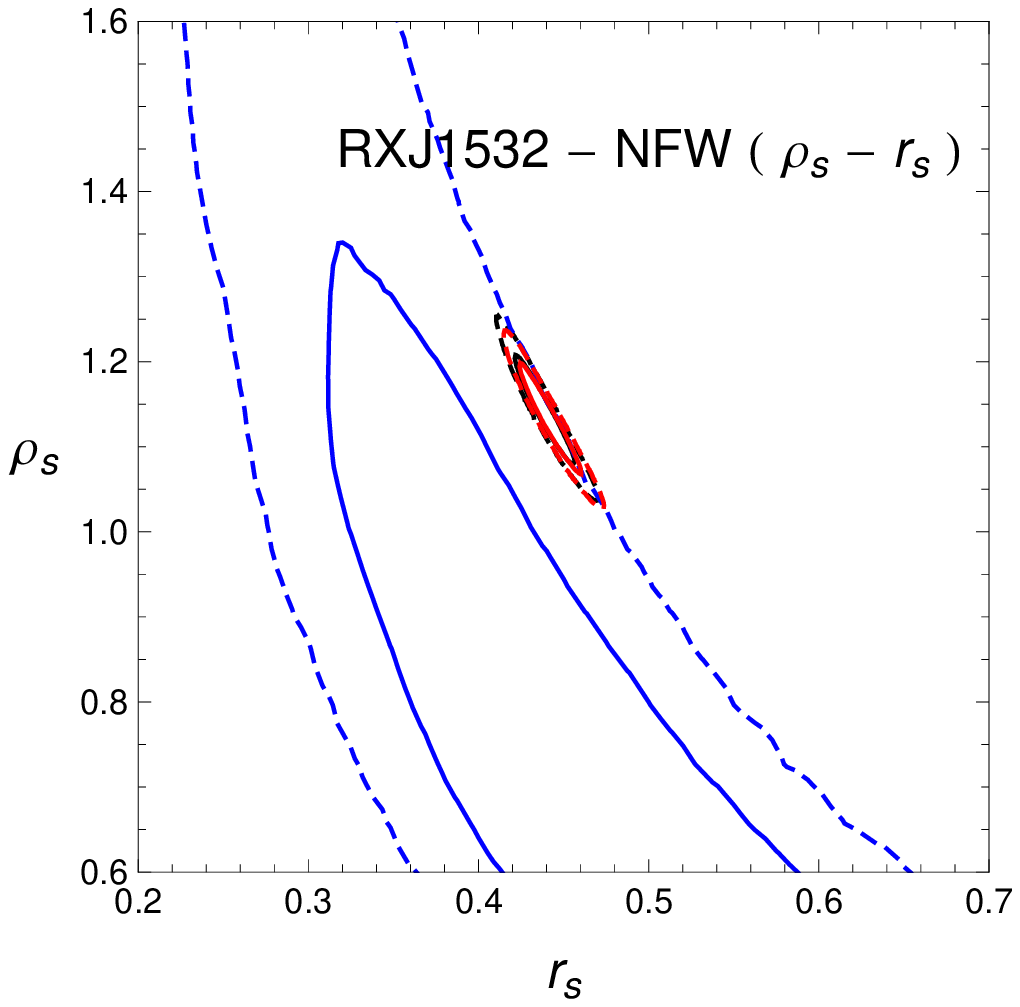}~~~
\includegraphics[width=5.5cm]{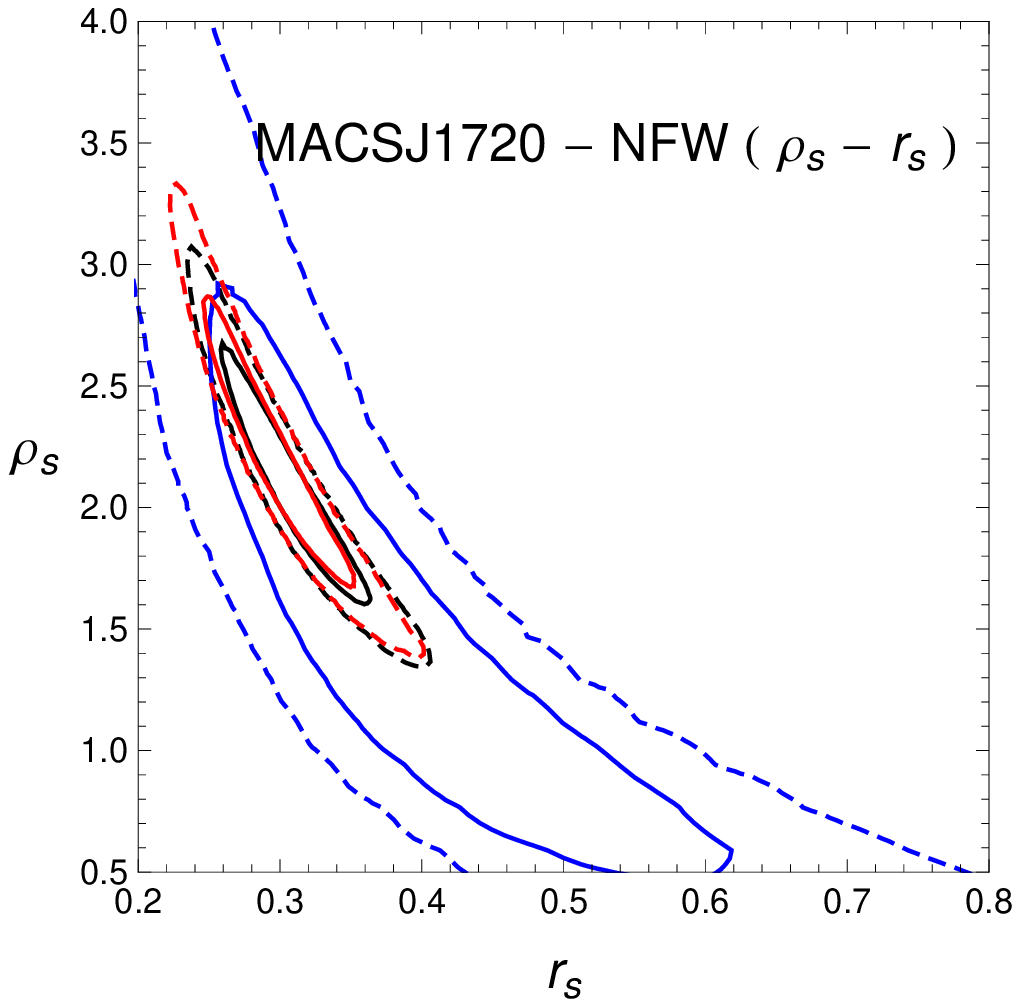}~~~
\includegraphics[width=5.5cm]{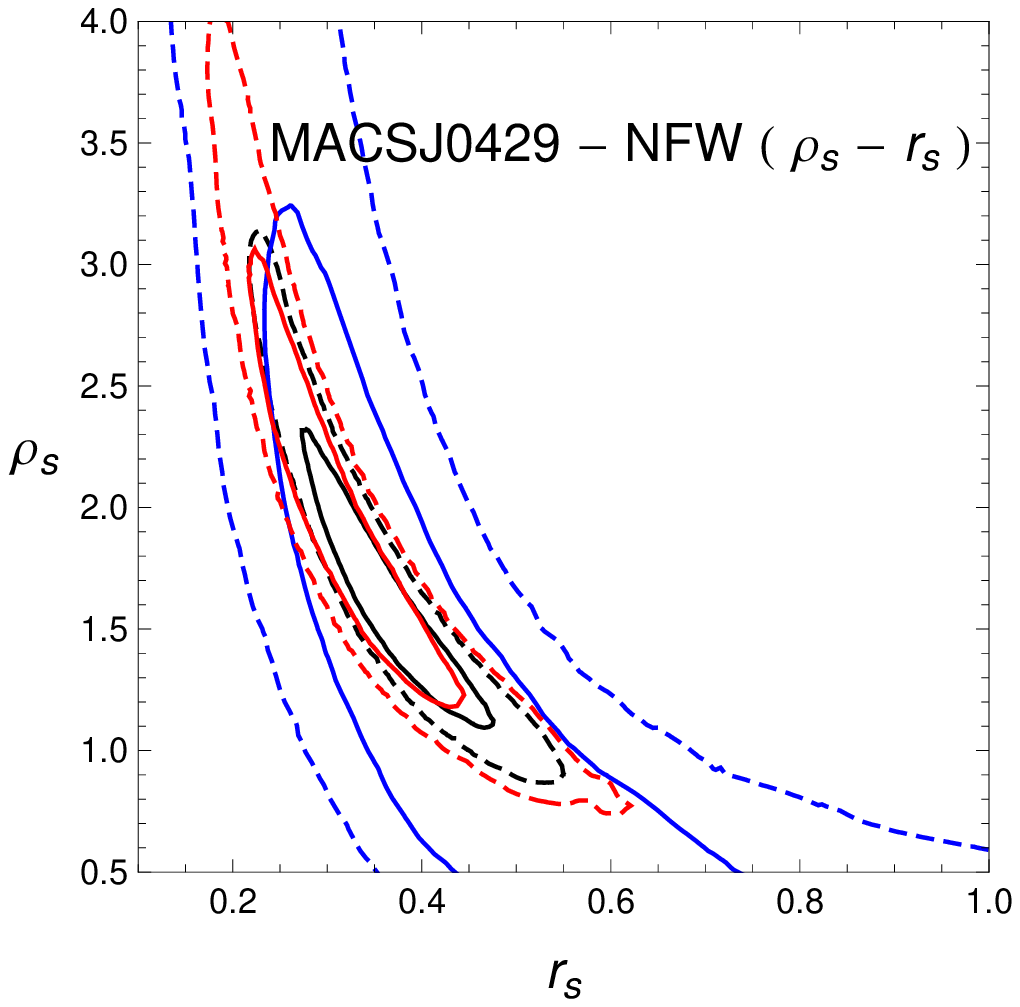}\\
~~~\\
\includegraphics[width=5.5cm]{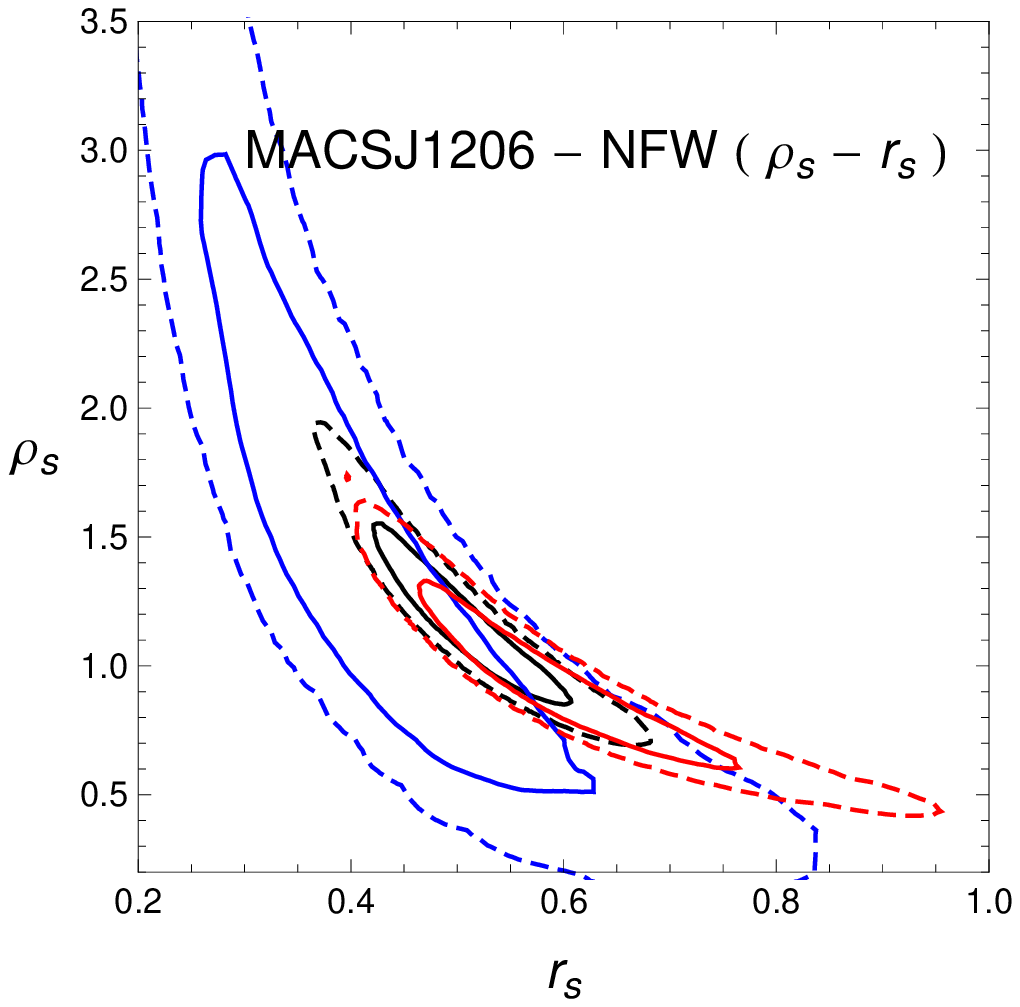}~~~
\includegraphics[width=5.5cm]{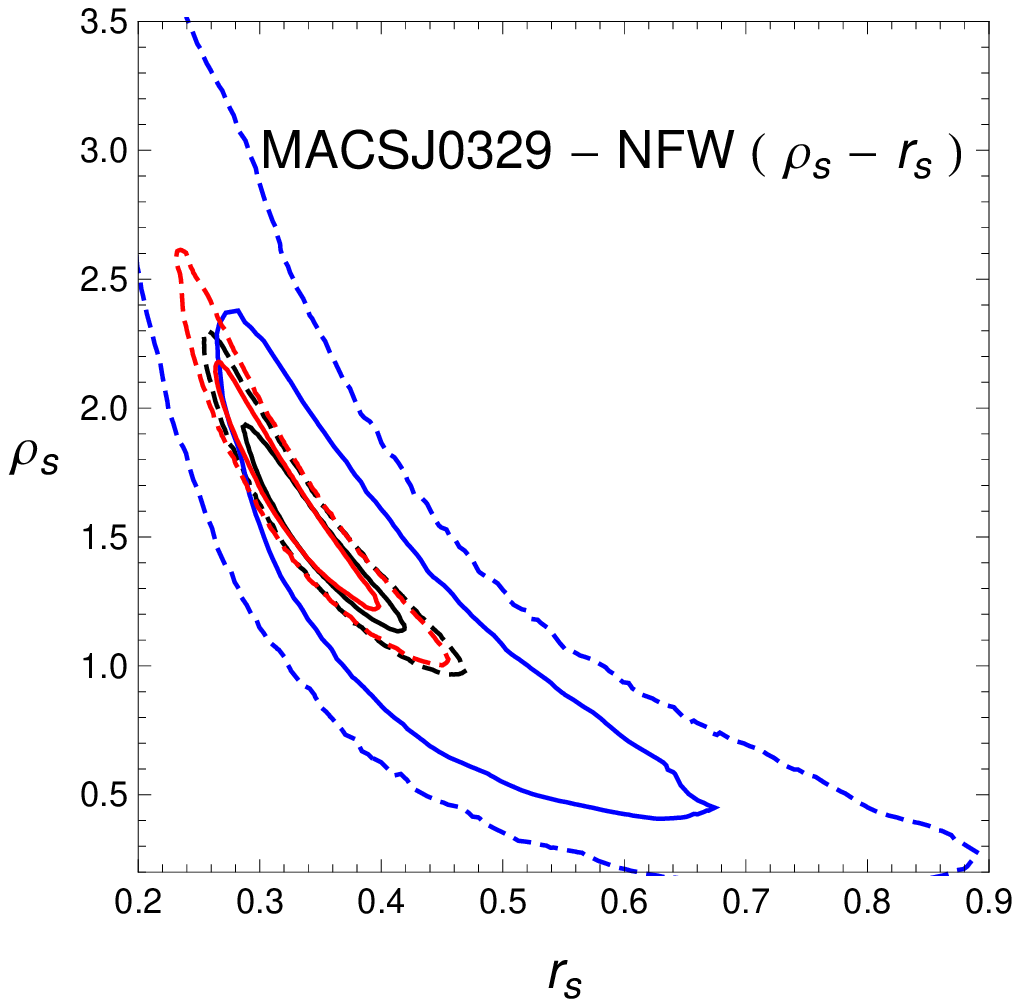}~~~
\includegraphics[width=5.5cm]{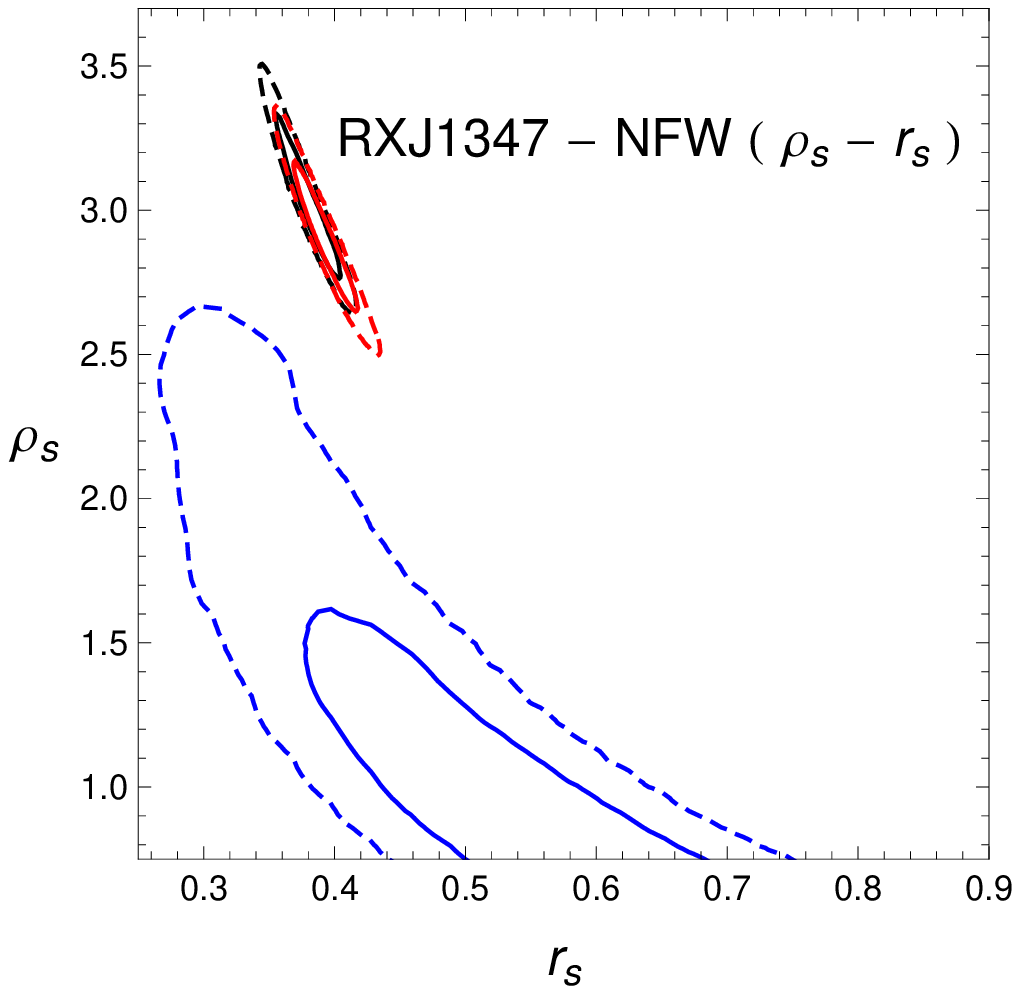}\\
~~~\\
\includegraphics[width=5.5cm]{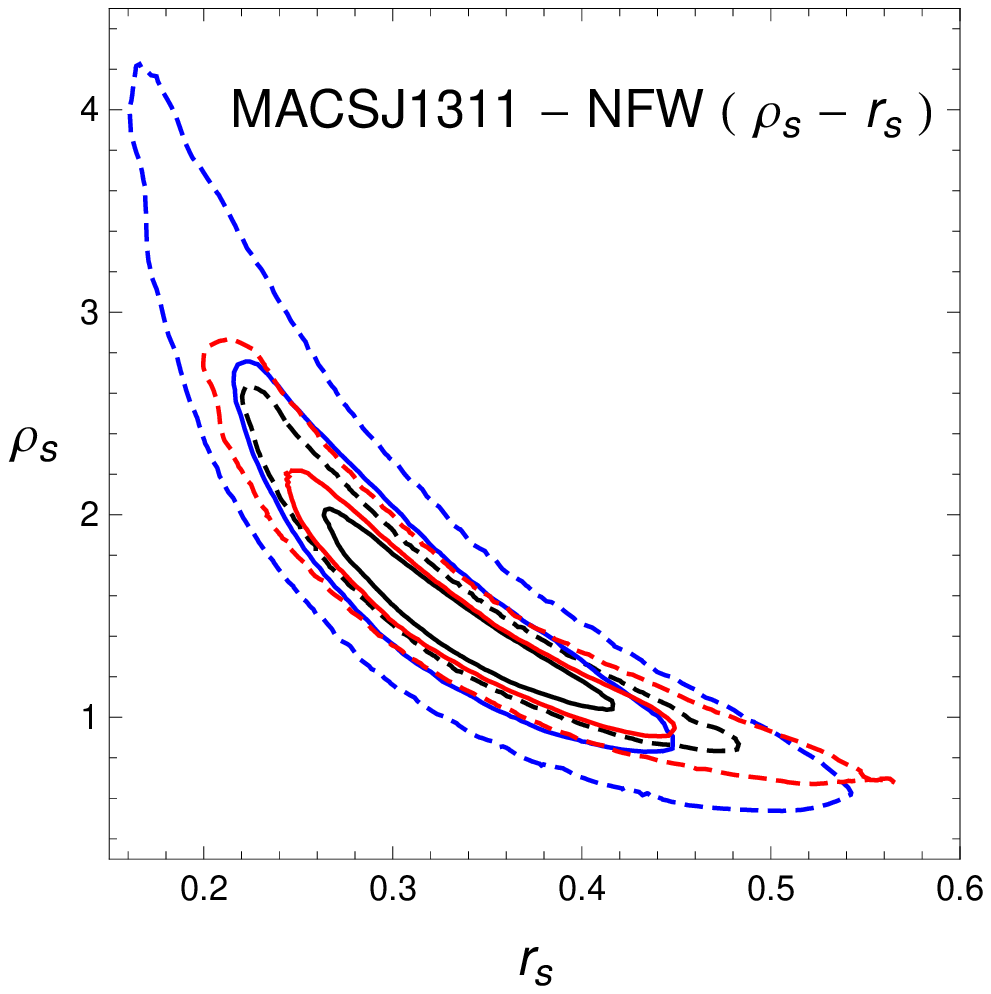}~~~
\includegraphics[width=5.5cm]{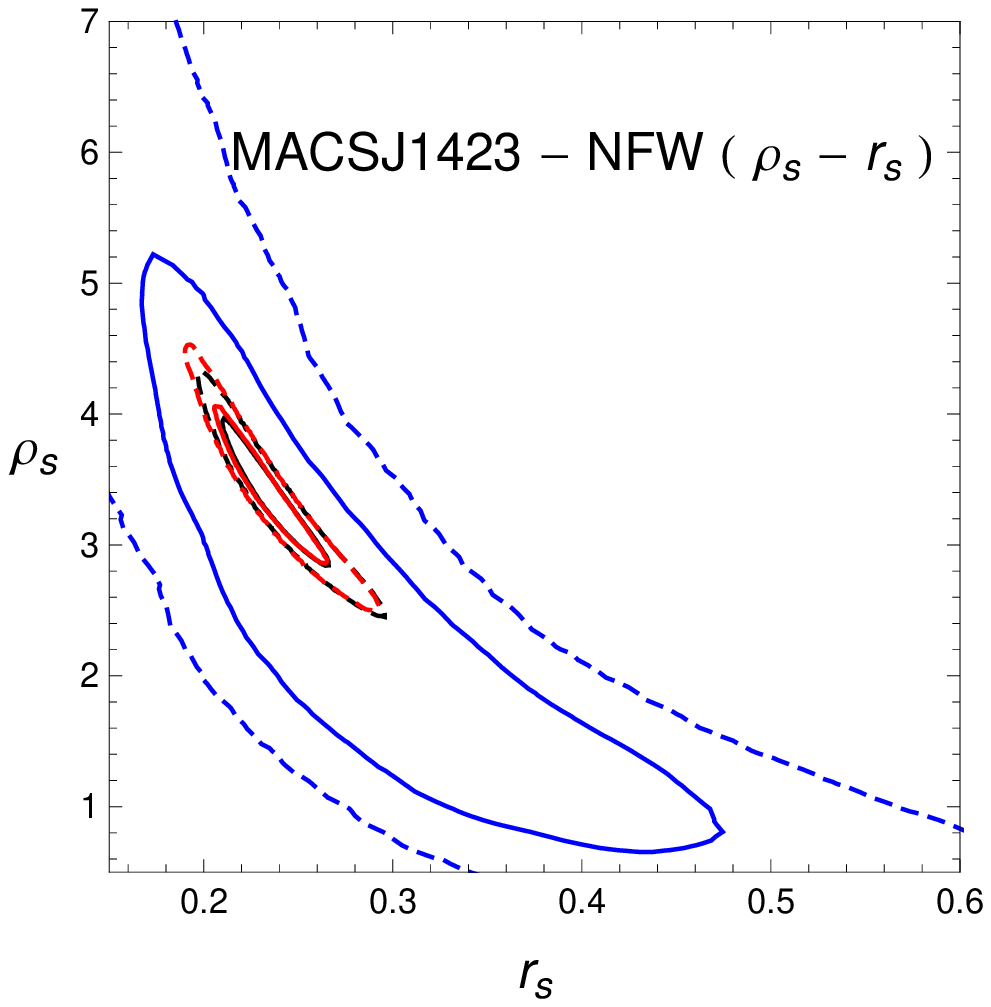}~~~
\includegraphics[width=5.5cm]{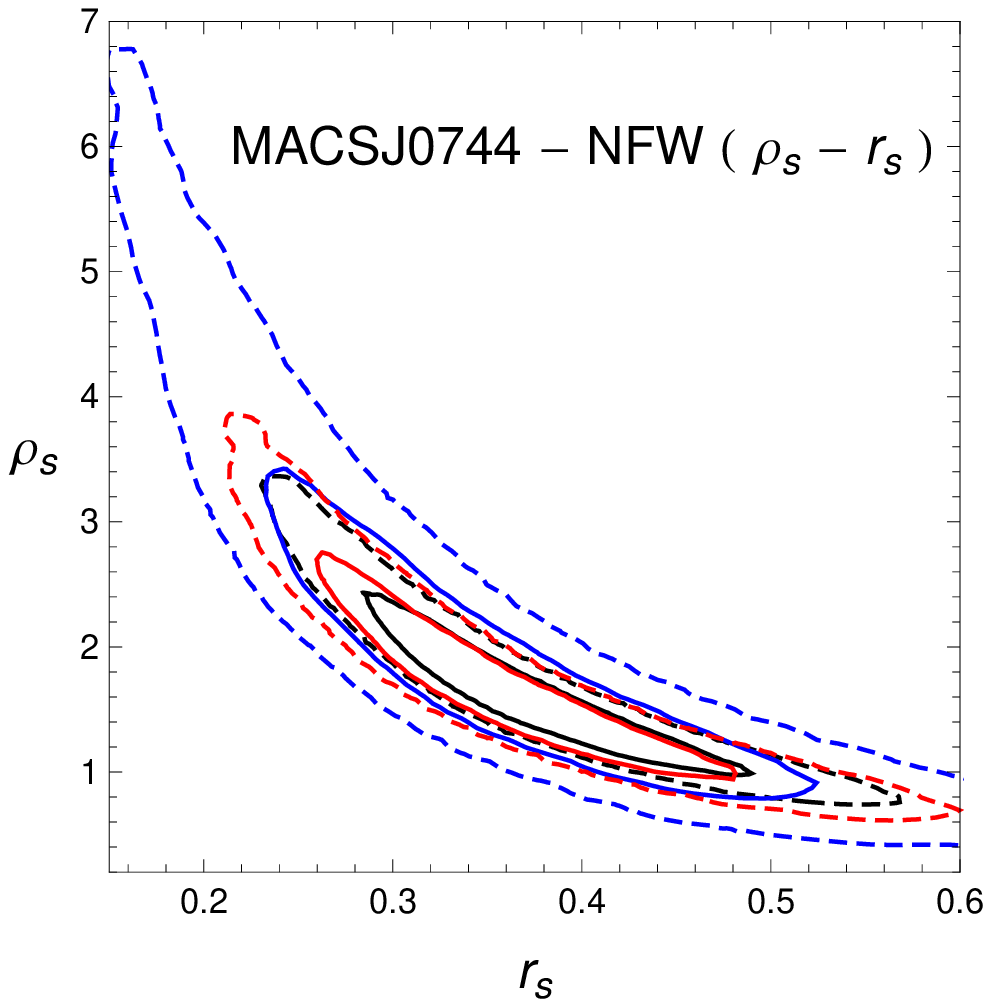}
\caption{NFW parameters likelihood, with scale $r_{s}$ in Mpc and density $\rho_{s}$ in $10^{15} \, M_{\odot}$ Mpc$^{-3}$. Blue: separate fit for lensing data; red: separate fit for X-ray gas data; black: joint analysis.}
%\caption{\textcolor{red}{``Normalized'' X-ray mass profiles}. On the left, thermal X-ray total mass estimation; on the right, gravitational lensing reconstruction. Color code:}\label{fig:lens_gas_2_Norm}
\end{figure*}

%\clearpage
%\newpage

\begin{figure*}[htbp]
\centering
\includegraphics[width=16.cm]{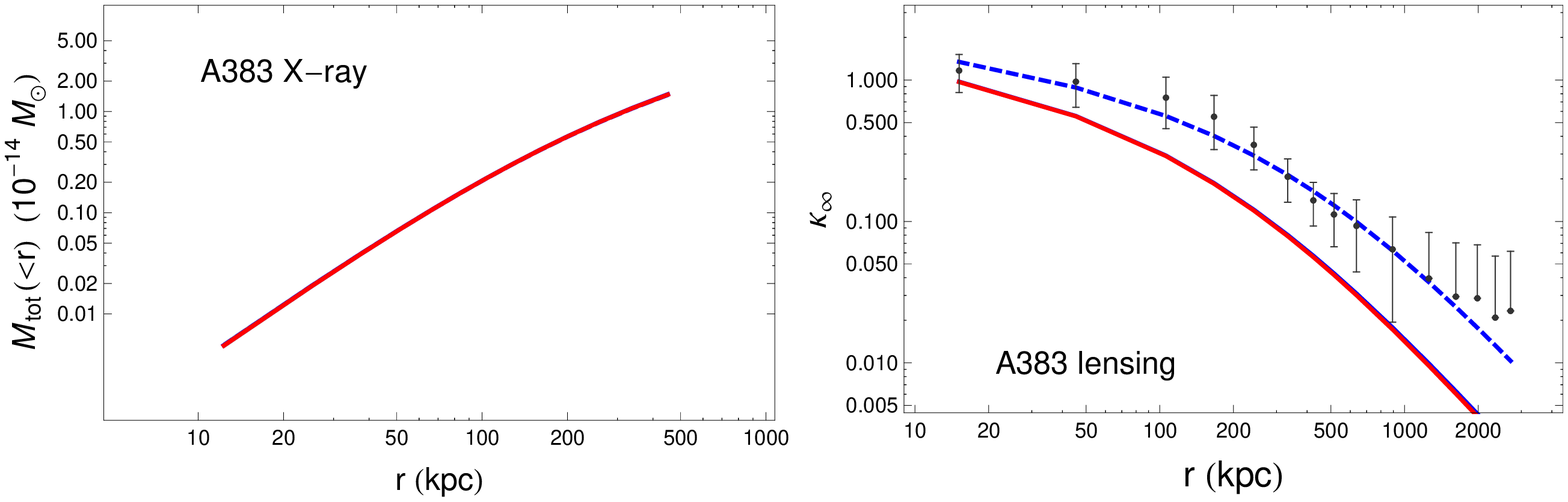} \\
\includegraphics[width=16.cm]{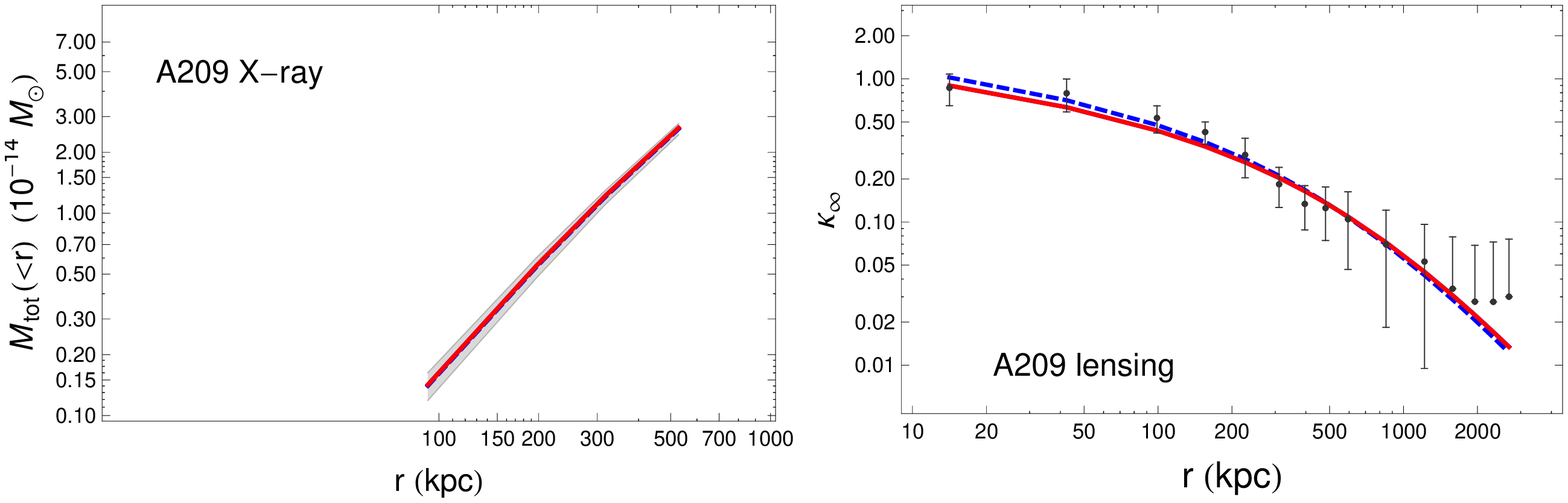} \\
\includegraphics[width=16.cm]{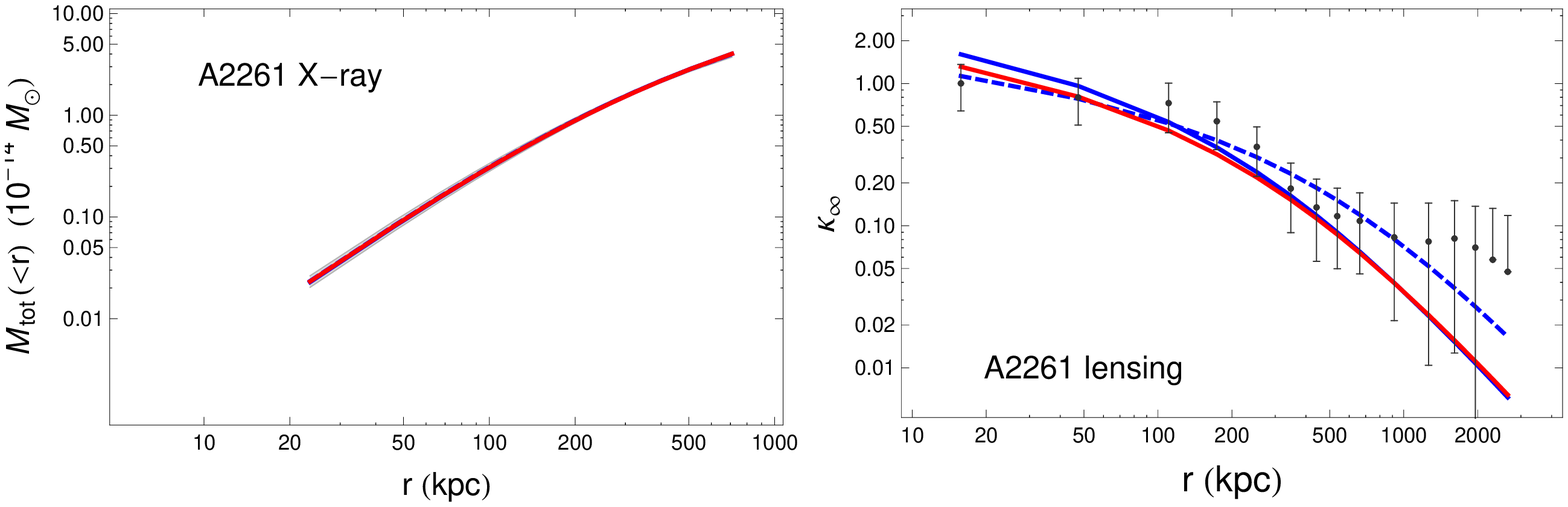} \\
\includegraphics[width=16.cm]{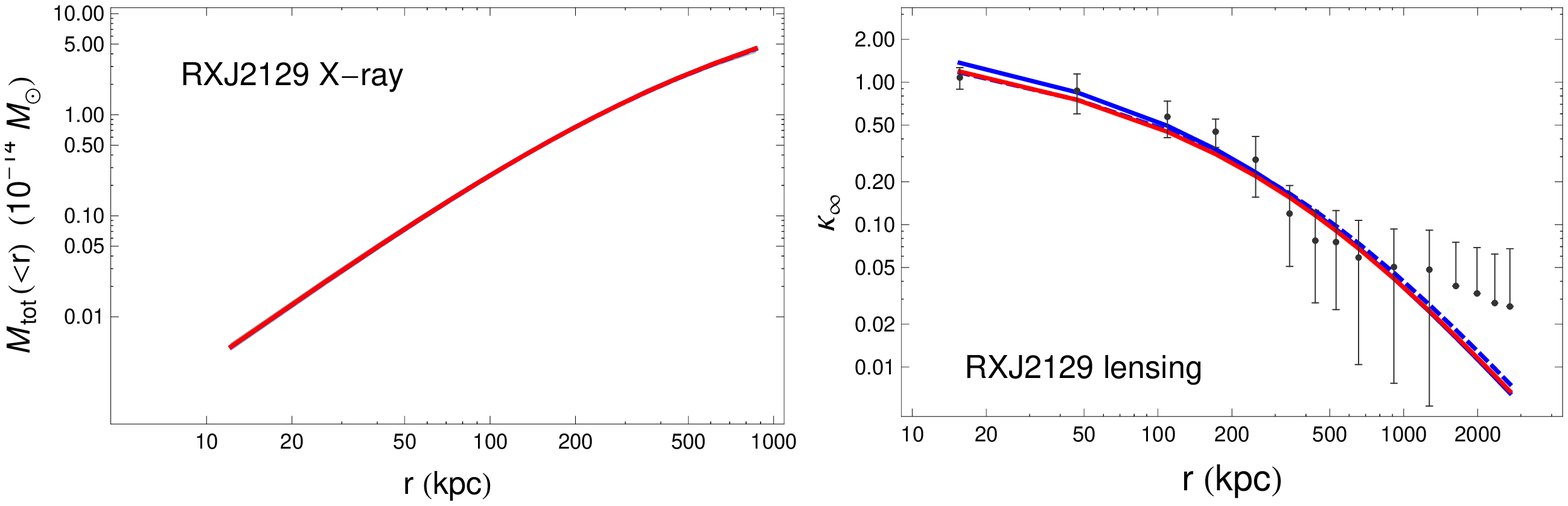}
\caption{Mass profiles from thermal X-ray gas (left) and gravitational lensing reconstruction (right). Color code: grey regions/points - observational data; dashed blue - NFW + GR fit from gas-only (right)/lensing-only (left); solid blue - NFW + GR from joint fit; solid red - NFW + galileon from joint fit.}\label{fig:final_plot}
\end{figure*}

\begin{figure*}[htbp]
\ContinuedFloat
%\captionsetup{list=off,format=cont}
\centering
\includegraphics[width=16.cm]{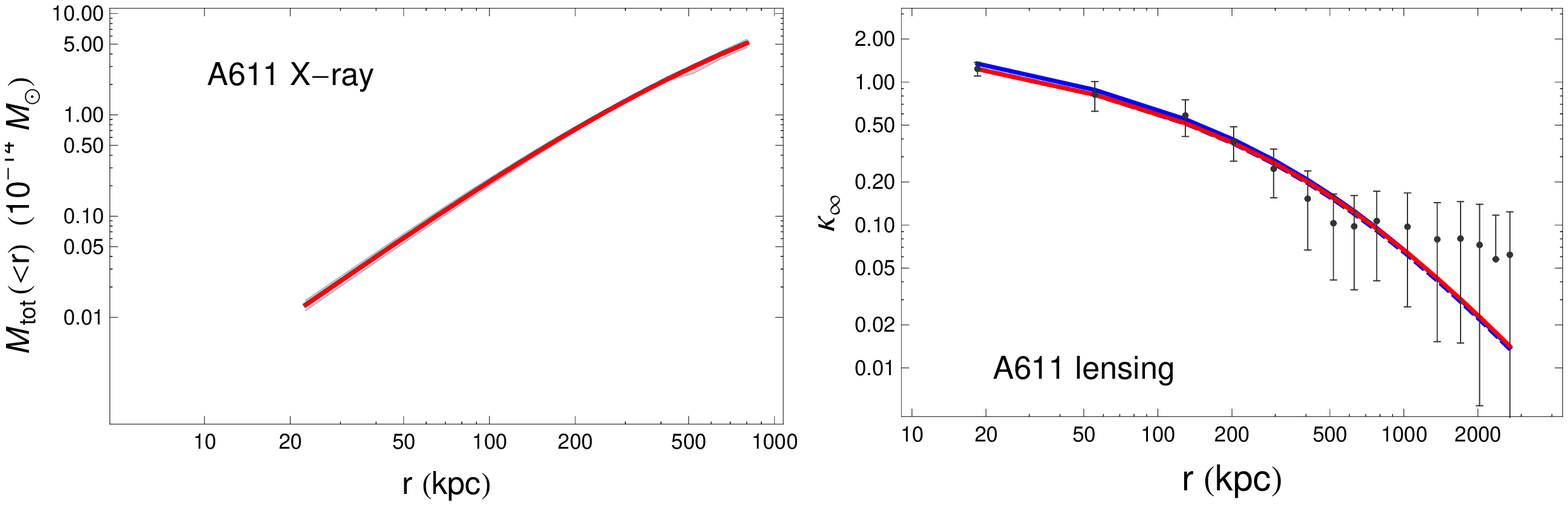} \\
\includegraphics[width=16.cm]{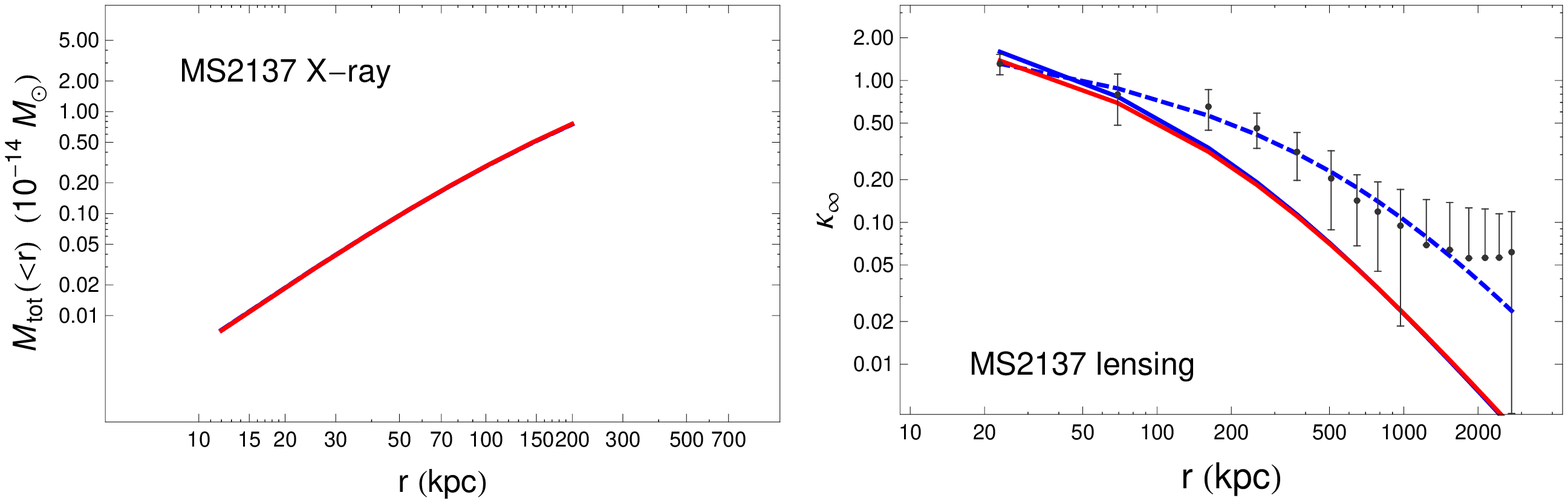} \\
\includegraphics[width=16.cm]{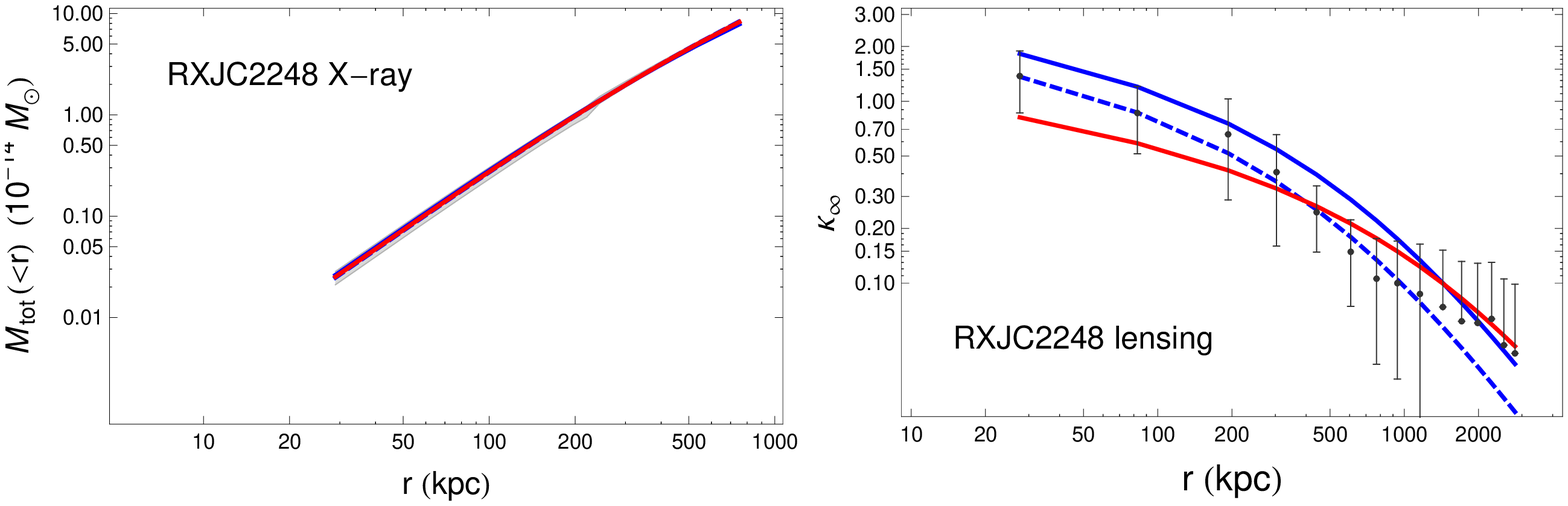} \\
\includegraphics[width=16.cm]{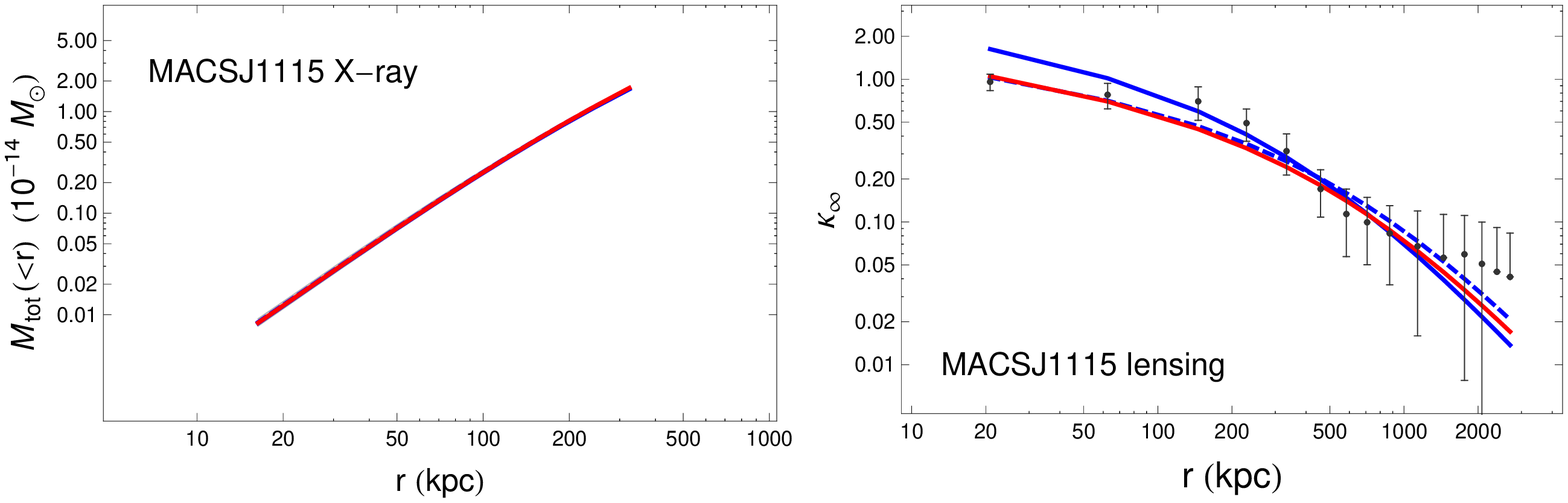}
\caption{Mass profiles from thermal X-ray gas (left) and gravitational lensing reconstruction (right). Color code: grey regions/points - observational data; dashed blue - NFW + GR fit from gas-only (right)/lensing-only (left); solid blue - NFW + GR from joint fit; solid red - NFW + galileon from joint fit.}
%\caption{\textcolor{red}{Non-``normalized'' X-ray mass profiles}. On the left, thermal X-ray total mass estimation; on the right, gravitational lensing reconstruction. Color code: grey regions/points - observational data; solid blue - density: NFW (in GR); dashed blue - density: NFW + gas (in GR); dotdashed blue - density: NFW + gas + BCG (in GR); solid red - density: NFW + galileon; dashed red: NFW + gas + galileon; dotdashed red: NFW + gas + BCG + galileon.}\label{fig:lens_gas_2}
\end{figure*}

\begin{figure*}[htbp]
\ContinuedFloat
%\captionsetup{list=off,format=cont}
\centering
\includegraphics[width=16.cm]{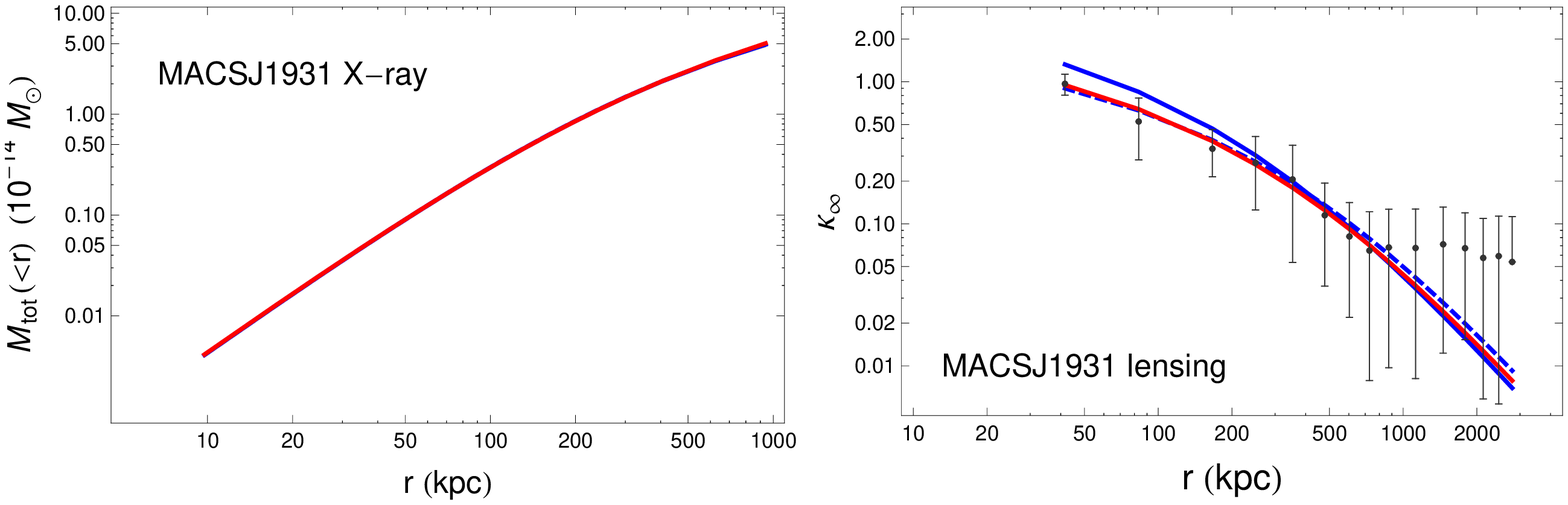} \\
\includegraphics[width=16.cm]{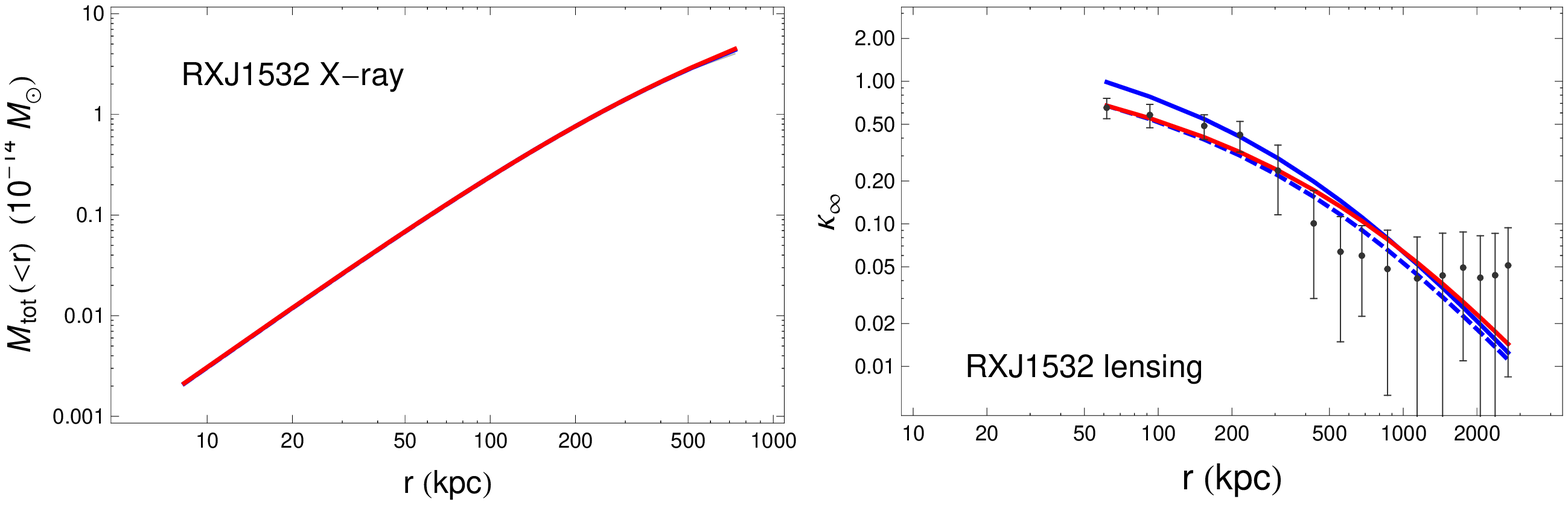} \\
\includegraphics[width=16.cm]{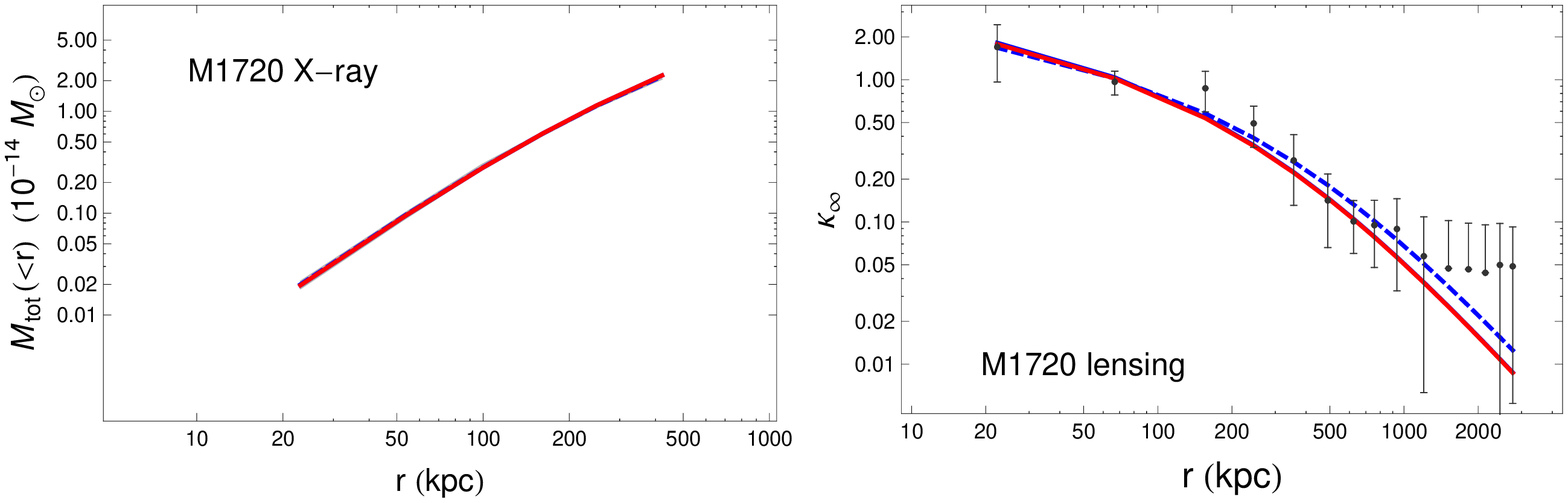} \\
\includegraphics[width=16.cm]{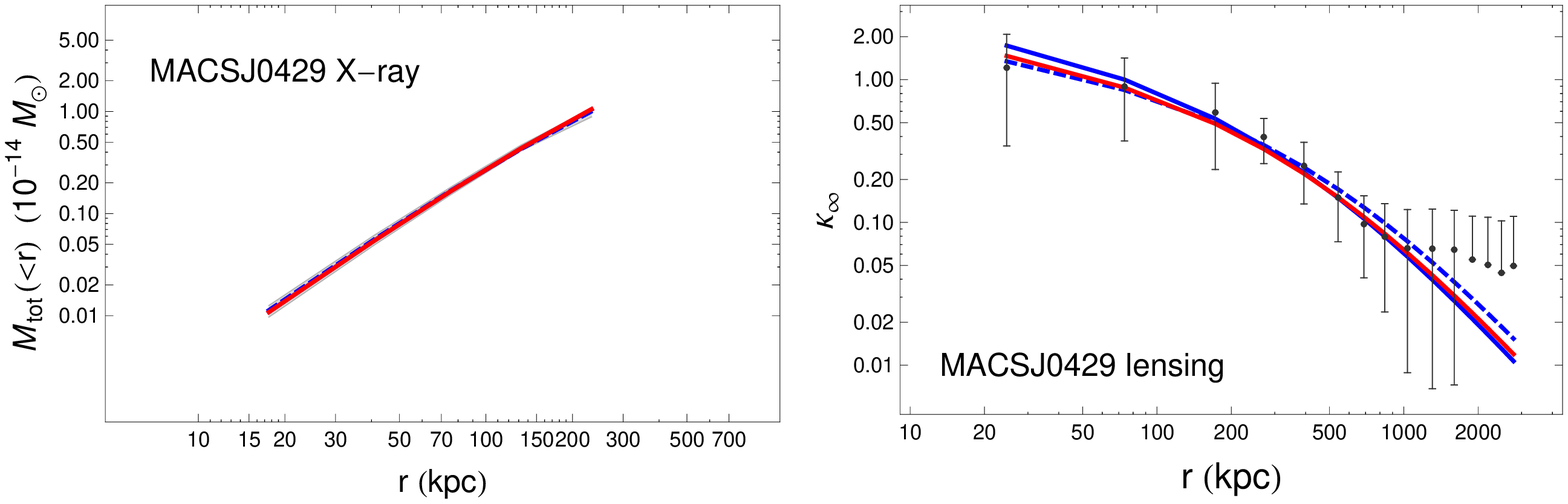}
\caption{Mass profiles from thermal X-ray gas (left) and gravitational lensing reconstruction (right). Color code: grey regions/points - observational data; dashed blue - NFW + GR fit from gas-only (right)/lensing-only (left); solid blue - NFW + GR from joint fit; solid red - NFW + galileon from joint fit.}
%\caption{\textcolor{red}{Non-``normalized'' X-ray mass profiles}. On the left, thermal X-ray total mass estimation; on the right, gravitational lensing reconstruction. Color code: grey regions/points - observational data; solid blue - density: NFW (in GR); dashed blue - density: NFW + gas (in GR); dotdashed blue - density: NFW + gas + BCG (in GR); solid red - density: NFW + galileon; dashed red: NFW + gas + galileon; dotdashed red: NFW + gas + BCG + galileon.}\label{fig:lens_gas_3}
\end{figure*}

\begin{figure*}[htbp]
\ContinuedFloat
%\captionsetup{list=off,format=cont}
\centering
\includegraphics[width=16.cm]{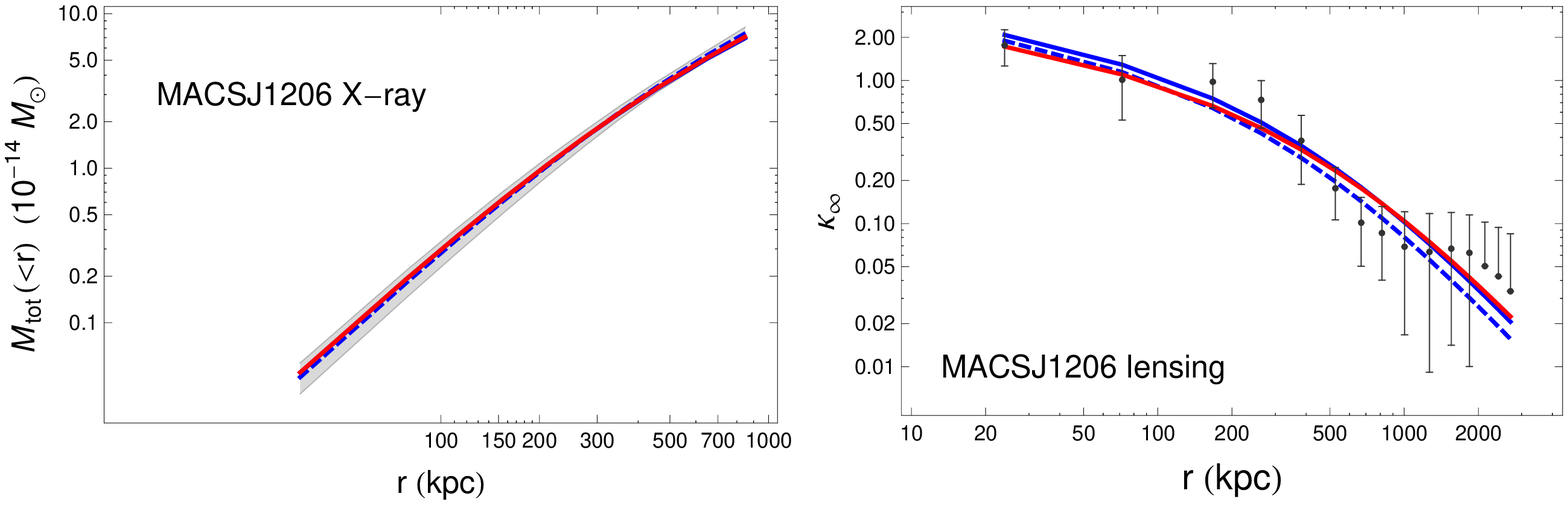} \\
\includegraphics[width=16.cm]{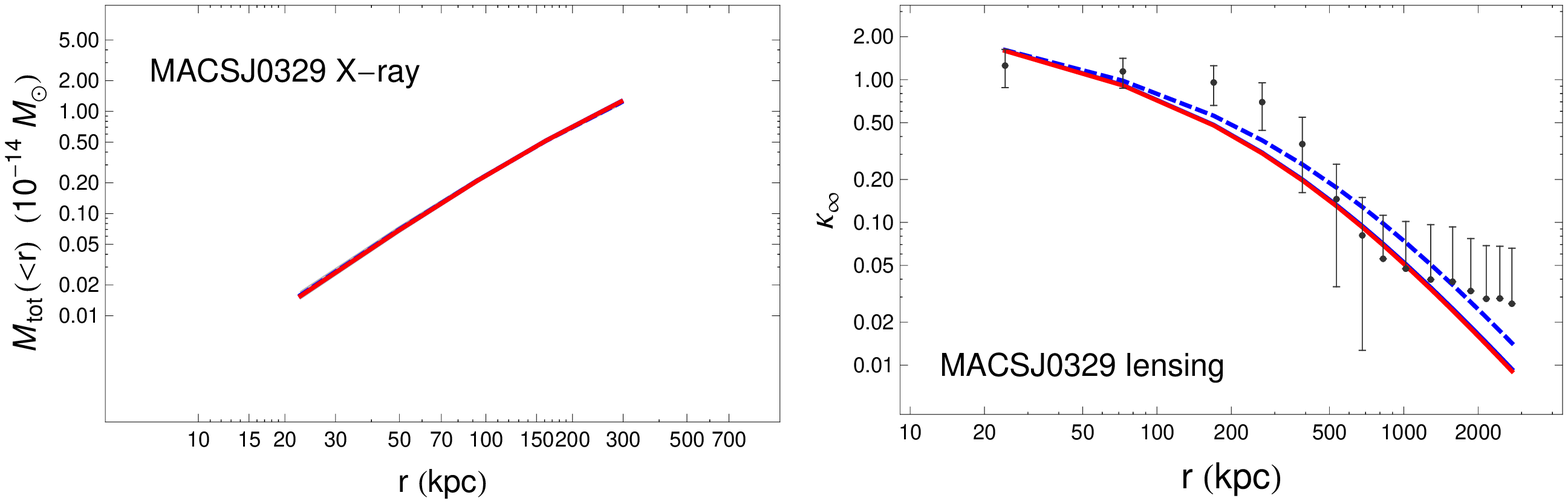} \\
\includegraphics[width=16.cm]{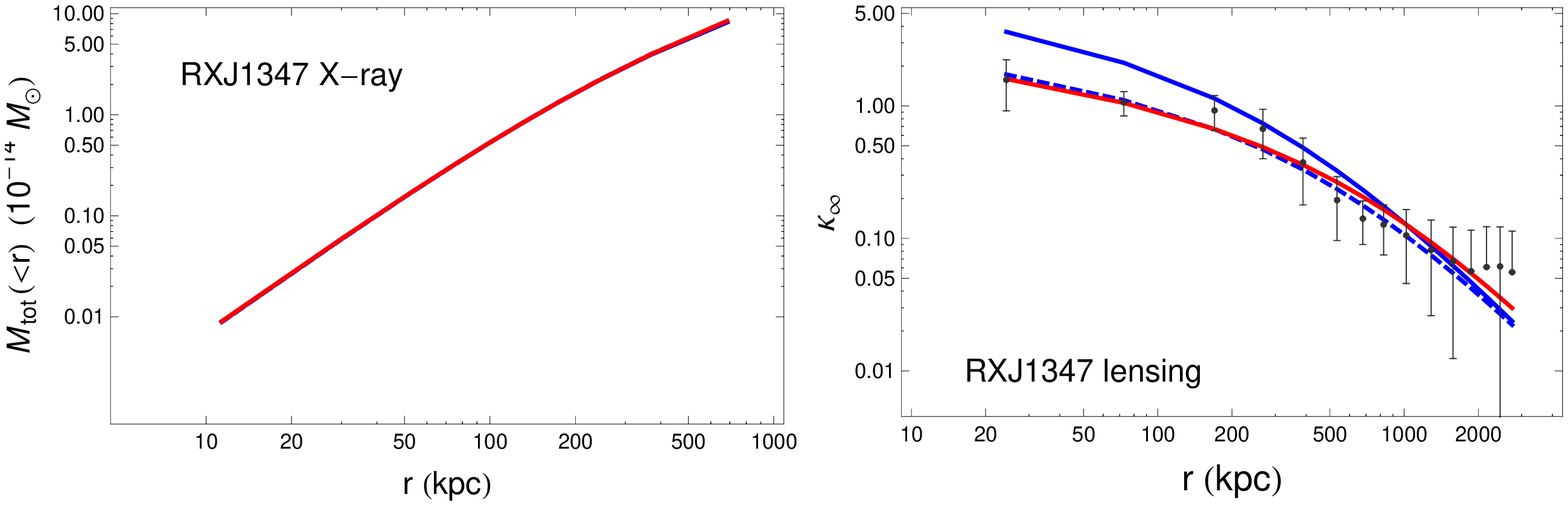} \\
\includegraphics[width=16.cm]{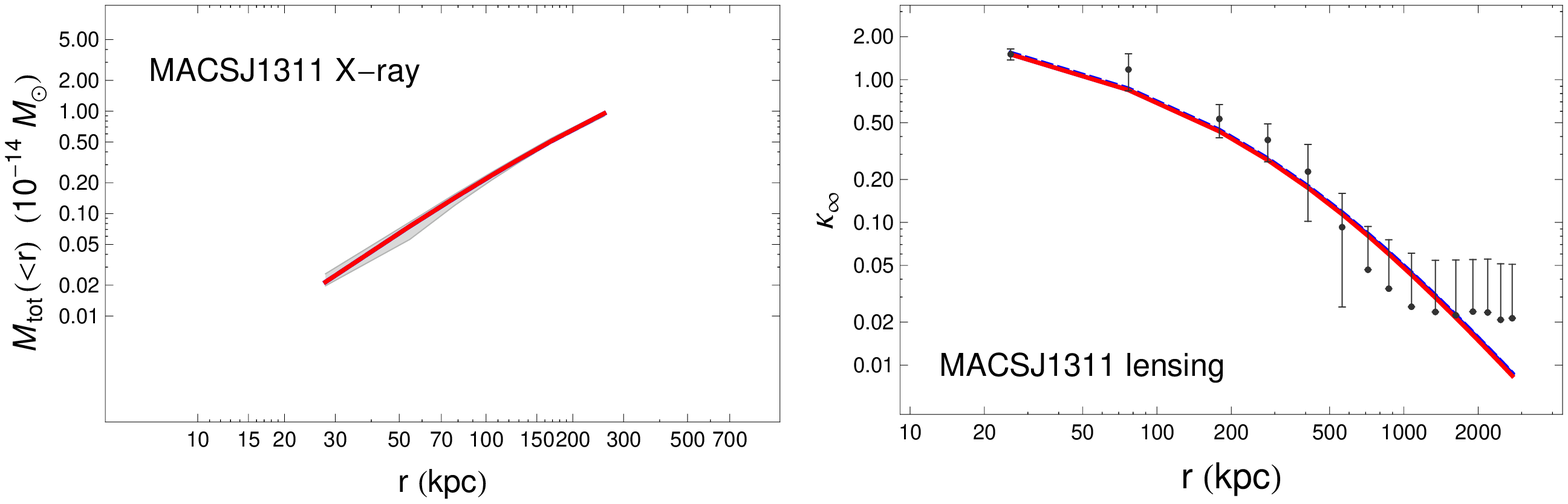}
\caption{Mass profiles from thermal X-ray gas (left) and gravitational lensing reconstruction (right). Color code: grey regions/points - observational data; dashed blue - NFW + GR fit from gas-only (right)/lensing-only (left); solid blue - NFW + GR from joint fit; solid red - NFW + galileon from joint fit.}
%\caption{\textcolor{red}{Non-``normalized'' X-ray mass profiles}. On the left, thermal X-ray total mass estimation; on the right, gravitational lensing reconstruction. Color code: grey regions/points - observational data; solid blue - density: NFW (in GR); dashed blue - density: NFW + gas (in GR); dotdashed blue - density: NFW + gas + BCG (in GR); solid red - density: NFW + galileon; dashed red: NFW + gas + galileon; dotdashed red: NFW + gas + BCG + galileon.}\label{fig:lens_gas_4}
\end{figure*}

\begin{figure*}[htbp]
\ContinuedFloat
%\captionsetup{list=off,format=cont}
\centering
\includegraphics[width=16.cm]{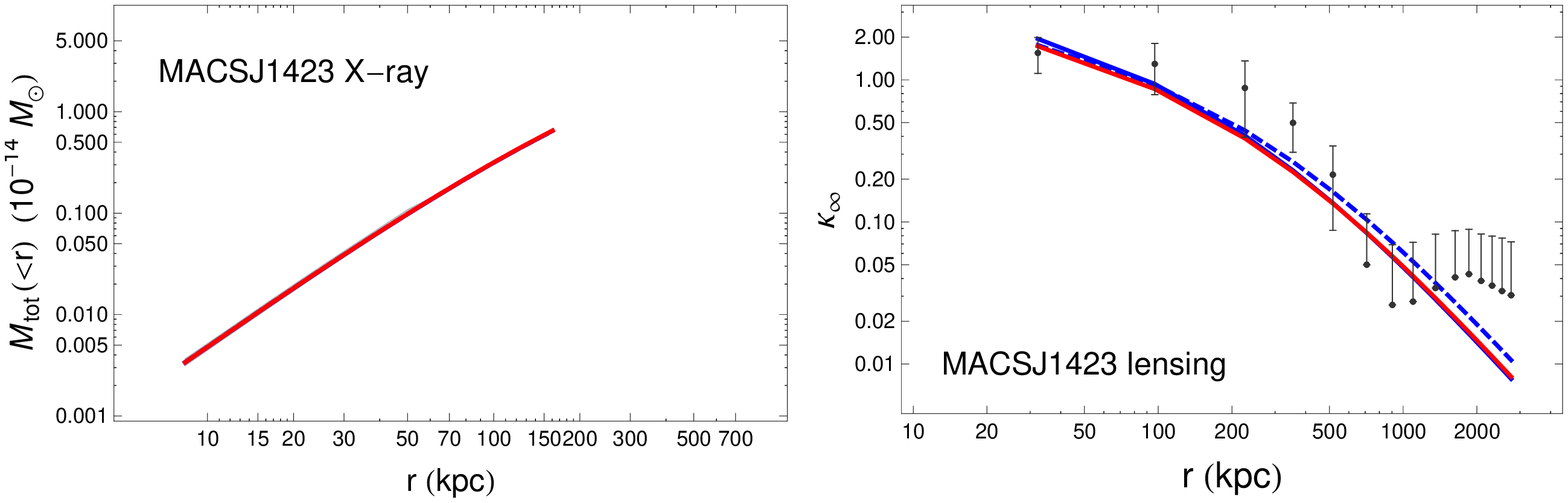} \\
\includegraphics[width=16.cm]{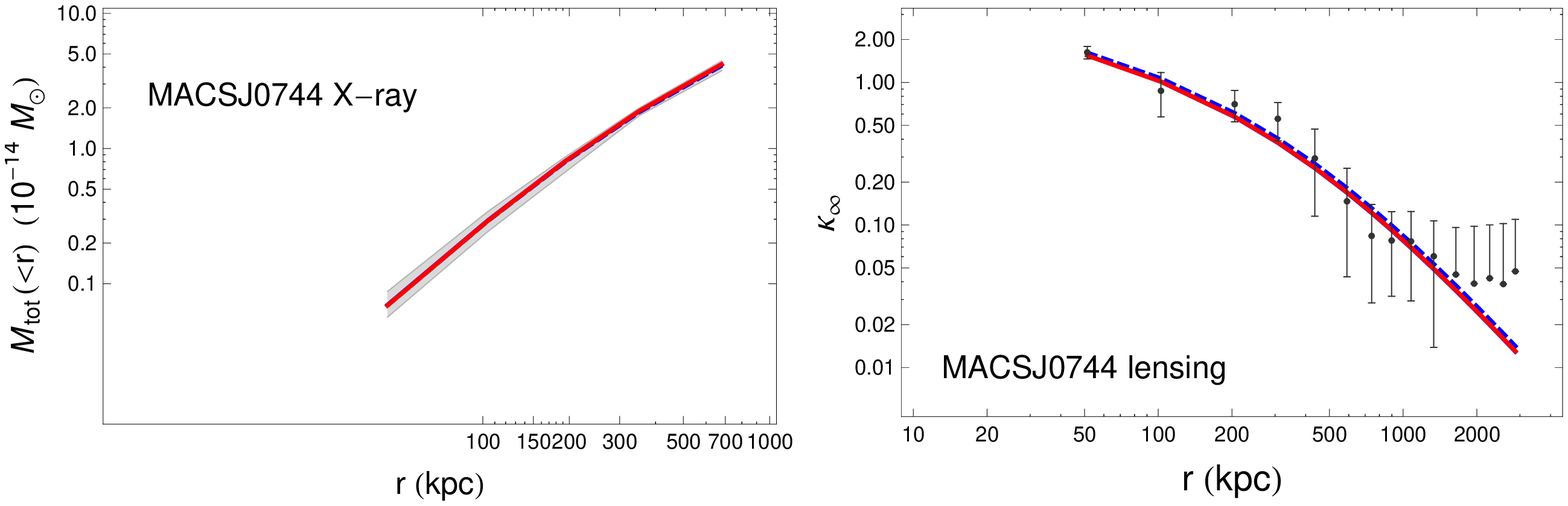}
\caption{Mass profiles from thermal X-ray gas (left) and gravitational lensing reconstruction (right). Color code: grey regions/points - observational data; dashed blue - NFW + GR fit from gas-only (right)/lensing-only (left); solid blue - NFW + GR from joint fit; solid red - NFW + galileon from joint fit.}
%\caption{\textcolor{red}{Non-``normalized'' X-ray mass profiles}. On the left, thermal X-ray total mass estimation; on the right, gravitational lensing reconstruction. Color code: grey regions/points - observational data; solid blue - density: NFW (in GR); dashed blue - density: NFW + gas (in GR); dotdashed blue - density: NFW + gas + BCG (in GR); solid red - density: NFW + galileon; dashed red: NFW + gas + galileon; dotdashed red: NFW + gas + BCG + galileon.}\label{fig:lens_gas_5}
\end{figure*}

%\begin{figure*}[htbp]
%\centering
%\includegraphics[width=9.cm]{likelihood.eps}
%\caption{Probability densities for $\Upsilon$. Grey: probability density for each single cluster; solid black: joint probability density for clusters from group on; vertical dot-dashed black line: limit on $\Upsilon$ from \citep{SaksteinPRL15}. All probability densities are normalized in order to have total area equal to one.}\label{fig:upsilon_joint}
%\end{figure*}

\begin{figure*}[htbp]
\centering
\includegraphics[width=17.cm]{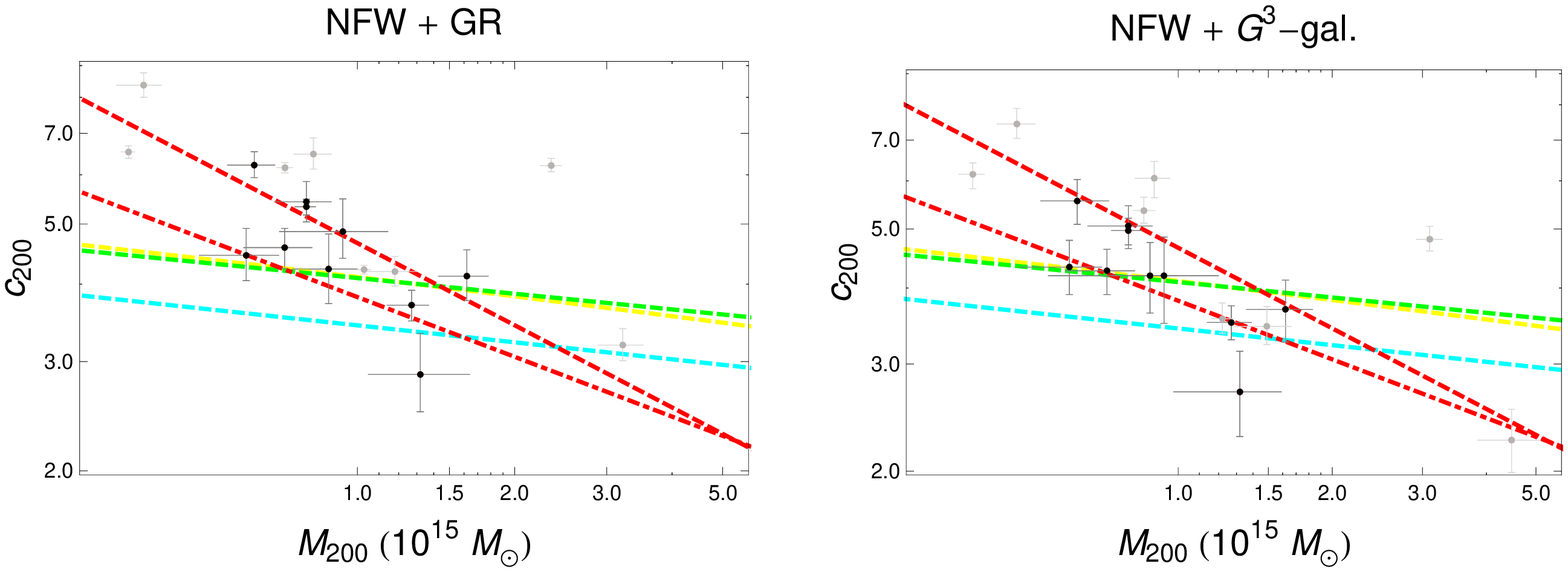}
\caption{Concentration and masses for CLASH clusters, derived from a NFW profile in GR (left panel) and $G^{3}$-galileon model (right panel). Black point are clusters from our group one; light gray point are clusters from groups two and three. Dashed colored lines are mass-concentration relations from numerical simulations for relaxed clusters: dashed cyan from \citep{Duffy08}; dashed yellow from \citep{Bhattacharya13}; dashed green from \citep{Meneghetti14}. Red lines are mass-concentration relation from lensing observations, fitting the $M_{200}-c_{200}$ relation when a NFW profile is used: dot-dashed from \citep{Merten15}; dashed from \citep{Umetsu15}. }\label{fig:Mass_concentratio}
\end{figure*}

\begin{figure*}[htbp]
\centering
\includegraphics[width=11.cm]{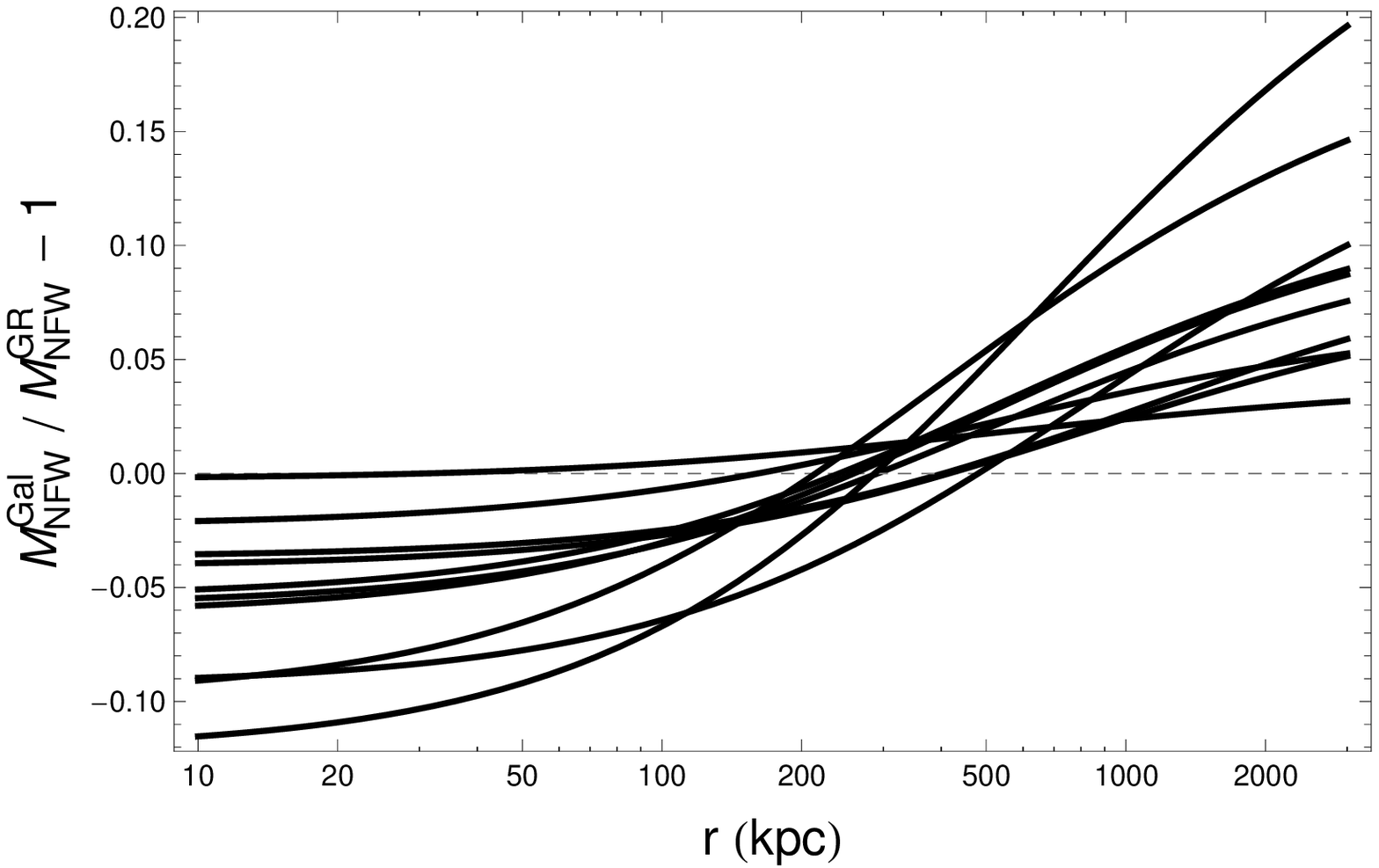}
\caption{Relative difference between GR and $G^{3}$-galileon for total NFW mass.)}\label{fig:Mass_ratio}
\end{figure*}

\end{document}